\documentclass[aps,prd,nofootinbib,groupedaddress,preprintnumbers]{revtex4}
\usepackage{dcolumn}
\usepackage{latexsym}
\usepackage{graphicx}
\usepackage{pstricks}
\usepackage{epsfig}
\usepackage{amsmath,amssymb,dsfont}

\input epsf

\newcommand{\be}{\begin{equation}}
\newcommand{\ee}{\end{equation}}
\newcommand{\bea}{\begin{eqnarray}}
\newcommand{\eea}{\end{eqnarray}}

\newcommand{\lbl}[1]{\label{eq:#1}}
\newcommand{ \rf}[1]{(\ref{eq:#1})}

\newcommand{\dl}{\langle\!\langle}
\newcommand{\dr}{\rangle\!\rangle}

\def\l{\langle}
\def\r{\rangle}

\newskip\humongous \humongous=0pt plus 1000pt minus 1000pt

\newif\ifdtup

\newcommand{\lapprox}{%
\mathrel{%
\setbox0=\hbox{$<$}
\raise0.6ex\copy0\kern-\wd0
\lower0.65ex\hbox{$\sim$}
}}
\newcommand{\gapprox}{%
\mathrel{%
\setbox0=\hbox{$>$}
\raise0.6ex\copy0\kern-\wd0
\lower0.65ex\hbox{$\sim$}
}}

\def\theequation{\arabic{section}.\arabic{equation}}

\begin{document}
\allowdisplaybreaks
\begin{flushright}
{\small
LMU-ASC~13/20\\ 
SI-HEP-2020-08\\
P3H-20-012\\ 
FERMILAB-PUB-20-140-T\\
April 2020\\
}
\end{flushright}

\vspace*{2cm}

\title{Higgs-Electroweak Chiral Lagrangian:\\
One-Loop Renormalization Group Equations}

\author{G. Buchalla$^a$, O. Cat\`a$^b$, A. Celis$^a$, 
M. Knecht$^c$, C. Krause$^d$}

\affiliation{\vspace*{0.4cm}
$^a$Ludwig-Maximilians-Universit\"at M\"unchen, 
Fakult\"at f\"ur Physik,\\
Arnold Sommerfeld Center for Theoretical Physics, 
D-80333 M\"unchen, Germany\\
\vspace*{-0.2cm}\\
$^b$Theoretische Physik 1, Universit\"at Siegen,\\
Walter-Flex-Stra\ss e 3, D-57068 Siegen, Germany\\
\vspace*{-0.2cm}\\
$^c$Centre de Physique Th\'{e}orique,\\ 
CNRS/Aix-Marseille Univ./Univ. du Sud Toulon-Var (UMR 7332)\\
CNRS-Luminy Case 907, 13288 Marseille Cedex 9, France\\
\vspace*{-0.2cm}\\
$^d$Theoretical Physics Department, Fermi National Accelerator Laboratory,\\
Batavia, IL, 60510, USA\\
\vspace*{0.5cm}}


\begin{abstract}
  Starting from the one-loop divergences we obtained previously, we work out the renormalization
  of the Higgs-Electroweak Chiral Lagrangian explicitly and in detail. This includes the
  renormalization of the lowest-order Lagrangian, as well as the decomposition of the remaining
  divergences into a complete basis of next-to-leading-order counterterms. We provide the list of the
  corresponding beta functions. We show how our results match the one-loop renormalization
  of some of the dimension-6 operators in SMEFT.
  We further point out differences with related work in the
  literature and discuss them. As an application of the obtained results, we evaluate the divergences
  of the vacuum expectation value of the Higgs field at one loop and show that they
  can be appropriately removed by the corresponding renormalization. We also work out the finite
  renormalization required to keep the no-tadpole condition on the Higgs field at one loop.
\end{abstract}

\maketitle

\vfill
\newpage


\section{Introduction}
\label{sec:intro}

Exploring the Higgs sector at the Large Hadron Collider (LHC) 
requires a parametrization of Higgs-boson properties, able to
account for effects beyond the Standard Model (SM) in a
well-defined way.
This is frequently done using the $\kappa$-framework, where 
SM couplings of the Higgs boson are rescaled by phenomenological
factors \cite{LHCHiggsCrossSectionWorkingGroup:2012nn}.
The electroweak chiral Lagrangian including a light Higgs 
boson (EWChL for short) provides us with a systematic 
effective field theory (EFT) formulation of this idea 
\cite{Buchalla:2015wfa} (see also \cite{deBlas:2018tjm} for a detailed discussion of the precise
relationship between the $\kappa$ parameters and EWChL coefficients).
The power counting of the EWChL is governed by chiral dimensions,
equivalent to an expansion in loop orders. The leading-order
Lagrangian ${\cal L}_2$, of chiral dimension 2, naturally
contains the dominant anomalous Higgs couplings.

Several authors have contributed to the development of the chiral
Lagrangian with a light Higgs boson as an EFT of electroweak
symmetry breaking \cite{Feruglio:1992wf,Bagger:1993zf,Koulovassilopoulos:1993pw,
Burgess:1999ha,Wang:2006im,Grinstein:2007iv,Contino:2010mh,Contino:2010rs,
Alonso:2012px,Alonso:2012pz},
which includes the complete classification of the next-to-leading 
order terms \cite{Buchalla:2013rka} and a systematic review of its
power counting \cite{Buchalla:2013eza}.
Motivated by the importance of the EWChL as an EFT of anomalous
Higgs couplings, we have computed its complete one-loop
renormalization in a previous article \cite{Buchalla:2017jlu}.
Similar work was reported shortly thereafter in \cite{Alonso:2017tdy}. 
The consideration of one-loop corrections is, in particular, needed
when treating subleading effects, which are of interest for their
impact on decay distributions.

Since the leading-order Lagrangian ${\cal L}_2$ of the EWChL is non-renormalizable,
the one-loop renormalization, besides affecting ${\cal L}_2$ itself, requires the
addition of counterterms with chiral dimension 4.  
In the present paper, we investigate the one-loop divergences
of the electroweak chiral Lagrangian in detail. 
This includes their decomposition into a basis of counterterms,
their renormalization, and the derivation of renormalization group
equations (RGEs) for all parameters in explicit terms.
This work parallels similar calculations done in the past in the
context of chiral perturbation theory for pions and kaons coupled 
to photons \cite{Urech:1994hd} and light fermions \cite{Knecht:1999ag},
extending the original results of \cite{Gasser:1983yg}.

The outline of the present paper is as follows.
We set up our notation in Section \ref{sec:notation}.
In Section \ref{sec:decomposition} we decompose the one-loop
divergences into the various classes of basis operators.
We work out the renormalization of the leading-order Lagrangian
and the next-to-leading order counterterms in Sections 
\ref{sec:LO_ren} and \ref{sec:NLO_ren}, respectively, deriving also 
the one-loop RGEs for all parameters of the theory.
Section \ref{sec:compsmeft} presents a comparison with the 
one-loop renormalization of SM Effective Field Theory (SMEFT),
which provides us with additional cross-checks of our results.
A brief survey of the literature on the renormalization of the
EWChL is given in Section \ref{sec:literature}.
As an application, we treat the one-loop corrections to the
vacuum expectation value (vev) of the Higgs field in 
Section \ref{sec:applications}, 
showing how it is renormalized and computing the finite terms.
We conclude in Section \ref{sec:concl}.
Three Appendices are included. Appendix~\ref{sec:eom} collects the equations of motion,
Appendix~\ref{sec:basis} the relevant basis operators at NLO and Appendix~\ref{sec:oneloopdiv} a 
summary of the one-loop divergences computed in \cite{Buchalla:2017jlu}.


\section{Notation and leading-order Lagrangian}
\label{sec:notation}
\setcounter{equation}{0}

In this paper we discuss the one-loop renormalization of the EWChL at lowest order,
which can be written as~\cite{Buchalla:2013rka,Buchalla:2013eza} 
\begin{eqnarray}\label{l2}
{\cal L}_2 &=& -\frac{1}{4} G_{\mu\nu}^\alpha G^{\alpha\mu\nu}
-\frac{1}{2}\langle W_{\mu\nu}W^{\mu\nu}\rangle 
-\frac{1}{4} B_{\mu\nu}B^{\mu\nu}
+\frac{v^2}{4}\ \langle L_\mu L^\mu \rangle \, F \left( \frac{h}{v} \right)+
\frac{1}{2} \partial_\mu h \partial^\mu h - V \left( \frac{h}{v} \right)
\nonumber\\
&& 
+\bar\psi i\!\not\!\! D\psi - \bar\psi m(h/v,U)\psi
,
\end{eqnarray}
starting from the one-loop divergences obtained in 
Ref.~\cite{Buchalla:2017jlu}. In this expression, $G^\alpha_{\mu}$, 
$W^a_{\mu}$ and $B_{\mu}$ are the gauge fields of $SU(3)_C$, $SU(2)_L$ and $U(1)_Y$, 
respectively, $h$ denotes the Higgs field and $v=246\,{\rm GeV}$ is the electroweak scale. 
The $SU(2)$ trace is denoted by $\langle\ldots\rangle$. 

The SM fermions are collected in the field 
$\psi=(u_i,d_i,\nu_i,e_i)^T$, where $u_i,d_i,\nu_i$ and $e_i$ are Dirac spinors and $i$ is the generation index. The covariant derivative on the fermion field reads
\begin{equation}\label{dcovpsi}
D_\mu \psi=\left(\partial_\mu +i g_s G_\mu + i g W_\mu P_L +
i g' B_\mu (Y_L P_L + Y_R P_R)\right)\psi 
,
\end{equation}
where $P_L$, $P_R$ are the left and right chiral projectors and the weak
hypercharge is described by the diagonal matrices
\begin{equation}
Y_L={\rm diag}(1/6,1/6,-1/2,-1/2)\, ,\qquad
Y_R={\rm diag}(2/3,-1/3,0,-1).
\end{equation}
We will also use the notation $\psi_L\equiv P_L\psi$, $\psi_R\equiv P_R\psi$
and $q=(u_i,d_i)^T$, $l=(\nu_i,e_i)^T$.

The electroweak Goldstone bosons are parametrized as $U=\exp(2i\varphi/v)$, 
where $\varphi=\varphi^a T^a$ and $T^a$ denote the generators of $SU(2)$, 
normalized as $\langle T^a T^b\rangle=\delta^{ab}/2$. It is convenient 
to define their covariant derivative as
\begin{equation}\label{dcovu}
L_\mu=i UD_\mu U^\dagger ,
\quad D_\mu U=\partial_\mu U + i g W_\mu U -i g' B_\mu \tau_L U ,
\quad \tau_L = U T^3 U^\dagger
.
\end{equation} 
The Yukawa term has been compactly expressed through
\begin{equation}\label{myukawa}
m(\eta,U)\equiv U {\cal M} (\eta) P_R 
+{\cal M}^\dagger (\eta) U^\dagger P_L
,
\end{equation}
where $\eta\equiv h/v$ and ${\cal M}$ is the block-diagonal mass matrix
\begin{equation}\label{mmdef}
{\cal M}={\rm diag}({\cal M}_u,{\cal M}_d,{\cal M}_\nu,{\cal M}_e) 
\end{equation}
acting on $\psi$. The entries ${\cal M}_f\equiv{\cal M}_f(\eta)$ 
are $h$-dependent matrices in generation space. It is understood 
that the right-handed neutrinos are absent when we assume SM particle content. 
Accordingly, we will take ${\cal M}_\nu=0$ in our calculation. 

The Higgs-dependent functions can be expanded as
\begin{equation}\label{fvdef}
F(\eta)=1+\sum^\infty_{n=1} F_{n} \eta^n\, ,\qquad 
V(\eta)=v^4\sum^\infty_{n=2} V_{n} \eta^n\, ,\qquad
{\cal M}(\eta)=\sum^\infty_{n=0} {\cal M}_{n} \eta^n
,
\end{equation}
where the potential is defined such that $V^\prime (0) = 0$.

In order to express our results in a compact way we will use the combinations
\begin{equation}\label{kapdef}
\kappa\equiv \frac{1}{2}F' F^{-1/2}\, ,\qquad 
{\cal B}\equiv -F^{-1/2}\kappa'=\frac{F'^2}{4 F^2}-\frac{F''}{2F}
.
\end{equation}
Here and in the following, a prime on $\eta$-dependent functions 
denotes differentiation with respect to this variable. Of particular
interest are the values of these functions for $\eta = 0$,
\begin{equation}
F (0) = 1 ,\quad \kappa (0) = F_1/2, \quad {\cal B} (0) = \frac{F_1^2}{4} - F_2
,\quad V(0) = V' (0) = 0  ,\quad V''(0) = 2 v^4 V_2 = v^2 m^2_h
,\quad {\cal M} (0) = {\cal M}_0
,
\end{equation}
and the deviations from these values,
\begin{equation}
{\bar F} (\eta) = F (\eta) - 1 , \quad {\bar\kappa} (\eta) = \kappa (\eta) - F_1/2
, \quad {\bar{\cal B}} = {\cal B} - \frac{F_1^2}{4} + F_2
,\quad {\bar{\cal M}} = {\cal M} - {\cal M}_0
.
\end{equation}


\section{Decomposition of the one-loop divergences on 
a basis of counterterms}\label{sec:decomposition}
\setcounter{equation}{0}

The first step will be to decompose the one-loop divergences worked out in
Ref. \cite{Buchalla:2017jlu} and project them onto the basis 
of one-loop counterterms derived in Ref. \cite{Buchalla:2013rka}.
Some of the divergences displayed 
in Ref. \cite{Buchalla:2017jlu} actually have the same structure as the
lowest-order effective Lagrangian ${\cal L}_2$, 
and are thus dealt with by means of a set of counterterms $\Delta {\cal L}_2$,
involving only the structures already present in ${\cal L}_2$.

It is convenient to decompose the one-loop divergent parts according to the
presence of spin-one field strengths, scalar fields or fermion fields as
\be
{\cal L}_{\rm div} = {\cal L}_{\rm div}^{(1)} + {\cal L}_{\rm div}^{(0)}
+ {\cal L}_{\rm div}^{(1/2)}
,
\lbl{Ldiv_decomp}
\ee
where the explicit expressions for each term on the right-hand side can be found in 
Eqs. \rf{Ldiv1}, \rf{Ldiv0}, and \rf{Ldiv1/2}, respectively, in Appendix \ref{sec:oneloopdiv}.
We proceed with each term in turn. 

In manipulating the one-loop divergent pieces, one is entitled to make use
of the equations of motion at leading order, which are collected in Appendix \ref{sec:eom}.
One reason for this is that the divergences
are expressed in terms of the classical background fields. One may thus
use the classical, lowest-order, equations of motion as long as the 
corresponding counterterms are inserted into tree-level diagrams only, which 
is the case for the computation of amplitudes at next-to-leading accuracy. 
More generally, the use of the equations of motion can also be implemented as 
a field redefinition, which does not change the S-matrix elements.
For a general discussion on this issue, see e.g. Ref. \cite{Politzer:1980me,Arzt:1993gz}.
As a practical consequence, it is therefore not necessary to introduce next-to-leading
counterterms corresponding to the field renormalizations for fermions and the Higgs boson.

In the results collected in Eqs. \rf{Ldiv1}, \rf{Ldiv0}, and \rf{Ldivpsi}, which are taken
from Ref. \cite{Buchalla:2017jlu}, the equations of motion had already been used for
the scalar fields $U$ and $h$. In Appendix \ref{sec:oneloopdiv} we extend this to fermions.

\subsection{Decomposition of ${\cal L}_{\rm div}^{(1)}$}
\label{subsec:decompl1}

From Eq. \rf{Ldiv1}, one has ($N_c$ stands for the number of colors,
$N_f$ for the number of quark flavors,
and $N_{\rm g}$ for the number of generations)
\bea
{\cal L}_{\rm div}^{(1)} &=&
- \frac{1}{16\pi^2} \frac{1}{d-4}\bigg\{
\frac{g^2_s}{2}\, G^{\alpha\mu\nu}G_{\mu\nu}^\alpha  \left(\frac{11}{3} N_c
  - \frac{2}{3} N_f   \right)
+ \frac{g^2}{2} \langle W^{\mu\nu}W_{\mu\nu} \rangle  \left[ \frac{43}{3}
  - \frac{2}{3} (N_c+1) N_{\rm g} - \frac{F_1^2 - 4}{24} \right]
\nonumber\\
&&
+\frac{g'^2}{2} \, B^{\mu\nu} B_{\mu\nu}  \left[ -\left(\frac{11}{27} N_c
    + 1\right) N_{\rm g} - \frac{1}{6} - \frac{F_1^2 - 4}{48} \right]
- \,  \frac{1}{24} {\bar\kappa} ({\bar\kappa} + F_1  ) \left[
g'^2  B^{\mu\nu} B_{\mu\nu} + 2 g^2 \langle W^{\mu\nu}W_{\mu\nu}\rangle
\right]
\nonumber\\
&&
- i \frac{{\bar\kappa} ({\bar\kappa} + F_1  )}{12}
\left(g\langle W^{\mu\nu}[L_\mu,L_\nu]\rangle
+ g' B^{\mu\nu} \langle\tau_L [L_\mu,L_\nu]\rangle+2igg' \langle\tau_L W^{\mu\nu}\rangle B_{\mu\nu}\right)
\nonumber\\
&&
- i \frac{F_1^2-4}{48}\left(g\langle W^{\mu\nu}[L_\mu,L_\nu]\rangle
+ g' B^{\mu\nu} \langle\tau_L [L_\mu,L_\nu]\rangle+2igg' \langle\tau_L W^{\mu\nu}\rangle B_{\mu\nu}\right)
\nonumber\\
&&
- \, \frac{1}{6}\partial^\mu (\kappa^2) \left(g\langle W_{\mu\nu}L^\nu\rangle
- g' B_{\mu\nu}\langle\tau_L L^\nu\rangle\right)\bigg\}
,
\eea
where we have separated explicitly terms with and without Higgs field dependence.

It is tempting to read off the one-loop beta functions for the gauge fields
directly from the coefficients of their kinetic terms, as given by the first
three terms above. This would however be incorrect
in the case of the $SU(2)$ and $U(1)$ couplings, due to the presence of the
terms in the fourth line. In order to reach the right result,
one may use, in the first two terms of the fourth line, the relations
\bea
\langle W^{\mu\nu}[L_\mu,L_\nu]\rangle &=&
ig \langle W^{\mu\nu}W_{\mu\nu}\rangle
- ig' B^{\mu\nu} \langle \tau_L W_{\mu\nu} \rangle
+ ig {\bar\psi}_L \not\!\!L \psi_L - ig \frac{v^2}{2} F (\eta)
\langle L^\nu L_\nu \rangle
,\\
B^{\mu\nu} \langle\tau_L [L_\mu,L_\nu]\rangle &=&
\frac{i g'}{2} B^{\mu\nu} B_{\mu\nu}
- ig B^{\mu\nu} \langle \tau_L W_{\mu\nu} \rangle 
- 2 ig' \langle \tau_L L^\nu \rangle
\left( {\bar\psi} \gamma_\nu (Y_L P_L + Y_R P_R) \psi
+ \frac{v^2}{2} F(\eta) \langle L_\nu \tau_L \rangle
\right),
\eea
which follow from the identities \cite{Buchalla:2013rka}
\bea
\lbl{identities}
&&
D_\mu L_\nu - D_\nu L_\mu =
g W_{\mu\nu} - g' B_{\mu\nu} \tau_L + i [ L_\mu , L_\nu ]
, \qquad\quad\quad
D_\mu \tau_L = i [ L_\mu , \tau_L ]
,
\nonumber\\
\\
&&
\langle T^a L^\nu \rangle \langle T^a L_\nu \rangle =
\frac{1}{2} \langle L^\nu L_\nu \rangle
, \qquad\quad\quad\quad\quad
\langle L^\nu T^a  \rangle {\bar\psi}_L \gamma_\nu T^a   \psi_L =
\frac{1}{2} {\bar\psi}_L \not\!\!L \psi_L
,
\nonumber
\eea
from the equations of motion given in Appendix \ref{sec:eom},
and from integrating by parts. Because of the last point, they can only be
used in the form given here if the operators on the left-hand side are not
multiplied by $h$-dependent functions.
The expression of ${\cal L}_{\rm div}^{(1)}$ then can be rewritten as
\bea
\lbl{divpartial}
{\cal L}_{\rm div}^{(1)} &=&
- \frac{1}{16\pi^2} \frac{1}{d-4}\bigg\{
\frac{g^2_s}{2}\, G^{\alpha\mu\nu}G_{\mu\nu}^\alpha  \left(\frac{11}{3} N_c
  - \frac{2}{3} N_f   \right)
+ \frac{g^2}{2} \langle W^{\mu\nu}W_{\mu\nu} \rangle  \left[ \frac{43}{3}
  - \frac{2}{3} (N_c+1) N_{\rm g} \right]
\nonumber\\
&&
+\frac{g'^2}{2} \, B^{\mu\nu} B_{\mu\nu}  \left[ -\left(\frac{11}{27} N_c
    + 1\right) N_{\rm g} - \frac{1}{6} \right]
- \,  \frac{1}{24} {\bar\kappa} ({\bar\kappa} + F_1  ) \left[
g'^2  B^{\mu\nu} B_{\mu\nu} + 2 g^2 \langle W^{\mu\nu}W_{\mu\nu}\rangle
\right]
\nonumber\\
&&
- i \frac{{\bar\kappa} ({\bar\kappa} + F_1  )}{12}
\left(g\langle W^{\mu\nu}[L_\mu,L_\nu]\rangle
+ g' B^{\mu\nu} \langle\tau_L [L_\mu,L_\nu]\rangle+2igg' \langle\tau_L W^{\mu\nu}\rangle B_{\mu\nu}\right)
\nonumber\\
&&
+  \frac{F_1^2-4}{48}\left[g^2 {\bar\psi}_L \not\!\!L \psi_L
  - g^2 \frac{v^2}{2} F (\eta) \langle L^\nu L_\nu \rangle
  - 2 g^{\prime 2} \langle \tau_L L^\nu \rangle
  \left( {\bar\psi} \gamma_\nu (Y_L P_L + Y_R P_R) \psi
+ \frac{v^2}{2} F(\eta) \langle L_\nu \tau_L \rangle
\right)
\right]
\nonumber\\
&&
-  \frac{1}{6}\partial^\mu (\kappa^2) \left(g\langle W_{\mu\nu}L^\nu\rangle
-  g' B_{\mu\nu}\langle\tau_L L^\nu\rangle\right)
\bigg\}
.
\eea
The one-loop beta functions of the gauge fields are now correctly given
by the first three terms between the curly brackets, and they coincide with
the expressions obtained in the SM.
We next consider the two terms proportional to $\partial^\mu (\kappa^2)$
on the last line of the above expression. 
At this stage, we will need the equations of motion of the electroweak gauge
fields, given in Eqs. (\ref{eomb}) and (\ref{eomw}), as well as the identities
of Eq. \rf{identities}. Up to total-derivative terms, which are discarded, we obtain
\bea
\partial^\mu (\kappa^2) \langle W_{\mu\nu} L^\nu\rangle &=&
- \frac{{\bar\kappa}({\bar\kappa} + F_1)}{2} g \left[ - \frac{v^2}{2} F(\eta)
  \langle L^\nu L_\nu \rangle + {\bar\psi}_L \not\!\!L \psi_L \right]
\nonumber\\
&&
- \frac{{\bar\kappa}({\bar\kappa} + F_1)}{2} \left[
  g \langle W^{\mu\nu} W_{\mu\nu} \rangle -
  g' B^{\mu\nu} \langle W_{\mu\nu} \tau_L \rangle
+ i \langle W_{\mu\nu} [ L^\mu , L^\nu ] \rangle
\right]
,
\eea
and
\bea
\partial^\mu (\kappa^2) B_{\mu\nu}\langle \tau_L L^\nu\rangle &=&
- {\bar\kappa}({\bar\kappa} + F_1)
g' \left[ \frac{v^2}{2} \langle L_\nu \tau_L \rangle F(\eta)
+ {\bar\psi} \gamma_\nu (Y_L P_L + Y_R P_R) \psi
\right] \langle \tau_L L^\nu\rangle
\nonumber\\
&&
+ \frac{{\bar\kappa}({\bar\kappa} + F_1)}{2} B_{\mu\nu}
\langle  i \tau_L [ L^\mu , L^\nu ]
- g \tau_L W^{\mu\nu} + g' \tau^2_L B^{\mu\nu} \rangle
.
\eea
Making the corresponding substitutions in the above expression of
${\cal L}_{\rm div}^{(1)}$, we find
\bea
{\cal L}_{\rm div}^{(1)} &=&
- \frac{1}{16\pi^2} \frac{1}{d-4}\bigg\{
\frac{g^2_s}{2}\, G^{\alpha\mu\nu}G_{\mu\nu}^\alpha  \left(\frac{11}{3} N_c
  - \frac{2}{3} N_f   \right) 
+\frac{g^2}{2} \langle W^{\mu\nu}W_{\mu\nu} \rangle \left[ \frac{43}{3}
  - \frac{2}{3} (N_c+1) N_{\rm g} \right]
\nonumber\\
&&
+ \frac{g'^2}{2} \, B^{\mu\nu} B_{\mu\nu}  \left[ - \left(\frac{11}{27} N_c
    + 1\right) N_{\rm g} - \frac{1}{6} \right]
- g^2 \frac{v^2}{24} (\kappa^2 - 1) F(\eta) \langle L^\nu L_\nu \rangle
\nonumber\\
&&
+ \frac{\kappa^2 - 1}{12} g^2 {\bar\psi}_L \not\!\!L \psi_L
- \frac{\kappa^2 - 1}{6} g'^2
\left[ \frac{v^2}{2} \langle L_\nu \tau_L \rangle F(\eta)
+ {\bar\psi} \gamma_\nu (Y_L P_L + Y_R P_R) \psi
\right] \langle \tau_L L^\nu\rangle
\bigg\}
\nonumber\\
&\equiv&
\Delta^{(1)} {\cal L}_2
+ \Delta^{(1)} {\cal L}_{\beta_1}
+ \Delta^{(1)} {\cal L}_{\psi^2UhD}
.
\eea 
The term
\bea
\Delta^{(1)} {\cal L}_2 &=&
- \frac{1}{16\pi^2} \frac{1}{d-4}\bigg\{
- \frac{g^2_s}{4}\, G^{\alpha\mu\nu}G_{\mu\nu}^\alpha
\left(- \frac{22}{3} N_c + \frac{4}{3} N_f   \right)
- \frac{g^2}{2} \langle W^{\mu\nu}W_{\mu\nu} \rangle  \left[ - \frac{44}{3}
  + \frac{2}{3} (N_c+1) N_{\rm g} + \frac{1}{3} \right]
\nonumber\\
&&
-\frac{g'^2}{4} \, B^{\mu\nu} B_{\mu\nu}  \left[ 2 \left(\frac{11}{27} N_c
    + 1\right) N_{\rm g} + \frac{1}{3} \right]
- g^2 \frac{v^2}{24} (\kappa^2 - 1) F (\eta) \langle L_\mu L^\mu \rangle
\bigg\}
\lbl{Delta1L2}
\eea
renormalizes the gauge-field and Goldstone-boson kinetic terms in ${\cal L}_2$.
The two last terms of ${\cal L}_{\rm div}^{(1)}$, 
to which we will return later, read
\be
\Delta^{(1)} {\cal L}_{\beta_1} = - \frac{1}{16\pi^2} \frac{1}{d-4}
\times g'^2 \frac{v^2}{2} F(\eta)
\frac{1-\kappa^2}{6}
\langle L^\nu \tau_L \rangle \langle L_\nu \tau_L \rangle
\ee
and
\be
\Delta^{(1)} {\cal L}_{\psi^2UhD} = - \frac{1}{16\pi^2} \frac{1}{d-4}
\frac{\kappa^2 - 1}{12}
\bigg\{
g^2 {\bar\psi}_L \not\!\!L \psi_L - 2 g'^2 \langle \tau_L L^\nu \rangle
{\bar\psi} \gamma_\nu (Y_L P_L + Y_R P_R) \psi
\bigg\}
.
\ee
$\Delta^{(1)} {\cal L}_{\beta_1}$ 
renormalizes the custodial-symmetry breaking operator ${\cal{L}}_{\beta_1}$, while 
$\Delta^{(1)} {\cal L}_{\psi^2UhD}$ renormalizes some of the counterterms of the 
class $\psi^2UhD$, respectively. We refer to Appendix~\ref{sec:basis} for 
the detailed definition of operators and their classes. We employ the 
nomenclature of Ref. \cite{Buchalla:2013rka}, which will be used throughout. 
Incidentally, we note that the divergences involving the next-to-leading order operators of
the classes $X^2Uh$ and $XUhD^2$ have canceled.

Finally, in order to bring all the terms in correspondence to the operator
basis retained in Ref. \cite{Buchalla:2013rka}, the quantities of the type
${\bar\psi}_{L} O_1 \! \not\!\!L O_2 \psi_{L}$ can be transformed as follows
($P_{12} = T^1 + i T^2$, $P_{21} = T^1 - i T^2$):
\bea
{\bar\psi}_{L} O_1 \! \not\!\!L O_2 \psi_{L} &=& 2 {\bar\psi}_{L} \gamma^\mu O_1 U T^a U^\dagger O_2 \psi_{L} \langle U T^a U^\dagger L_\mu \rangle
\nonumber\\
&=&
{\bar\psi}_{L} \gamma^\mu O_1 U P_{12} U^\dagger O_2 \psi_{L} \langle U P_{21} U^\dagger L_\mu \rangle
+
{\bar\psi}_{L} \gamma^\mu O_1 U P_{21} U^\dagger O_2 \psi_{L} \langle U P_{12} U^\dagger L_\mu \rangle
\nonumber\\
&&
+ 2 {\bar\psi}_{L} \gamma^\mu O_1 \tau_L O_2 \psi_{L} \langle \tau_L L_\mu \rangle
.
\lbl{identity_slashL}
\eea
Using the previous relation, we find
\bea
\Delta^{(1)} {\cal L}_{\psi^2UhD} &=& - \frac{1}{16\pi^2} \frac{1}{d-4}
\frac{\kappa^2 - 1}{12}
\bigg\{
2 g^2 {\bar\psi}_L \tau_L \gamma^\mu \psi_L \langle \tau_L L_\mu \rangle
\nonumber\\
&&
+ g^2 {\bar\psi}_L U P_{12} U^\dagger \gamma^\mu \psi_L \langle U P_{21} U^\dagger L_\mu \rangle
+ g^2 {\bar\psi}_L U P_{21} U^\dagger \gamma^\mu \psi_L \langle U P_{12} U^\dagger L_\mu \rangle
\nonumber\\
&&
- 2 g'^2 \langle \tau_L L^\nu \rangle  {\bar\psi}_L \gamma_\nu Y_L \psi_L
- \frac{4}{3} g'^2 \langle \tau_L L^\nu \rangle  {\bar u}_R \gamma_\nu  u_R
+ \frac{2}{3} g'^2 \langle \tau_L L^\nu \rangle  {\bar d}_R \gamma_\nu  d_R
+ 2 g'^2 \langle \tau_L L^\nu \rangle  {\bar e}_R \gamma_\nu  e_R
\bigg\}
.~~~~~
\eea

\subsection{Decomposition of ${\cal L}_{\rm div}^{(0)}$}

We consider next the divergent pieces involving spin-zero fields given in Eq. \rf{Ldiv0},
which we rewrite as
\be
{\cal L}_{\rm div}^{(0)} = \Delta^{(0)} {\cal L}_2 +
\Delta^{(0)} {\cal L}_{UhD^4}+ \Delta^{(0)} {\cal L}_{\beta_1},
\ee
where ($\dl\ldots\dr$ denotes a trace over isospin, as well as
generations and color)
\bea
\Delta^{(0)} {\cal L}_2 &=&  
- \frac{1}{16\pi^2} \frac{1}{d-4}\bigg\{
\frac{v^2}{4} \left[ - \frac{g'^2}{8} (F_1^2 + 12)
  - \frac{3 g^2}{4} (F_1^2 + 4) + V_2 ( F_1^2 - 4 F_2)
+ \frac{4}{v^2} \dl {\cal M}_0^\dagger {\cal M}_0\dr 
\right.
\nonumber\\
&&
\left.
\ \, \qquad
- \, \frac{g'^2}{8} (F_1^2 + 12) {\bar F} 
- \frac{g'^2}{2} {\bar\kappa}({\bar\kappa} + F_1) F  - \frac{3 g^2}{4} (F_1^2 + 4) {\bar F}
- 3 g^2 {\bar\kappa}({\bar\kappa} + F_1) F 
\right.
\nonumber\\
&&
\left.
\ \, \qquad
- 2 \, (\kappa^2 - 1) \frac{F'V'}{Fv^4} 
+ 2 \frac{(V'' - 2 v^4 V_2)}{v^4} F{\cal B} + 4  V_2 ({\cal B} {\bar F} + {\bar{\cal B}})
+ \frac{4}{v^2} \dl {\overline {\cal M}}^\dagger {\cal M}_0 
+  {\cal M}_0^\dagger {\overline{\cal M}} +
{\overline {\cal M}}^\dagger {\overline{\cal M}}\dr \right] 
\langle L^\mu L_\mu\rangle
\nonumber\\
&&
+ \, \frac{1}{2} \left[ - \frac{3 F_1^2 + 4 F_2}{8} (3g^2+g'^2) +
  \frac{4}{v^2} \dl {\cal M}_1^\dagger {\cal M}_1 \dr 
\right.
\nonumber\\
&&
\left.
\  \qquad
+ \, \frac{1}{2} (3g^2+g'^2) ( {\bar F} {\cal B} + {\bar{\cal B}} -
4 {\bar\kappa}^2 - 4 F_1 {\bar\kappa} )
+ 3 \frac{F'V'}{F v^4} {\cal B} + \frac{4}{v^2} 
\dl {\overline{{\cal M}'}}^\dagger {\cal M}_1 +
{\cal M}_1^\dagger {\overline{{\cal M}'}} +   
{\overline{{\cal M}'}}^\dagger {\overline{{\cal M}'}} \dr \right]
\partial^\mu h \partial_\mu h
\nonumber\\
&&
+\frac{3}{2}(3 g^4 + 2 g^2 g'^2 + g'^4)\frac{v^4}{16} F^2  
+\frac{3 g^2 + g'^2}{8}F'V'+
\frac{3}{8}\left(\frac{F'V'}{F v^2}\right)^2
+ \frac{1}{2}\left(\frac{V''}{v^2}\right)^2
- 2 \dl ({\cal M}^\dagger {\cal M})^2\dr
\nonumber\\
&&
+ 4i \langle \tau_L L_\mu \rangle \,
\dl ( \partial^\mu {\mathcal M}^\dagger {\mathcal M} -
{\mathcal M}^\dagger \partial^\mu {\mathcal M}) T^3 \dr
\bigg\}
\lbl{Delta0L2}
,
\eea
\bea
\Delta^{(0)} {\cal L}_{UhD^4} &=& 
- \frac{1}{16\pi^2} \frac{1}{d-4}\bigg\{
\frac{(\kappa^2-1)^2}{6} \langle L_\mu L_\nu\rangle \langle L^\mu L^\nu\rangle
+\left(\frac{(\kappa^2-1)^2}{12}+\frac{F^2{\cal B}^2}{8}\right)
\langle L^\mu L_\mu\rangle^2 
+\frac{2}{3}\kappa'^2\langle L_\mu L_\nu\rangle \partial^\mu\eta\partial^\nu\eta
\nonumber\\
&&-\left((\kappa^2-1){\cal B}+\frac{\kappa'^2}{6}\right)
\langle L^\mu L_\mu\rangle \partial^\nu\eta\partial_\nu\eta
+\frac{3}{2}{\cal B}^2(\partial^\mu\eta\partial_\mu\eta)^2
\bigg\}
,
\\
\Delta^{(0)} {\cal L}_{\beta_1} &=&  
- \frac{1}{16\pi^2} \frac{1}{d-4}\times
\frac{3}{4} g'^2 v^2 (1-\kappa^2)F\,\langle\tau_LL^\mu\rangle\langle\tau_LL_\mu\rangle
.
\eea
The two last terms renormalize the counterterms of the class $UhD^4$ and the custodial-symmetry
breaking operator ${\cal{L}}_{\beta_1}$, respectively. $\Delta^{(0)} {\cal L}_{UhD^4}$ comes already 
fully expressed in terms of the basis elements displayed in Ref. \cite{Buchalla:2013rka}.

The last term of Eq. \rf{Delta0L2} does not naturally appear as a renormalization of ${\mathcal L}_2$,
but can be shown to renormalize the Yukawa term.
Using the equation of motion for $B_\mu$ in (\ref{eomb}), one may write
\be
4i \langle \tau_L L_\mu \rangle \,
\dl ( \partial^\mu {\mathcal M}^\dagger {\mathcal M} - {\mathcal M}^\dagger \partial^\mu {\mathcal M}) T^3 \dr =
\frac{8i}{v^2 F} \left[ \frac{1}{g'} \partial^\nu B_{\nu\mu} -  {\bar\psi} \gamma_\mu (Y_L P_L + Y_R P_R) \psi  \right]
\dl ( \partial^\mu {\mathcal M}^\dagger {\mathcal M} - {\mathcal M}^\dagger \partial^\mu {\mathcal M}) T^3 \dr
.
\ee
The first term between square brackets on the right-hand side of this relation leads
to a total derivative,
\bea
F^{-1} \partial^\nu B_{\nu\mu} 
\dl ( \partial^\mu {\mathcal M}^\dagger {\mathcal M} - {\mathcal M}^\dagger \partial^\mu {\mathcal M}) T^3 \dr
&=&
\partial^\nu \left[ F^{-1} B_{\nu\mu} 
\dl ( \partial^\mu {\mathcal M}^\dagger {\mathcal M} - {\mathcal M}^\dagger \partial^\mu {\mathcal M}) T^3 \dr \right]
\nonumber\\
&&\!\!\!\!\!
- B_{\nu\mu} \partial^\nu \left[ F^{-1} 
\dl ( \partial^\mu {\mathcal M}^\dagger {\mathcal M} - {\mathcal M}^\dagger \partial^\mu {\mathcal M}) T^3 \dr \right]
,
\eea
since the second term in the above relation vanishes, being proportional either to $B_{\mu\nu} (\partial^\mu \eta) (\partial^\nu \eta)$
or to $B_{\mu\nu} (\partial^\mu \partial^\nu \eta)$. One thus ends up with
\bea
4i \langle \tau_L L_\mu \rangle \, 
\dl ( \partial^\mu {\mathcal M}^\dagger {\mathcal M} - {\mathcal M}^\dagger \partial^\mu {\mathcal M}) T^3 \dr
&=&
\frac{8i}{v^2} 
\dl ( F^{-1/2} {\mathcal M}^\dagger \partial^\mu (F^{-1/2} {\mathcal M}) - \partial^\mu (F^{-1/2} {\mathcal M}^\dagger) F^{-1/2} {\mathcal M}  ) T^3 \dr
\nonumber\\
&&\qquad \times
{\bar\psi} \gamma_\mu (Y_L P_L + Y_R P_R) \psi
\nonumber\\
&=&
\frac{8i}{v^2} 
\dl ( F^{-1/2} {\mathcal M}^\dagger (F^{-1/2} {\mathcal M})' - (F^{-1/2} {\mathcal M}^\dagger)' F^{-1/2} {\mathcal M}  ) T^3 \dr
\nonumber\\
&&\qquad \times
(\partial^\mu \eta) {\bar\psi} \gamma_\mu (Y_L P_L + Y_R P_R) \psi
\nonumber\\
&=&
\frac{8i}{v^2} \partial^\mu \!\!\int_0^\eta \!\! ds 
\,
\dl ( F^{-1/2} {\mathcal M}^\dagger (F^{-1/2} {\mathcal M})' -
(F^{-1/2} {\mathcal M}^\dagger)' F^{-1/2} {\mathcal M}  ) T^3 \dr
\nonumber\\
&&\qquad \times
{\bar\psi} \gamma_\mu (Y_L P_L + Y_R P_R) \psi
\nonumber\\
&=&
\frac{8}{v^2} \!\!\int_0^\eta \!\! ds 
\,
\dl F^{-1/2} {\mathcal M}^\dagger (F^{-1/2} {\mathcal M})' T^3 - (F^{-1/2} {\mathcal M}^\dagger)' F^{-1/2} {\mathcal M} T^3 \dr
\nonumber\\
&&\qquad \times
{\bar\psi} [U {\cal M} (Y_R-Y_L) P_R - (Y_R-Y_L)  {\cal M}^\dagger U^\dagger P_L ] \psi
.
\eea
Here the integration variable $s$ has been introduced, which denotes
the dependence of the integrands on $h/v=s$.
In the last step, an integration by parts has been performed, the resulting 
total-derivative term has been dropped, and the equations of motion for the 
fermionic fields have been used. Notice that $Y_R-Y_L = T^3$.
Objects like
\be
\int_0^\eta ds \left[ F^{-1/2} {\cal M}^\dagger (F^{-1/2} {\cal M})' -
  (F^{-1/2} {\cal M}^\dagger)' F^{-1/2} {\cal M}  \right]
\ee
are perfectly well defined as formal power series in $\eta$, obtained upon
multiplication of the formal series defining, e.g.,
$F^{-1/2} (s) {\cal M}^\dagger (s)$ and $(F^{-1/2} (s) {\cal M} (s))'$,
and term-by-term integration.

\subsection{Decomposition of ${\cal L}_{\rm div}^{(1/2)}$}

Turning finally to the divergences involving also fermionic
fields, Eq. \rf{Ldiv1/2} gives
\bea
{\cal L}_{\rm div}^{(1/2)} &=& 
- \frac{1}{16\pi^2} \frac{1}{d-4} \bigg\{
\bar\psi_L\left(\frac{3}{2}g^2+2 g'^2 Y^2_L\right)i\!\not\!\! D\psi_L
+\bar\psi_R\, 2g'^2 Y^2_R i\!\not\!\! D\psi_R\nonumber\\
&&
+ 2 g^2_s C_F \, \bar q\left(i\!\not\!\! D 
-4(U{\cal M}_q P_R+{\cal M}^\dagger_q U^\dagger P_L)\right)q
\nonumber\\
&&+\frac{V''}{v^4}\left(\bar\psi_LU {\cal M}'' \psi_R +{\rm h.c.}\right)
-8 g'^2\left(\bar\psi_L Y_L U{\cal M}Y_R\psi_R +{\rm h.c.}\right)
\nonumber\\
&&
+\left((3g^2+g'^2)\frac{v^2}{4} F+\frac{3}{2}\frac{F'V'}{F v^2}\right)
\frac{F^{-1}}{v^2}
\left(
\bar\psi_L U\left(\frac{F'}{2}{\cal M}'-{\cal M}\right)\psi_R +{\rm h.c.}
\right)
\nonumber\\
&&
+ \frac{3}{v^2}F^{-1}\left(\bar\psi_L U {\cal M}{\cal M}^\dagger {\cal M}\psi_R + {\rm h.c.}\right)
-\frac{3}{v^2}F^{-1}\left(\bar\psi_L U \langle{\cal M}{\cal M}^\dagger \rangle {\cal M}\psi_R + {\rm h.c.}\right)
\nonumber\\
&&
+
\frac{2}{v^2}\left(\bar\psi_L U{\cal M}'{\cal M}^\dagger{\cal M}'\psi_R + {\rm h.c.}\right)
+
\frac{1}{2 v^2} \left( \bar\psi_L U {\cal M} {\cal M}'^\dagger {\cal M}' \psi_R + {\rm h.c.} \right)
+
\frac{1}{2 v^2} \left(
\bar\psi_L U {\cal M}' {\cal M}'^\dagger {\cal M} \psi_R + {\rm h.c.} \right)
\nonumber\\
&&
+ 
\frac{3 F^{-1}}{2 v^2}\bar\psi_R 
\left( {\cal M}^\dagger i\!\not\! \partial {\cal M} - i\!\not\! \partial {\cal M}^\dagger {\cal M} \right) \psi_R
+
\frac{1}{2 v^2} \bar\psi_R \left( {\cal M}'^\dagger i\!\not\! \partial {\cal M}'
- i\!\not\! \partial {\cal M}'^\dagger {\cal M}' \right) \psi_R
\nonumber\\
&&
- \frac{F^{-1}}{2 v^2} \bar\psi_L U \left( {\cal M} i \! \not\!\partial {\cal M}^\dagger
- i \! \not\!\partial {\cal M} {\cal M}^\dagger\right) U^\dagger \psi_L
+ \frac{1}{2 v^2}  \bar\psi_L U \left( {\cal M}' i\!\not\! \partial {\cal M}'^\dagger 
- i\!\not\! \partial{\cal M}' {\cal M}'^\dagger \right) U^\dagger \psi_L
\nonumber\\
&&
+ \frac{F^{-1}}{v^2} \bar\psi_L \langle {\cal M} i \! \not\!\partial {\cal M}^\dagger 
- i \! \not\!\partial {\cal M} {\cal M}^\dagger \rangle \psi_L
-\frac{F^{-1}}{v^2}\bar\psi_R{\cal M}^\dagger 
 U^\dagger \not\!\! L U{\cal M}\psi_R 
-\frac{1}{v^2}\bar\psi_R{\cal M}'^\dagger 
 U^\dagger \not\!\! L U{\cal M}'\psi_R
\nonumber\\
&&
+\frac{\kappa}{v^2}F^{-1/2}\left(\bar\psi_R{\cal M}^\dagger U^\dagger
   \not\!\! L U{\cal M}'\psi_R +{\rm h.c.}\right)
-\frac{\kappa}{v^2}F^{-1/2}\left(\bar\psi_L U {\cal M}'{\cal M}^\dagger 
   U^\dagger\not\!\! L\psi_L +{\rm h.c.}\right)
\nonumber\\
&&
+
\frac{1}{2 v^2} 
\bar\psi_L \left(
U {\cal M}'{\cal M}'^\dagger U^\dagger \not\!\! L + \not\!\! L U {\cal M}'{\cal M}'^\dagger U^\dagger
\right) \psi_L
+ \frac{F^{-1}}{2 v^2} \bar\psi_L \left( \not\!\!L U {\cal M} {\cal M}^\dagger U^\dagger
+ U {\cal M} {\cal M}^\dagger U^\dagger \not\!\!L \right) \psi_L
\bigg\}
\nonumber\\
&&
 + \Delta^{(1/2)} {\cal L}_{\psi^2 Uh D^2} + \Delta^{(1/2)} {\cal L}_{\psi^4 Uh}
,
\eea
with
\bea
\Delta^{(1/2)} {\cal L}_{\psi^2 Uh D^2} &=& 
- \frac{1}{16\pi^2} \frac{1}{d-4} \bigg\{ \langle L^\mu L_\mu\rangle
\left[\frac{F{\cal B}}{2v^2}\bar\psi_L U{\cal M}''\psi_R
-\frac{\kappa^2-1}{Fv^2}
\bar\psi_L U\left(\frac{F'}{2}{\cal M}'-{\cal M}\right)\psi_R +{\rm h.c.}
\right]
\\
&&+\frac{2\kappa'}{v^2}\partial^\mu\eta\left(
i\bar\psi_L L_\mu U(F^{-1/2}{\cal M})'\psi_R +{\rm h.c.}\right)
+\frac{3{\cal B}}{Fv^2} \partial^\mu\eta\partial_\mu\eta\left(\bar\psi_L U
\left(\frac{F'}{2}{\cal M}'-{\cal M}\right)\psi_R +{\rm h.c.}\right)
\!\bigg\}
,
\nonumber\\
\Delta^{(1/2)} {\cal L}_{\psi^4 Uh} &=& 
- \frac{1}{16\pi^2} \frac{1}{d-4} \bigg\{\frac{3F^{-2}}{2v^4}\left(\bar\psi_L U
\left(\frac{F'}{2}{\cal M}'-{\cal M}\right)\psi_R +{\rm h.c.}\right)^2
+\frac{1}{2v^4}\left(\bar\psi_L U{\cal M}''\psi_R +{\rm h.c.}\right)^2
\nonumber\\
&&+\frac{4}{v^4}\left(i\bar\psi_L U T^a
\left(F^{-1/2} {\cal M}\right)'\psi_R +{\rm h.c.}\right)^2 
\bigg\}
.
\eea
The first six lines in the above expression of ${\cal L}_{\rm div}^{(1/2)}$
correspond to structures already present in the lowest-order Lagrangian ${\cal L}_2$.
The five lines that follow are of the type $\psi^2 Uh D$, but are not present in
the basis considered in Ref. \cite{Buchalla:2013rka}. They need therefore to be
transformed, proceeding as for the last term in $\Delta^{(0)} {\cal L}_2$. Considering the first term of this type, one may write
\bea
\frac{3 F^{-1}}{2 v^2}\bar\psi_R 
\left( {\cal M}^\dagger i\!\not\! \partial {\cal M} - i\!\not\! \partial {\cal M}^\dagger {\cal M} \right) \psi_R
&=&
\frac{3}{2 v^2}\bar\psi_R 
\left[ F^{-1/2} {\cal M}^\dagger i\!\not\! \partial (F^{-1/2} {\cal M}) - i\!\not\! \partial (F^{-1/2} {\cal M}^\dagger) F^{-1/2} {\cal M} \right] \psi_R
\nonumber\\
&=&
\frac{3}{2 v^2} i (\partial_\mu \eta)
\bar\psi_R \gamma^\mu \left[ F^{-1/2} {\cal M}^\dagger (F^{-1/2} {\cal M})' -  (F^{-1/2} {\cal M}^\dagger)' F^{-1/2} {\cal M} \right] \psi_R
\nonumber\\
&=&
\frac{3}{2 v^2} \bar\psi_R i\!\not\! \partial 
\int_0^\eta ds \left[ F^{-1/2} {\cal M}^\dagger (F^{-1/2} {\cal M})' -  (F^{-1/2} {\cal M}^\dagger)' F^{-1/2} {\cal M}  \right]  \psi_R
\nonumber\\
&=&
- \frac{3}{2 v^2} \bar\psi_R 
\int_0^\eta ds \left[ F^{-1/2} {\cal M}^\dagger (F^{-1/2} {\cal M})' -  (F^{-1/2} {\cal M}^\dagger)' F^{-1/2} {\cal M}  \right] {\cal M}^\dagger U^\dagger  \psi_L
\nonumber\\
&&
+ \frac{3}{2 v^2} \bar\psi_L U {\cal M}
\int_0^\eta ds \left[ F^{-1/2} {\cal M}^\dagger (F^{-1/2} {\cal M})' -  (F^{-1/2} {\cal M}^\dagger)' F^{-1/2} {\cal M}  \right] \psi_R
.
\nonumber\\
\eea
In the last step, an integration by parts has been performed, the resulting total-derivative term
has been dropped, and the lowest-order equations of motion for the fermionic fields have been used.
The other structures of this type can be handled in a similar manner to obtain:
\bea
\frac{1}{2 v^2}\bar\psi_R 
\left( {\cal M}'^\dagger i\!\not\! \partial {\cal M}' - i\!\not\! \partial {\cal M}'^\dagger {\cal M}' \right) \psi_R
&=&
\frac{1}{2 v^2} \bar\psi_R 
\int_0^\eta ds \left[ {\cal M}''^\dagger  {\cal M}' - {\cal M}'^\dagger {\cal M}'' \right] {\cal M}^\dagger U^\dagger  \psi_L
\nonumber\\
&&\!\!\!\!\!
+ \, \frac{1}{2 v^2} \bar\psi_L U {\cal M}
\int_0^\eta ds \left[ {\cal M}'^\dagger {\cal M}'' -  {\cal M}''^\dagger  {\cal M}'  \right] \psi_R
,
\nonumber\\
\frac{F^{-1}}{2 v^2}\bar\psi_L U
\left( {\cal M} i\!\not\! \partial {\cal M}^\dagger - i\!\not\! \partial {\cal M} {\cal M}^\dagger \right) U^\dagger \psi_L
&=&
+ \frac{1}{2 v^2} \bar\psi_R {\cal M}^\dagger
\int_0^\eta ds \left[ F^{-1/2} {\cal M} (F^{-1/2} {\cal M}^\dagger)' - (F^{-1/2} {\cal M})' F^{-1/2} {\cal M}^\dagger \right] U^\dagger  \psi_L
\nonumber\\
&&
+ \frac{1}{2 v^2} \bar\psi_L U 
\int_0^\eta ds \left[ (F^{-1/2} {\cal M})' F^{-1/2} {\cal M}^\dagger - F^{-1/2} {\cal M}  (F^{-1/2} {\cal M}^\dagger)'  \right] {\cal M} \psi_R
\nonumber\\
&&
+ \frac{1}{2 v^2} \bar\psi_L U 
\int_0^\eta ds \left[ (F^{-1/2} {\cal M})' F^{-1/2} {\cal M}^\dagger - F^{-1/2} {\cal M}  (F^{-1/2} {\cal M}^\dagger)' \right] 
U^\dagger \not\!\!L \psi_L
\nonumber\\
&&
+ \frac{1}{2 v^2} \bar\psi_L \not\!\!L U 
\int_0^\eta ds \left[ F^{-1/2} {\cal M} (F^{-1/2} {\cal M}^\dagger)' -  (F^{-1/2} {\cal M})' F^{-1/2} {\cal M}^\dagger  \right] U^\dagger \psi_L
,
\nonumber\\
\frac{1}{2 v^2}\bar\psi_L U 
\left( {\cal M}' i\!\not\! \partial {\cal M}'^\dagger - i\!\not\! \partial {\cal M}' {\cal M}'^\dagger \right) U^\dagger \psi_L
&=&
+ \frac{1}{2 v^2} \bar\psi_R {\cal M}^\dagger
\int_0^\eta ds \left[ {\cal M}' {\cal M}''^\dagger   - {\cal M}'' {\cal M}'^\dagger \right] U^\dagger  \psi_L
\nonumber\\
&&
+ \frac{1}{2 v^2} \bar\psi_L U
\int_0^\eta ds \left[ {\cal M}''  {\cal M}'^\dagger - {\cal M}' {\cal M}''^\dagger  \right] {\cal M} \psi_R
\nonumber\\
&&
+ \frac{1}{2 v^2} \bar\psi_L U 
\int_0^\eta ds \left[ {\cal M}''  {\cal M}'^\dagger - {\cal M}' {\cal M}''^\dagger \right] 
U^\dagger \not\!\!L \psi_L
\nonumber\\
&&
+ \frac{1}{2 v^2} \bar\psi_L \not\!\!L U 
\int_0^\eta ds \left[ {\cal M}' {\cal M}''^\dagger   - {\cal M}'' {\cal M}'^\dagger  \right] U^\dagger \psi_L
,
\nonumber\\
\frac{F^{-1}}{v^2}\bar\psi_L 
\langle {\cal M} i\!\not\! \partial {\cal M}^\dagger - i\!\not\! \partial {\cal M} {\cal M}^\dagger \rangle \psi_L
&=&
- \frac{1}{v^2} \bar\psi_R {\cal M}^\dagger U^\dagger
\int_0^\eta ds \left[ \langle
(F^{-1/2} {\cal M})' F^{-1/2} {\cal M}^\dagger - F^{-1/2} {\cal M} (F^{-1/2} {\cal M}^\dagger)' 
\rangle \right]   \psi_L
\nonumber\\
&&
- \frac{1}{v^2} \bar\psi_L 
\int_0^\eta ds \left[\langle 
F^{-1/2} {\cal M} (F^{-1/2} {\cal M}^\dagger)' -  (F^{-1/2} {\cal M})' F^{-1/2} {\cal M}^\dagger  
\rangle \right] U {\cal M} \psi_R
\nonumber
,\\
\eea
which reduce to structures already present in ${\cal L}_2$. The structures of the form 
$\psi_{I} O_1 \!\!\! \not\!\!\!L O_2 \psi_{I}$, where $I=L,R$, can be handled upon
using the identity \rf{identity_slashL}, such that the whole structure can be expressed
in terms of the basis of Ref. \cite{Buchalla:2013rka}.
One thus obtains the decomposition
\be
{\cal L}_{\rm div}^{(1/2)} = \Delta^{(1/2)} {\cal L}_2 + \Delta^{(1/2)} {\cal L}_{\psi^2 Uh D}
+ \Delta^{(1/2)} {\cal L}_{\psi^2 Uh D^2} + \Delta^{(1/2)} {\cal L}_{\psi^4 Uh}
,
\ee
with (the lowest-order equations of motion of the fermion fields
have been applied)
\bea
\Delta^{(1/2)} {\cal L}_2 &=&
- \frac{1}{16\pi^2} \frac{1}{d-4} \bigg\{
\frac{3}{4}g^2 \bar\psi_L U {\cal M} \psi_R + 
g'^2 \bar\psi_L Y^2_L U {\cal M} \psi_R 
+ g'^2 \bar\psi_L\,  U {\cal M} Y^2_R \psi_R 
\nonumber\\
&&
- 6 g^2_s C_F \, \bar q U{\cal M}_q P_R q 
+\frac{V''}{v^4} \bar\psi_LU {\cal M}'' \psi_R
-8 g'^2 \bar\psi_L Y_L U{\cal M}Y_R\psi_R 
\nonumber\\
&&
+\left((3g^2+g'^2)\frac{v^2}{4} F+\frac{3}{2}\frac{F'V'}{F v^2}\right)
\frac{F^{-1}}{v^2}
\bar\psi_L U\left(\frac{F'}{2}{\cal M}'-{\cal M}\right)\psi_R
\nonumber\\
&&
+ \frac{3}{v^2}F^{-1}\, \bar\psi_L U {\cal M}{\cal M}^\dagger {\cal M}\psi_R
-\frac{3}{v^2}F^{-1}\, 
\bar\psi_L U \langle{\cal M}{\cal M}^\dagger \rangle {\cal M}\psi_R
\nonumber\\
&&
+
\frac{2}{v^2} \bar\psi_L U{\cal M}'{\cal M}^\dagger{\cal M}'\psi_R
+
\frac{1}{2 v^2} \bar\psi_L U {\cal M} {\cal M}'^\dagger {\cal M}' \psi_R
+
\frac{1}{2 v^2} \bar\psi_L U {\cal M}' {\cal M}'^\dagger {\cal M} \psi_R
\nonumber\\
&&
+ \frac{3}{2 v^2} \bar\psi_L U {\cal M}
\int_0^\eta ds \left[ F^{-1/2} {\cal M}^\dagger (F^{-1/2} {\cal M})' -  
(F^{-1/2} {\cal M}^\dagger)' F^{-1/2} {\cal M}  \right] \psi_R
\nonumber\\
&&
- \frac{1}{2 v^2} \bar\psi_L U 
\int_0^\eta ds \left[ (F^{-1/2} {\cal M})' F^{-1/2} {\cal M}^\dagger - 
F^{-1/2} {\cal M}  (F^{-1/2} {\cal M}^\dagger)'  \right] {\cal M} \psi_R
\nonumber\\
&&
+ \frac{1}{2 v^2} \bar\psi_L U {\cal M}
\int_0^\eta ds \left[ {\cal M}'^\dagger {\cal M}'' -  
{\cal M}''^\dagger {\cal M}'  \right] \psi_R
+ \frac{1}{2 v^2} \bar\psi_L U 
\int_0^\eta ds \left[ {\cal M}'' {\cal M}'^\dagger -  
{\cal M}' {\cal M}''^\dagger  \right] {\cal M} \psi_R
\nonumber\\
&&
- \frac{1}{v^2} \bar\psi_L U
\int_0^\eta ds \langle F^{-1/2} {\cal M} (F^{-1/2} {\cal M}^\dagger)' -  
(F^{-1/2} {\cal M})' F^{-1/2} {\cal M}^\dagger  \rangle {\cal M} \psi_R
+ {\rm h.c.}\bigg\}
,
\lbl{Delta1/2L2}
\eea
and
\bea
\Delta^{(1/2)} {\cal L}_{\psi^2 Uh D} &=&
- \frac{1}{16\pi^2} \frac{1}{d-4} \bigg\{
-\frac{F^{-1}}{v^2}\bar\psi_R{\cal M}^\dagger 
 U^\dagger \not\!\! L U{\cal M}\psi_R 
-\frac{1}{v^2}\bar\psi_R{\cal M}'^\dagger 
 U^\dagger \not\!\! L U{\cal M}'\psi_R
\nonumber\\
&&
+\frac{\kappa}{v^2}F^{-1/2}\left(\bar\psi_R{\cal M}^\dagger U^\dagger
   \not\!\! L U{\cal M}'\psi_R +{\rm h.c.}\right)
-\frac{\kappa}{v^2}F^{-1/2}\left(\bar\psi_L U {\cal M}'{\cal M}^\dagger 
   U^\dagger\not\!\! L\psi_L +{\rm h.c.}\right)
\nonumber\\
&&
+
\frac{1}{2 v^2} 
\left( \bar\psi_L 
U {\cal M}'{\cal M}'^\dagger U^\dagger \not\!\! L \psi_L + {\rm h.c.} \right)
+ \frac{F^{-1}}{2 v^2} \left( \bar\psi_L 
U {\cal M} {\cal M}^\dagger U^\dagger \not\!\!L \psi_L + {\rm h.c.} \right)
\nonumber\\
&&
- \frac{1}{2 v^2} \left(\bar\psi_L U 
\int_0^\eta ds \left[ (F^{-1/2} {\cal M})' F^{-1/2} {\cal M}^\dagger - 
F^{-1/2} {\cal M}  (F^{-1/2} {\cal M}^\dagger)' \right] 
U^\dagger \not\!\!L \psi_L + {\rm h.c.} \right)
\nonumber\\
&&
+ \frac{1}{2 v^2} \left( \bar\psi_L U 
\int_0^\eta ds \left[ {\cal M}'' {\cal M}'^\dagger -   
{\cal M}' {\cal M}''^\dagger \right] 
U^\dagger \not\!\!L \psi_L + {\rm h.c.} \right)
\bigg\}
.
\eea

\subsection{Summary}

In this section, the divergences at next-to-leading order have been decomposed
into the basis of counterterms given in Ref. \cite{Buchalla:2013rka}. 
The result of this procedure is summarized by
\bea
{\cal L}_{\rm div} &=&
\left[ \Delta^{(0)} {\cal L}_2 +  \Delta^{(1)} {\cal L}_2 +
  \Delta^{(1/2)} {\cal L}_2  \right] + \left[ \Delta^{(0)} {\cal L}_{\beta_1}
  +  \Delta^{(1)} {\cal L}_{\beta_1}  \right]
  + \Delta^{(0)} {\cal L}_{UhD^4} + \left[ \Delta^{(1)} {\cal L}_{\psi^2 UhD}
  + \Delta^{(1/2)} {\cal L}_{\psi^2 UhD} \right]
\nonumber\\
&&
+ \Delta^{(1/2)} {\cal L}_{\psi^2 UhD^2}  + \Delta^{(1/2)} {\cal L}_{\psi^4 Uh}
,
\label{ldivsummary}
\eea
where the expressions for each term are given above.


\section{Renormalization of the leading-order Lagrangian}
\label{sec:LO_ren}
\setcounter{equation}{0}

The renormalization of the parameters in ${\cal L}_2$ is derived
from the first bracket in (\ref{ldivsummary}),
\begin{equation}\label{dell2}
\Delta {\cal L}_2 \equiv
\Delta^{(0)} {\cal L}_2 + \Delta^{(1)} {\cal L}_2 + \Delta^{(1/2)} {\cal L}_2 
\, ,
\end{equation}
The expressions of  $\Delta^{(1)} {\cal L}_2$, $\Delta^{(0)} {\cal L}_2$,
and $\Delta^{(1/2)} {\cal L}_2$ are given in Eqs. \rf{Delta1L2},
\rf{Delta0L2}, and \rf{Delta1/2L2}, respectively.
We rewrite (\ref{dell2}) as
\begin{equation}\label{l2gs}
\Delta {\cal L}_2  = \Delta {\cal L}_{2,gauge}  + \Delta {\cal L}_{2,scalar} 
\end{equation}  
where
\bea
32\pi^2\varepsilon\, \Delta {\cal L}_{2,gauge} &=&
- \frac{g^2_s}{4}\, G^{\alpha\mu\nu}G_{\mu\nu}^\alpha
\left(- \frac{22}{3} N_c + \frac{4}{3} N_f   \right)
- \frac{g^2}{2} \langle W^{\mu\nu}W_{\mu\nu} \rangle  \left[ - \frac{44}{3}
  + \frac{2}{3} (N_c+1) N_{\rm g} + \frac{1}{3} \right]
\nonumber\\
&&
-\frac{g'^2}{4} \, B^{\mu\nu} B_{\mu\nu}  \left[ 2 \left(\frac{11}{27} N_c
    + 1\right) N_{\rm g} + \frac{1}{3} \right]
\label{l2gauge}\\
32\pi^2\varepsilon\, \Delta {\cal L}_{2,scalar} &=&
  \frac{v^2}{4}\ \langle L_\mu L^\mu \rangle\, A_F
  +\partial_\mu h \partial^\mu h\,  A_h + A_V
  - \left(\bar\psi U A_{{\cal M}} P_R \psi + {\rm h.c.}\right).
\label{l2scalar}  
\eea  
Here and in the following, we define for the dimension of space-time
\begin{equation}\label{d4eps}
d\equiv 4 - 2\varepsilon .
\end{equation}  
The various functions introduced on the right-hand side of (\ref{l2scalar})
read
\bea
A_F(\eta) &=&
- \frac{g'^2}{2} (\kappa^2 + 3) F - \frac{g^2}{6} (19 \kappa^2 + 17) F
- \, 2 (\kappa^2 - 1) \frac{F'V'}{Fv^4}
+ \, 2 \frac{V''}{v^4} {\cal B} F +
\frac{4}{v^2} \dl {\cal M}^\dagger {\cal M}\dr
,\\
\label{afeta}
A_h(\eta)  &=& 
  \frac{1}{4} (3g^2+g'^2) ( F {\cal B} - 4 \kappa^2)
+ \frac{3}{2} \frac{F'V'}{F v^4} {\cal B}
+ \frac{2}{v^2} \dl {{\cal M}'}^{\dagger} {\cal M}' \dr 
,\\
\label{aheta}
A_V (\eta) &=& \frac{3}{2}(3 g^4 + 2 g^2 g'^2 + g'^4)\frac{v^4}{16} F^2  
+\frac{3 g^2 + g'^2}{8}F'V'+\frac{3}{8}\left(\frac{F'V'}{F v^2}\right)^2
+ \, \frac{1}{2}\left(\frac{V''}{v^2}\right)^2
- 2 \dl ({\cal M}^\dagger {\cal M})^2\dr
,\\
\label{aveta}
A_{{\cal M}}(\eta) &=&
- \frac{3}{4}g^2  {\cal M} - g'^2 ({\cal M} Y^2_L + {\cal M} Y^2_R )
+ 6 g_s^2 C_F {\cal M}_q  + 8 g'^2 Y_L {\cal M} Y_R
\nonumber\\
&&
- \left[(3g^2+g'^2)\frac{v^2}{4} F+\frac{3}{2}\frac{F'V'}{F v^2}\right]
\frac{F^{-1}}{v^2} \left(\frac{F'}{2}{\cal M}'-{\cal M}\right)
- \frac{V''}{v^4} {\cal M}''
\nonumber\\
&&
- \frac{3}{v^2}F^{-1} {\cal M}{\cal M}^\dagger {\cal M}
+ \frac{3}{v^2}F^{-1} \langle{\cal M}{\cal M}^\dagger \rangle {\cal M}
- \frac{2}{v^2}{\cal M}'{\cal M}^\dagger{\cal M}'
- \frac{1}{2 v^2} {\cal M} {\cal M}'^\dagger {\cal M}'
- \frac{1}{2 v^2} {\cal M}' {\cal M}'^\dagger {\cal M}
\nonumber\\
&&
- \frac{3}{2 v^2} {\cal M}
\int_0^\eta \!\! ds \left[ F^{-1/2} {\cal M}^\dagger (F^{-1/2} {\cal M})'
  - (F^{-1/2} {\cal M}^\dagger)' F^{-1/2} {\cal M}  \right] 
\nonumber\\
&&
+ \frac{1}{2 v^2} 
\int_0^\eta \!\! ds \left[ (F^{-1/2} {\cal M})' F^{-1/2} {\cal M}^\dagger
  - F^{-1/2} {\cal M}  (F^{-1/2} {\cal M}^\dagger)'  \right] {\cal M} 
\nonumber\\
&&
- \frac{1}{2 v^2} {\cal M}
\int_0^\eta \!\! ds
\left[ {\cal M}'^\dagger {\cal M}'' -  {\cal M}''^\dagger {\cal M}'  \right] 
- \frac{1}{2 v^2} 
\int_0^\eta ds \left[ {\cal M}'' {\cal M}'^\dagger
  - {\cal M}' {\cal M}''^\dagger  \right] {\cal M} 
\nonumber\\
&&
+ \frac{1}{v^2} \int_0^\eta \!\! ds 
\langle F^{-1/2} {\cal M} (F^{-1/2} {\cal M}^\dagger)'  -  
(F^{-1/2} {\cal M})' F^{-1/2} {\cal M}^\dagger  \rangle {\cal M} 
\nonumber\\
&&
+ \frac{8}{v^2}  {\cal M} T^3
\int_0^\eta \!\! ds \, 
\dl (F^{-1/2} {\cal M}^\dagger)' F^{-1/2} {\cal M} T^3
- F^{-1/2} {\cal M}^\dagger (F^{-1/2} {\cal M})' T^3 \dr
.
\label{ameta}
\eea
In the expression for $A_{{\cal M}}$ it is understood that the contribution proportional
to $g_s^2$ only affects the quark fields.

The renormalization that cancels $\Delta {\cal L}_{2,gauge}$ is standard and
will be summarized below. In order to treat $\Delta {\cal L}_{2,scalar}$, it
is convenient to first apply suitable redefinitions of the Higgs field $h$,
which bring $\Delta {\cal L}_{2,scalar}$ to a canonical form.
The renormalization of ${\cal L}_2$ then proceeds in three steps.
First, we eliminate the term $\partial_\mu h \partial^\mu h\, A_h$
from $\Delta {\cal L}_{2,scalar}$ using integration by parts and the lowest-order equation
of motion for $h$. Second, we shift the Higgs field, $\eta\to\eta +\delta$,
and fix $\delta$ such that the minimum of the potential remains at $\eta =0$.
In a third step, we renormalize the fields and parameters in ${\cal L}_2$.
The resulting counterterms are then determined in the usual way from
the requirement that they cancel the divergences of $\Delta {\cal L}_2$.

We remark that for the field shift $\delta$, only the divergent part,
relevant for the UV renormalization, is considered in the present section.
However, $\delta$ also includes a finite piece, which must be 
chosen such as to preserve the condition $V'(0) =0$. Minimal subtraction is 
therefore not sufficient here. The finite counterterm that is required in 
this case is computed in Section \ref{sec:applications}.

\subsection{Redefinitions of the Higgs field}

In a first step, we eliminate  $\partial_\mu h \partial^\mu h\, A_h$
from (\ref{l2scalar}) by writing
\begin{equation}\label{elimah}
  \partial_\mu h \partial^\mu h\, A_h(\eta) =
  -v\, \partial^2 h\, \int^\eta_0 ds\, A_h(s) =
  \left[-F' \frac{v^2}{4} \langle L_\mu L^\mu \rangle + V' +
  {\bar\psi} m'(\eta , U) \psi\right]\, \int^\eta_0 ds\, A_h(s)\, ,
\end{equation}  
where we used an integration by parts and the equation of motion
of the Higgs field (\ref{eomh}). Inserting (\ref{elimah}) in (\ref{l2scalar})
we obtain
\bea
32\pi^2\varepsilon\, \Delta {\cal L}_{2,scalar} &=&
\frac{v^2}{4}\ \langle L_\mu L^\mu \rangle\,
\left[A_F - F' \int^\eta_0 ds\, A_h(s)\right]+A_V + V'\int^\eta_0 ds\, A_h(s)
\nonumber\\
&&- \left(\bar\psi U
  \left[A_{{\cal M}}-{\cal M}' \int^\eta_0 ds\, A_h(s)\right] P_R \psi
  + {\rm h.c.}\right).
\label{l2sah}
\eea
This step is equivalent to a field redefinition for $h$.
The form of (\ref{l2sah}) has the advantage that the Higgs kinetic term
$\partial h\partial h$ no longer receives a (divergent) one-loop correction.
Consequently, there is no need to renormalize $h$.
This is fully analogous to our treatment of the fermions,
where we have also used the equations of motion to dispense with
their field renormalization.

In a second step, we shift $\eta\to\eta +\delta$ to ensure a minimum
of the potential at $\eta=0$. Since $V'(0)=0$ at leading order,
$\delta$ is of the order of a one-loop contribution.
The shift then leads to an additional one-loop term in the Lagrangian:
\be
\label{l2delta}
{\cal L}_2 \to {\cal L}_2 +
\frac{v^2}{4}\ \langle L_\mu L^\mu \rangle\, F' \delta - V' \delta
- \left(\bar\psi U {\cal M}' \delta P_R \psi + {\rm h.c.}\right).
\ee
The divergent Lagrangian $\Delta {\cal L}_{2,scalar}$ then becomes
\bea
32\pi^2\varepsilon\, \Delta {\cal L}_{2,scalar} &=&
\frac{v^2}{4}\ \langle L_\mu L^\mu \rangle\,
\left[A_F - F' \int^\eta_0 ds\, A_h(s) + F'\tilde\delta \right]
+A_V + V'\int^\eta_0 ds\, A_h(s) - V' \tilde\delta
\nonumber\\
&&- \left(\bar\psi U \left[
    A_{{\cal M}}-{\cal M}' \int^\eta_0 ds\, A_h(s) + {\cal M}' \tilde\delta
  \right] P_R \psi + {\rm h.c.}\right),
\label{l2sshift}
\eea
where
\be\label{deltatil}
\tilde\delta \equiv 32 \pi^2\varepsilon\, \delta
\ee
The condition that the minimum of the effective potential remains
at $\eta=0$ implies for the divergent part of $\delta$
\be
\label{deltacond}
\left[\left(A_V + V'\int^\eta_0 ds\, A_h(s)\right)'
  - V'' \tilde\delta\right]_{\eta=0} = 0
,
\ee
and accordingly
\be\label{deltatil2}
\tilde\delta =\frac{3 F_1}{32 V_2}\left(3 g^4 + 2 g^2 g'^2 + g'^4\right)
+ \frac{3g^2 + g'^2}{8} F_1 + 6 V_3 -\frac{2}{v^4 V_2}
\dl({\cal M}^\dagger_1 {\cal M}_0 +{\cal M}^\dagger_0 {\cal M}_1)
{\cal M}^\dagger_0 {\cal M}_0\dr
.
\ee

\subsection{Renormalization of ${\cal L}_2$}
\label{subsec:l2ren}

We start from the leading order Lagrangian (\ref{l2}), where we
consider all fields and parameters $X$ as unrenormalized, denoted by
$\stackrel{{\mbox{\tiny{o}}}}{X}$.
As discussed above, no renormalization is needed for the Higgs field and
the fermions, and we thus take $\stackrel{{\mbox{\tiny{o}}}}{h}=h$,
$\stackrel{{\mbox{\tiny{o}}}}{\psi}=\psi$.
For the remaining quantities we introduce renormalization constants
in the form
\bea
\stackrel{{\mbox{\tiny{o}}}}{G}_\mu  { }^{\!\!\!\!\!\!\alpha} =
Z_G^{1/2} G^\alpha_\mu
,
\qquad
\stackrel{{\mbox{\tiny{o}}}}{W}_\mu  { }^{\!\!\!\!\!\!a} &=& Z_W^{1/2} W^a_\mu
,
\qquad
\stackrel{{\mbox{\tiny{o}}}}{B}_\mu = Z_B^{1/2} B_\mu
,
\nonumber\\
\nonumber\\
\stackrel{{\mbox{\tiny{o}}}}{g}_s = Z_G^{-1/2} g_s \mu^\varepsilon
,
\qquad
\stackrel{{\mbox{\tiny{o}}}}{g} &=& Z_W^{-1/2} g \mu^\varepsilon
,
\qquad
{\stackrel{{\mbox{\tiny{o}}}}{g}}' = Z_B^{-1/2} g' \mu^\varepsilon
,
\nonumber\\
\nonumber\\
\stackrel{{\mbox{\tiny{o}}}}{\varphi} { }^{\!\!a} = Z_\varphi^{1/2} \varphi^a
,
 & &  \qquad
\stackrel{{\mbox{\tiny{o}}}}{v} = Z_\varphi^{1/2} v \mu^{-\varepsilon}
,
\nonumber\\
\nonumber\\
\stackrel{{\mbox{\tiny{o}}}}{F}_n = Z_{F_n} F_n
,
\qquad
\stackrel{{\mbox{\tiny{o}}}}{V}_n &=& Z_{V_n} V_n\, \mu^{2\varepsilon}
,
\qquad
{\stackrel{{\mbox{\tiny{o}}}}{{\cal M}}}_n = {\cal M}_n +\Delta {\cal M}_n
.
\label{zfactors}
\eea
Notice that the renormalization of the gauge fields and of the
corresponding gauge couplings involve the same renormalization constants.
This is because we are using the background field gauge, with
the effect that the Slavnov-Taylor identities boil down to the simple
QED-type Ward-Takahashi identities \cite{Denner:1994xt}. 

Since
\be
\frac{v^2}{4}\ \langle L_\mu L^\mu \rangle =
\frac{1}{2}\partial_\mu \varphi^a \partial^\mu \varphi^a +
\frac{v^2}{8}\left(g W^a_\mu - g' B_\mu \delta^{a3}\right)^2 +\ldots\, ,
\ee
we observe that the Goldstone kinetic term and the gauge-boson mass term
receive the same divergent one-loop corrections. In addition, the
term $g W^a_\mu - g' B_\mu \delta^{a3}$ is not renormalized within our scheme.
It follows that the renormalization factors for $\varphi^a$ and $v$
are identical, a fact which we have already used in (\ref{zfactors}).
As a consequence, $\varphi^a/v$ and the Goldstone matrix $U$ are
not renormalized.

It will be convenient to write the one-loop $Z$-factor for a quantity $X$
in the minimal subtraction (MS) scheme as
\be
\label{zxms}
Z_X=1+\Delta Z_X = 1 +\frac{\Delta \tilde Z_X}{32\pi^2\varepsilon}\, ,
\qquad \Delta \tilde Z_X \equiv 32\pi^2\varepsilon\, \Delta Z_X\, ,
\quad \Delta \tilde {\cal M}_n \equiv 32\pi^2\varepsilon\, \Delta {\cal M}_n\, .
\ee
In the ${\overline{\rm MS}}$ scheme, the pole term is replaced by
\be
\label{msbarpole}
\frac{1}{\varepsilon} \equiv \frac{2}{4-d}\to 
\frac{2}{4-d} - \gamma_E + \ln 4\pi \, .
\ee

Employing (\ref{zfactors}) in the unrenormalized version of (\ref{l2}), and
subtracting the renormalized Lagrangian, we obtain the counterterm 
${\cal L}_{2,CT} = \stackrel{{\mbox{\tiny{o}}}}{\cal L}_2 - {\cal L}_2$,
where
\bea
\lbl{L_ctr}
{\cal L}_{2,CT} &=&
-\frac{1}{4} (Z_G - 1) G_{\mu\nu}^\alpha G^{\alpha\mu\nu}
-\frac{1}{2} (Z_W - 1) \langle W_{\mu\nu}W^{\mu\nu}\rangle 
-\frac{1}{4} (Z_B - 1)  B_{\mu\nu}B^{\mu\nu}
\nonumber\\
&&
+ \, \frac{v^2}{4}\ \langle L_\mu L^\mu \rangle\,
\sum^\infty_{n=0}\left(Z^{1-n/2}_\varphi Z_{F_n} -1\right) F_n \eta^n
-\sum^\infty_{n=2}\left(Z^{2-n/2}_\varphi Z_{V_n} -1\right) v^4 V_n \eta^n
\nonumber\\
&&  
-\left( \bar\psi U \sum^\infty_{n=0}
  \left( Z^{-n/2}_\varphi ({\cal M}_n +\Delta {\cal M}_n) - {\cal M}_n\right)
  \eta^n \, P_R\psi + {\rm h.c.}\right)
\nonumber\\
&=&
-\frac{1}{4} \Delta Z_G\,  G_{\mu\nu}^\alpha G^{\alpha\mu\nu}
-\frac{1}{2} \Delta Z_W\,  \langle W_{\mu\nu}W^{\mu\nu}\rangle 
-\frac{1}{4} \Delta Z_B\,  B_{\mu\nu}B^{\mu\nu}
\nonumber\\
&&
+ \, \frac{v^2}{4}\ \langle L_\mu L^\mu \rangle\,
\sum^\infty_{n=0}
\left(\left(1-\frac{n}{2}\right)\Delta Z _\varphi +\Delta Z_{F_n}\right)
F_n \eta^n
-v^4 \sum^\infty_{n=2}
\left(\left(2-\frac{n}{2}\right) \Delta Z_\varphi +\Delta Z_{V_n}\right)
V_n \eta^n
\nonumber\\
&&  
-\left( \bar\psi U \sum^\infty_{n=0}
  \left( -\frac{n}{2} \Delta Z_\varphi {\cal M}_n +\Delta {\cal M}_n \right)
  \eta^n \, P_R\psi + {\rm h.c.}\right)
.
\eea
The approximation in the second equality holds to one-loop order.

The renormalization constants are fixed by requiring
$-{\cal L}_{2,CT} = \Delta {\cal L}_{2,gauge} +\Delta {\cal L}_{2,scalar}$,
with $\Delta {\cal L}_{2,gauge}$ in (\ref{l2gauge}) and
$\Delta {\cal L}_{2,scalar}$ in the redefined form of (\ref{l2sshift}).
This implies
\bea
\Delta \tilde Z_G &=& 
-\left(- \frac{22}{3} N_c + \frac{4}{3} N_f   \right) g_s^2
,
\nonumber\\
\Delta \tilde Z_W &=& 
-\left[-\frac{44}{3}  + \frac{2}{3} (N_c+1) N_{\rm g} +\frac{1}{3}\right] g^2
,
\nonumber\\
\Delta \tilde Z_B &=& 
-\left[2\left(\frac{11}{27} N_c + 1\right) N_{\rm g} +\frac{1}{3}\right] g'^2
\label{delzg}
\eea
and 
\bea
\sum^\infty_{n=0} \left(\left(1-\frac{n}{2}\right)
  \Delta \tilde Z _\varphi +\Delta \tilde Z_{F_n}\right)
F_n \eta^n &=& -A_F + F'\, \int^\eta_0 ds\, A_h(s) - F' \tilde\delta
,
\label{delzaf}
\\
\sum^\infty_{n=2} \left(\left(2-\frac{n}{2}\right)
  \Delta \tilde Z_\varphi +\Delta \tilde Z_{V_n}\right)
v^4 V_n \eta^n &=& A_V + V'\, \int^\eta_0 ds\, A_h(s) - V' \tilde\delta
,
\label{delzav}
\\
\sum^\infty_{n=0}\left( -\frac{n}{2}
\Delta \tilde Z_\varphi {\cal M}_n +\Delta \tilde{\cal M}_n \right)\eta^n &=&
  -A_{\cal M} + {\cal M}'\, \int^\eta_0 ds\, A_h(s) - {\cal M}' \tilde\delta
.
\label{delzam}
\eea
$\Delta \tilde Z_\varphi$ follows from the $n=0$ term in (\ref{delzaf}),
noting that $F_0\equiv 1$ and $\Delta \tilde Z_{F_0}\equiv 0$.
We find
\bea
\Delta \tilde Z_\varphi &=&
\left(\frac{F^2_1}{4}\frac{5}{3}+\frac{17}{6}\right)g^2 +\frac{3}{2} g'^2
-V_2 F^2_1 + 4 V_2 F_2 - 6 F_1 V_3
-\frac{4}{v^2}\dl {\cal M}^\dagger_0 {\cal M}_0\dr
\nonumber\\
&&-\frac{3 F^2_1}{32 V_2}\left( 3g^4 + 2 g^2 g'^2 + g'^4\right)
+\frac{2 F_1}{v^4 V_2}
\dl ({\cal M}^\dagger_1 {\cal M}_0 +{\cal M}^\dagger_0 {\cal M}_1)
{\cal M}^\dagger_0 {\cal M}_0\dr
.
\label{delzphi}
\eea
Eqs. (\ref{delzg}) -- (\ref{delzam}), together with (\ref{delzphi}),
fix the complete set of one-loop counterterms that renormalize the
parameters contained in the leading-order Lagrangian (\ref{l2}).


\subsection{Renormalization group equations for
the coefficients of ${\cal L}_2$}

Knowing the renormalization constants, it is straightforward to
derive the renormalization-group beta functions.
For any parameter $X$ we define
\be
\label{betadef}
\beta_X = 16\pi^2 \frac{dX}{d\ln\mu}
.\ee
Recalling (\ref{zxms}), we have
\be
\label{betax}
\stackrel{{\mbox{\tiny{o}}}}{X}=Z_X X \mu^{r\varepsilon}\qquad
\Rightarrow\qquad
\beta_X=-\frac{1}{2\varepsilon}\frac{d\Delta\tilde Z_X}{d\ln\mu} X
= \Delta\tilde Z_X\, X
.
\ee
In the last step, we have used
\be
\label{delzxmu}
\frac{d\Delta\tilde Z_X}{d\ln\mu} = -2\varepsilon\, \Delta\tilde Z_X
,
\ee
which holds for $X=g_s, g, g', v^2, F_n, V_n$, 
and also for $\Delta\tilde Z_X \to \Delta\tilde{\cal M}_n$.
In all these cases, $\Delta\tilde Z_X$ is a homogeneous function
of degree 2 in the weak couplings $k=g_s, g, g', \sqrt{V_n}, {\cal M}_n/v$.
This implies (\ref{delzxmu}), since at tree level in $d=4-2\varepsilon$
dimensions
\be
\label{kmutree}
\frac{dk}{d\ln\mu} = -\varepsilon k,\qquad
{\rm whereas}\qquad
\frac{d F_n}{d\ln\mu} = \frac{d{\cal M}_n}{d\ln\mu} = 0
.
\ee

Using (\ref{betax}) and the results of section \ref{subsec:l2ren},
we finally obtain
\be
\label{betagg}
\beta_{g_s}=-\frac{1}{2}\Delta\tilde Z_G\, g_s ,\qquad
\beta_{g}=-\frac{1}{2}\Delta\tilde Z_W\, g ,\qquad
\beta_{g'}=-\frac{1}{2}\Delta\tilde Z_B\, g' , 
\ee
\bea
\sum^\infty_{n=0} \beta_{F_n} \eta^n &=&
-A_F + F'\, \int^\eta_0 ds\, A_h(s) - F' \tilde\delta
- \left( F - \frac{\eta}{2} F'\right) \Delta \tilde Z_\varphi
,
\label{betaf}
\\
v^4 \sum^\infty_{n=2} \beta_{V_n} \eta^n &=&
A_V + V'\, \int^\eta_0 ds\, A_h(s) - V' \tilde\delta
- \left( 2 V - \frac{\eta}{2} V'\right) \Delta \tilde Z_\varphi
,
\label{betav}
\\
\sum^\infty_{n=0} \beta_{{\cal M}_n} \eta^n &=&
-A_{\cal M} + {\cal M}'\, \int^\eta_0 ds\, A_h(s) - {\cal M}' \tilde\delta
+ \frac{\eta}{2} {\cal M}' \Delta \tilde Z_\varphi
,
\label{betam}
\eea
\be
\label{betav2}
\beta_{v^2} = \Delta\tilde Z_\varphi\, v^2
.
\ee
The functions $A_F$, $A_h$, $A_V$ and $A_{\cal M}$ are defined in
(\ref{afeta}) -- (\ref{ameta}). 

Eqs. (\ref{betagg}) -- (\ref{betav2}) summarize the one-loop
beta functions in the MS (or ${\overline{\rm MS}}$) scheme
for all the parameters of the leading-order electroweak chiral Lagrangian
(\ref{l2}).
We emphasize that the beta functions for the gauge couplings
(\ref{betagg}) are identical with their SM expressions, 
i.e., the contributions to the gauge-beta functions from the scalar sector only depend 
on the Goldstone modes and are independent of the Higgs couplings 
to gauge bosons.
We have checked that (\ref{betaf}) -- (\ref{betav2}) are in agreement
with the SM results, when the corresponding limit is taken.
In particular, the function on the right-hand side of (\ref{betaf})
vanishes in this limit, as it should.

\subsection{Running of the $hVV$ coupling $F_1$}
\label{subsec:f1run}

As an example for the RGE running of anomalous Higgs couplings
within the EWChL we consider $F_1$, the coupling of $h$ to
a pair of vector bosons. From (\ref{betaf}) we find
\begin{align}
  \beta_{F_1}&=\frac{3}{64V_2}F_1(F_1^2-4F_2)(3g^4+2g^2g'^2+g'^4)-
               \frac{g^2}{12}F_1\left[\frac{37}{4}(F_1^2-4F_2)+17(F_2-1)\right]
               -\frac{3}{16}g'^2F_1(F_1^2-4)\nonumber\\
             &+V_2\left[F_1\left(\frac{5}{2}(F_1^2-4F_2)+4(F_2-1)\right)+12F_3\right]
   -\frac{F_1^2-4F_2}{v^4V_2}\dl {\cal{M}}_0^{\dagger}{\cal{M}}_0({\cal{M}}_0^{\dagger}{\cal{M}}_1
   +{\cal{M}}_1^{\dagger}{\cal{M}}_0)\dr\nonumber\\
       &+2\frac{F_1-2}{v^2}\dl{\cal{M}}_0^{\dagger}{\cal{M}}_0+{\cal{M}}_1^{\dagger}{\cal{M}}_1\dr
       +\frac{4}{v^2}\dl({\cal{M}}_1^{\dagger}-{\cal{M}}_0^{\dagger})({\cal{M}}_1-{\cal{M}}_0)\dr
\label{betaf1}
\end{align}
In the limit of large top mass and Yukawa coupling this becomes
\be
\beta_{F_1}\approx (4 F_2 - F^2_1) 4  N_c \frac{m^4_t}{v^2 m^2_h}
.
\label{bf1mtop}
\ee
Therefore
\be
\frac{F_1(\mu)}{2} \approx \frac{F_1(v)}{2}
+ \frac{\beta_{F_1}}{32\pi^2} \ln\frac{\mu}{v}
\approx \frac{F_1(v)}{2} + 0.036 (4 F_2 - F^2_1) \ln\frac{\mu}{v}
\approx \frac{F_1(v)}{2} + 0.125 (4 F_2 - F^2_1)
,
\label{bf1num}
\ee
where in the last expression we have taken $\mu = 8\, {\rm TeV}$
as a representative cut-off scale for the EWChL.
Experimentally $F_1/2$ is close to 1 (within 10\%), but
$F_2$ may still deviate significantly from its SM value $F_2=1$.
Eq. (\ref{bf1num}) indicates that the difference 
between $F_1$ at the electroweak scale $v$ and at the high scale $\mu$
may be appreciable.


\section{Renormalization of the next-to-leading-order counterterms}
\label{sec:NLO_ren}
\setcounter{equation}{0}

Since ${\cal L}_2$ is not renormalizable, one also needs to introduce a
set ${\cal L}_4 $ of new operators in order to absorb all the divergences generated at one loop.
The structure of ${\cal L}_4$ is entirely dictated by the symmetries of ${\cal L}_2$ and by power
counting. A complete set of counterterms at the one-loop level is already
available \cite{Buchalla:2013rka} and we will stick to the operator basis displayed there.

In the previous Section we dealt with the divergences that can be absorbed 
through renormalization of the leading-order Lagrangian ${\cal L}_2$. The remaining
divergences have to be eliminated through the renormalization of the coun\-ter\-terms
at next-to-leading order. These remaining divergences are given by
\be
{\cal L}_{\rm div} - \Delta {\cal L}_2 =
\Delta {\cal L}_{\beta_1} 
+ \Delta {\cal L}_{UhD^4}
+ \Delta {\cal L}_{\psi^2 UhD} 
+ \Delta {\cal L}_{\psi^2 UhD^2} + \Delta {\cal L}_{\psi^4 Uh}
,
\ee
with
\bea
\Delta {\cal L}_{\beta_1} &\equiv & \Delta^{(0)} {\cal L}_{\beta_1} + \Delta^{(1)} {\cal L}_{\beta_1}
~ = ~
- \frac{1}{16\pi^2} \frac{1}{d-4}\left(
- \frac{5}{6} \right)  g'^2 v^2 (\kappa^2 - 1)F\,\langle\tau_L L^\mu\rangle \langle\tau_L L_\mu\rangle
\\
\nonumber\\
\Delta {\cal L}_{UhD^4} &\equiv & \Delta^{(0)} {\cal L}_{UhD^4} 
~=~
- \frac{1}{16\pi^2} \frac{1}{d-4}\bigg\{
\left(\frac{(\kappa^2-1)^2}{12}+\frac{F^2{\cal B}^2}{8}\right)
\langle L^\mu L_\mu\rangle^2 
+
\frac{(\kappa^2-1)^2}{6} \langle L_\mu L_\nu\rangle \langle L^\mu L^\nu\rangle
\nonumber\\
&&
\qquad\qquad\qquad\quad
-\left((\kappa^2-1){\cal B}+\frac{\kappa'^2}{6}\right)
\langle L^\mu L_\mu\rangle \partial^\nu\eta\partial_\nu\eta
+\frac{2}{3}\kappa'^2\langle L_\mu L_\nu\rangle \partial^\mu\eta\partial^\nu\eta
+\frac{3}{2}{\cal B}^2(\partial^\mu\eta\partial_\mu\eta)^2
\bigg\}
,
\nonumber\\
\\
\Delta {\cal L}_{\psi^2 Uh D} & \equiv &
\Delta^{(1/2)} {\cal L}_{\psi^2 UhD} + \Delta^{(1)} {\cal L}_{\psi^2 UhD}
\nonumber\\ 
&=& 
- \frac{1}{16\pi^2} \frac{1}{d-4} \bigg\{
{\bar\psi}_L U {\mathfrak L} ({\cal M} , {\cal M}^\dagger ) U^\dagger \not\!\! L \psi_L
+
{\bar\psi}_L \not\!\! L U {\mathfrak L} ({\cal M} , {\cal M}^\dagger )^\dagger  U^\dagger \psi_L
\nonumber\\
&&
+
\frac{\kappa^2 - 1}{12} \big[
2 g'^2 \langle \tau_L L^\nu \rangle  {\bar e}_R \gamma_\nu  e_R
- 2 g'^2 \langle \tau_L L^\nu \rangle  {\bar\psi}_L \gamma_\nu Y_L \psi_L
- \frac{4}{3} g'^2 \langle \tau_L L^\nu \rangle  {\bar u}_R \gamma_\nu  u_R
+ \frac{2}{3} g'^2 \langle \tau_L L^\nu \rangle  {\bar d}_R \gamma_\nu  d_R
\big]
\nonumber\\
&&
-\frac{F^{-1}}{v^2}\bar\psi_R{\cal M}^\dagger 
 U^\dagger \not\!\! L U{\cal M}\psi_R 
-\frac{1}{v^2}\bar\psi_R{\cal M}'^\dagger 
 U^\dagger \not\!\! L U{\cal M}'\psi_R
 +\frac{\kappa}{v^2}F^{-1/2}\left(\bar\psi_R{\cal M}^\dagger U^\dagger
   \not\!\! L U{\cal M}'\psi_R +{\rm h.c.}\right)
   \bigg\}
   ,
\nonumber\\
\\
\Delta {\cal L}_{\psi^2 Uh D^2} &\equiv & \Delta^{(1/2)} {\cal L}_{\psi^2 Uh D^2}
\nonumber\\
&=&
- \frac{1}{16\pi^2} \frac{1}{d-4} \bigg\{ \langle L^\mu L_\mu\rangle
\left[\frac{F{\cal B}}{2v^2}\bar\psi_L U{\cal M}''\psi_R
-\frac{\kappa^2-1}{Fv^2}
\bar\psi_L U\left(\frac{F'}{2}{\cal M}'-{\cal M}\right)\psi_R +{\rm h.c.}
\right]
\nonumber\\
&&
+\frac{2\kappa'}{v^2}\partial^\mu\eta\left(
i\bar\psi_L L_\mu U(F^{-1/2}{\cal M})'\psi_R +{\rm h.c.}\right)
+\frac{3{\cal B}}{Fv^2} \partial^\mu\eta\partial_\mu\eta\left(\bar\psi_L U
\left(\frac{F'}{2}{\cal M}'-{\cal M}\right)\psi_R +{\rm h.c.}\right)
\!\bigg\}
,
\\
\nonumber\\
\Delta {\cal L}_{\psi^4 Uh} &\equiv & \Delta^{(1/2)} {\cal L}_{\psi^4 Uh}
\nonumber\\
&=&
- \frac{1}{16\pi^2} \frac{1}{d-4} \bigg\{\frac{3F^{-2}}{2v^4}\left(\bar\psi_L U
\left(\frac{F'}{2}{\cal M}'-{\cal M}\right)\psi_R +{\rm h.c.}\right)^2
+\frac{1}{2v^4}\left(\bar\psi_L U{\cal M}''\psi_R +{\rm h.c.}\right)^2
\nonumber\\
&&+\frac{4}{v^4}\left(i\bar\psi_L U T^a
\left(F^{-1/2} {\cal M}\right)'\psi_R +{\rm h.c.}\right)^2 
\bigg\}
.
\label{ldivnlo}
\eea
In the expression of $\Delta {\cal L}_{\psi^2 Uh D}$, the quantity
\bea
{\mathfrak L}  ({\cal M} , {\cal M}^\dagger ) &=&
g^2 \frac{\kappa^2 - 1}{24} + \frac{1}{2 v^2} \left[ {\cal M}'{\cal M}'^\dagger + F^{-1} {\cal M}{\cal M}^\dagger  \right]
- \frac{\kappa}{v^2} F^{-1/2}  {\cal M}'{\cal M}^\dagger
\nonumber\\
&&
- \frac{1}{2 v^2} \int_0^{\eta} ds \left[ {\cal M}'{\cal M}''^\dagger - {\cal M}''{\cal M}'^\dagger 
- F^{-1} {\cal M}  {\cal M}^{\dagger\prime} + F^{-1} {\cal M}' {\cal M}^\dagger\right]
\lbl{lmmdef}
\eea
has been introduced.

\subsection{Renormalization of the counterterm $\beta_1$}

Let us start with the elimination of the divergent term $\Delta {\cal L}_{\beta_1}$.
It requires the counterterm ${\cal L}_{\beta_1}$ given in \cite{Buchalla:2013rka}, 
namely (note the slight change in notation as compared to this reference)
\be
{\cal L}_{\beta_1} = - v^2 \langle\tau_L L^\mu\rangle \langle\tau_L L_\mu\rangle
\left[ \beta_1 + F_{\beta_1} (\eta)  \right]
,\quad 
F_{\beta_1} (\eta) = \sum_{n\ge 1} f_{\beta_1,n} \eta^n
.
\ee
In order to perform the renormalization (for instance, in the 
${\overline{\rm MS}}$ scheme), one interprets the coefficients as unrenormalized ones and writes
\be
\stackrel{{\mbox{\tiny{o}}}}{f}_{\beta_1,n} = f_{\beta_1,n} (\mu) 
+ \frac{\gamma_{\beta_1,n}}{16\pi^2} \mu^{d-4} \left[ \frac{1}{d-4} - \frac{1}{2} (\ln 4\pi - \gamma_E) \right]
,
\ \stackrel{{\mbox{\tiny{o}}}}{\beta}_1 = \beta_1 (\mu) 
+ \frac{\gamma_{\beta_1}}{16\pi^2} \mu^{d-4} \left[ \frac{1}{d-4} - \frac{1}{2} (\ln 4\pi - \gamma_E) \right]
,
\ee
i.e.
\be
\stackrel{{\mbox{\tiny{o}}}}{F}_{\beta_1} (\eta) = F_{\beta_1} (\eta ; \mu)  
+ \frac{\Gamma_{\beta_1} (\eta)}{16\pi^2} \mu^{d-4} \left[ \frac{1}{d-4} - \frac{1}{2} (\ln 4\pi - \gamma_E) \right]
,
\ee
with
\be
\label{gambet1}
\gamma_{\beta_1} =
\frac{5}{24}  g'^2  (F_1^2 - 4)
,
\quad 
\Gamma_{\beta_1} (\eta) = \frac{5}{6} g'^2 
\left[ {\bar\kappa} ({\bar\kappa} + F_1) + {\bar F} (\kappa^2 - 1)  \right]
.
\ee

\subsection{Renormalization of the counterterms 
in the class $UhD^4$}

The elimination of the divergences contained in $\Delta^{(0)} {\cal L}_{UhD^4}$
requires five counterterms of the class $UhD^4$,
\be
{\cal L}_{UhD^4} = \sum_{i=1}^{15} {\cal O}_{Di} \left[ c_{Di} + F_{Di} (\eta) \right]
,\quad 
F_{Di} (\eta) = \sum_{n\ge 1} f_{Di,n} \eta^n
,
\ee
namely ${\cal O}_{Di}$ for $i= 1, 2, 7, 8, 11$.
Notice that in contrast to Ref. \cite{Buchalla:2013rka}, we have not written
the overall factor $v^2/\Lambda^2$, 
and we have introduced the specific couplings
$c_{Di}$ in a different way. The renormalization proceeds as previously,
\be
\stackrel{{\mbox{\tiny{o}}}}{c}_{Di} = c_{Di} ( \mu)  
+ \frac{\gamma_{Di}}{16\pi^2} \mu^{d-4} \left[ \frac{1}{d-4} - 
\frac{1}{2} (\ln 4\pi - \gamma_E) \right]
, \quad
\stackrel{{\mbox{\tiny{o}}}}{F}_{Di} (\eta) = F_{Di} (\eta ; \mu)  
+ \frac{\Gamma_{Di} (\eta)}{16\pi^2} \mu^{d-4} \left[ \frac{1}{d-4} - 
\frac{1}{2} (\ln 4\pi - \gamma_E) \right]
,
\ee
with
\bea
\gamma_{D1} &=& \frac{1}{12} \left( \frac{F_1^2}{4} - 1 \right)^2 + \frac{1}{8} \left( \frac{F_1^2}{4} - F_2 \right)^2
,~~
\gamma_{D2} \ =\ \frac{1}{6} \left( \frac{F_1^2}{4} - 1 \right)^2
,
\nonumber\\
\gamma_{D7} &=& - \frac{1}{6} \left( \frac{F_1^2}{4} - F_2 \right) \left( \frac{7}{4} F_1^2 - F_2 - 6  \right) 
,~~
\gamma_{D8} \ =\ \frac{2}{3} \left( \frac{F_1^2}{4} - F_2 \right)^2
,~~
\gamma_{D11} \ =\ \frac{3}{2} \left( \frac{F_1^2}{4} - F_2 \right)^2
,~~
\label{gamd1}
\eea
and
\bea
\nonumber\\
\Gamma_{D1} &=& \frac{1}{12} {\bar\kappa} ({\bar\kappa} + F_1) 
\left[ {\bar\kappa} ({\bar\kappa} + F_1) + \frac{F_1^2}{2} - 2 \right]  
+ \frac{1}{8} \left( F {\cal B} - \frac{F_1^2}{4} + F_2  \right) \left( F {\cal B} + \frac{F_1^2}{4} - F_2  \right)
,
\nonumber\\
\Gamma_{D2} &=& \frac{1}{6} {\bar\kappa} ({\bar\kappa} + F_1) 
\left[ {\bar\kappa} ({\bar\kappa} + F_1) + \frac{F_1^2}{2} - 2 \right]
,
\nonumber\\
\Gamma_{D7} &=& - {\bar{\cal B}} \left( \frac{F_1^2}{4} - 1 \right) - {\cal B} {\bar\kappa} ({\bar\kappa} + F_1)
- \frac{{\bar F}}{6} \left( \frac{F_1^2}{4} - F_2 \right)^2 
- \frac{F}{6} {\bar{\cal B}} \left( {\bar{\cal B}} + \frac{F_1^2}{2} - 2 F_2  \right)
,
\nonumber\\
\Gamma_{D8} &=& \frac{2}{3} F {\bar{\cal B}} \left( {\bar{\cal B}} + \frac{F_1^2}{2} - 2 F_2 \right)
+ \frac{2}{3} {\bar F} \left( \frac{F_1^2}{4} - F_2 \right)^2  
,
\nonumber\\
\Gamma_{D11} &=& \frac{3}{2} {\bar{\cal B}} \left( {\bar{\cal B}} + \frac{F_1^2}{2} - 2 F_2 \right)
.
\label{gamd2}
\eea

\subsection{Renormalization of the counterterms in the class $\psi^2 UhD$}

In order to proceed with the renormalization of the operators of the class $\psi^2 UhD$,
we first need to express ${\cal L}_{\psi^2 UhD}$ in terms of the 
basis operators given in Ref. \cite{Buchalla:2013rka}. Recalling that ${\cal M}$ and
therefore also ${\mathfrak L}$ of \rf{lmmdef} are diagonal in $SU(2)$ space, one has
\be\label{mathfracl}
{\mathfrak L} = \frac{1}{2} \langle {\mathfrak L} \rangle + 2 T^3 \langle T^3 {\mathfrak L} \rangle
,
\ee
so that
\be\label{mathfracl2}
{\bar\psi}_L U {\mathfrak L}  U^\dagger \! \not\!\! L \psi_L
=
\frac{1}{2} {\bar\psi}_L \langle {\mathfrak L} \rangle \! \not\!\! L \psi_L
+ 2 {\bar\psi}_L \langle {\mathfrak L} T^3 \rangle \tau_L \! \not\!\! L \psi_L
.
\ee
Note that traces are over $SU(2)$ only, i.e., the expressions on the right-hand side of
(\ref{mathfracl}) are still matrices in flavor space. 

Upon using the identity \rf{identity_slashL}, (\ref{mathfracl2}) becomes 
($P_\pm = \frac{1}{2} \pm T^3$, $P_{12} = T^1 + i T^2$, $P_{21} = T^1 - i T^2$)
\bea
{\bar\psi}_L U {\mathfrak L}  U^\dagger \! \not\!\! L \psi_L
&=&
{\bar\psi}_L \gamma^\mu \langle {\mathfrak L} \rangle \tau_L \psi_L \langle \tau_L L_\mu \rangle
+ \frac{1}{2} {\bar\psi}_L \gamma^\mu \langle {\mathfrak L} \rangle U P_{12} U^\dagger \psi_L \langle U P_{21} U^\dagger  L_\mu \rangle
+ \frac{1}{2} {\bar\psi}_L \gamma^\mu \langle {\mathfrak L} \rangle U P_{21} U^\dagger \psi_L \langle U P_{12} U^\dagger  L_\mu \rangle
\nonumber\\
&&\!\!\!\!\!
+ \, {\bar\psi}_L \gamma^\mu \langle {\mathfrak L} T^3 \rangle \psi_L \langle \tau_L L_\mu \rangle
+ {\bar\psi}_L \gamma^\mu \langle {\mathfrak L} T^3 \rangle U P_{12} U^\dagger \psi_L \langle U P_{21} U^\dagger  L_\mu \rangle
- {\bar\psi}_L \gamma^\mu \langle {\mathfrak L} T^3 \rangle U P_{21} U^\dagger \psi_L \langle U P_{12} U^\dagger  L_\mu \rangle
\nonumber\\
&=&
{\bar\psi}_L \gamma^\mu \langle {\mathfrak L} T^3 \rangle \psi_L \langle \tau_L L_\mu \rangle
+ {\bar\psi}_L \gamma^\mu \langle {\mathfrak L} \rangle \tau_L \psi_L \langle \tau_L L_\mu \rangle
\nonumber\\
&&\!\!\!\!\!
+ {\bar\psi}_L \gamma^\mu \langle {\mathfrak L} P_+ \rangle U P_{12} U^\dagger \psi_L \langle U P_{21} U^\dagger  L_\mu \rangle
+ {\bar\psi}_L \gamma^\mu \langle {\mathfrak L} P_- \rangle U P_{21} U^\dagger \psi_L \langle U P_{12} U^\dagger  L_\mu \rangle
,
\label{ululpsil}
\eea
due to the relations $T^3 P_{12} = + P_{12}/2$ and $T^3 P_{21} = - P_{21}/2$.

The terms involving the right-handed fermion fields can essentially be handled
along similar lines. One first establishes an identity similar to \rf{identity_slashL},
\bea
{\bar\psi}_{R} O_1  U^\dagger \! \not\!\!L U O_2 \psi_{R} &=& 2 {\bar\psi}_{R} \gamma^\mu O_1 T^a O_2 \psi_{R} \langle U T^a U^\dagger L_\mu \rangle
\nonumber\\
&=&
{\bar\psi}_{R} \gamma^\mu O_1 P_{12} O_2 \psi_{R} \langle U P_{21} U^\dagger L_\mu \rangle
+
{\bar\psi}_{R} \gamma^\mu O_1 P_{21} O_2 \psi_{R} \langle U P_{12} U^\dagger L_\mu \rangle
\nonumber\\
&&
+ 2 {\bar\psi}_{R} \gamma^\mu O_1 T^3 O_2 \psi_{R} \langle \tau_L L_\mu \rangle
.
\lbl{identity_slashR}
\eea
Then one obtains, for instance,
\bea
\bar\psi_R {\cal M}^\dagger U^\dagger \! \not\!\! L U {\cal M}\psi_R 
&=&
2 {\bar\psi}_{R} \gamma^\mu {\cal M}^\dagger T^3 {\cal M} \psi_{R} \langle \tau_L L_\mu \rangle
+ {\bar\psi}_{R} \gamma^\mu {\cal M}^\dagger P_{12} {\cal M} \psi_{R} \langle U P_{21} U^\dagger L_\mu \rangle
+ {\bar\psi}_{R} \gamma^\mu {\cal M}^\dagger P_{21} {\cal M} \psi_{R} \langle U P_{12} U^\dagger L_\mu \rangle
\nonumber\\
&=&
{\bar u_R} \gamma^\mu {\cal M}_u^\dagger {\cal M}_u u_R \langle \tau_L L_\mu \rangle
- {\bar d_R} \gamma^\mu {\cal M}_d^\dagger {\cal M}_d d_R \langle \tau_L L_\mu \rangle
- {\bar e_R} \gamma^\mu {\cal M}_e^\dagger {\cal M}_e e_R \langle \tau_L L_\mu \rangle
\nonumber\\
&&
+ {\bar u_R} \gamma^\mu {\cal M}_u^\dagger {\cal M}_d d_R  \langle U P_{21} U^\dagger L_\mu \rangle
+ {\bar d_R} \gamma^\mu {\cal M}_d^\dagger {\cal M}_u u_R  \langle U P_{12} U^\dagger L_\mu \rangle
.
\eea
These divergences can now be removed through the renormalization of the 
operators
${\cal O}_{\psi Vi}$ (and ${\cal O}_{\psi Vi}^\dagger$) of ${\cal L}_{\psi^2 UhD}$.
Explicitly, one has (notice that the functions $F_{\psi Vi}$ are actually 
matrices in generation space)
\be 
\stackrel{{\mbox{\tiny{o}}}}{F}_{\psi Vi} (\eta) = F_{\psi Vi} (\eta ; \mu)  
+ \frac{\Gamma_{\psi Vi} (\eta)}{16\pi^2} \mu^{d-4} \left[ \frac{1}{d-4} - \frac{1}{2} (\ln 4\pi - \gamma_E) \right]
,
\ee
with
\bea
\Gamma_{\psi V1} &=& - \frac{1}{2} ( {\mathfrak L}_u - {\mathfrak L}_d  + {\rm h.c.} ) + g'^2 \frac{\kappa^2 - 1}{36} ,
\nonumber\\
\Gamma_{\psi V2} &=& - {\mathfrak L}_u - {\mathfrak L}_d + {\rm h.c.} ,
\nonumber\\
\Gamma_{\psi V3} &=& - {\mathfrak L}_u - {\mathfrak L}_d^\dagger ,
\nonumber\\
\Gamma_{\psi V4} &=&  + \frac{1}{v^2} \left[ F^{-1} {\cal M}_u^\dagger {\cal M}_u + {\cal M}_u'^\dagger {\cal M}_u' 
- \kappa F^{-1/2} \left( {\cal M}_u^\dagger {\cal M}_u' + {\cal M}_u'^\dagger {\cal M}_u \right) \right] + g'^2 \frac{\kappa^2 - 1}{9} ,
\nonumber\\
\Gamma_{\psi V5} &=&  - \frac{1}{v^2} \left[ F^{-1} {\cal M}_d^\dagger {\cal M}_d + {\cal M}_d'^\dagger {\cal M}_d'
- \kappa F^{-1/2} \left( {\cal M}_d^\dagger {\cal M}_d' + {\cal M}_d'^\dagger {\cal M}_d \right) \right] - g'^2 \frac{\kappa^2 - 1}{18} ,
\nonumber\\
\Gamma_{\psi V6} &=&  + \frac{1}{v^2} \left[ F^{-1} {\cal M}_u^\dagger {\cal M}_d + {\cal M}_u'^\dagger {\cal M}_d'
- \kappa F^{-1/2} \left( {\cal M}_u^\dagger {\cal M}_d' + {\cal M}_u'^\dagger {\cal M}_d \right) \right] ,
\nonumber\\
\Gamma_{\psi V7} &=& + \frac{1}{2} ({\mathfrak L}_e  + {\rm h.c.} ) - g'^2 \frac{\kappa^2 - 1}{12} ,
\nonumber\\
\Gamma_{\psi V8} &=& - {\mathfrak L}_e   + {\rm h.c.} ,
\nonumber\\
\Gamma_{\psi V9} &=& - {\mathfrak L}_e^\dagger , 
\nonumber\\
\Gamma_{\psi V10} &=&  - \frac{1}{v^2} \left[ F^{-1} {\cal M}_e^\dagger {\cal M}_e + {\cal M}_e'^\dagger {\cal M}_e'
  - \kappa F^{-1/2} \left( {\cal M}_e^\dagger {\cal M}_e' + {\cal M}_e'^\dagger {\cal M}_e \right) \right] - g'^2 \frac{\kappa^2 - 1}{6} .
\label{gampsiv}
\eea
In these expressions, 
${\mathfrak L}_j \equiv {\mathfrak L} ({\cal M}_j , {\cal M}_j^\dagger)$ with $j=u,d,e$.

\subsection{Renormalization of the counterterms in the class $\psi^2 UhD^2$}

Recalling that the matrix ${\cal M}$ is $SU(2)$-diagonal, one has the
following relations:
\bea
{\bar\psi}_L U {\cal F} ({\cal M}) \psi_R &=& {\bar q}_L U P_+ {\cal F} ({\cal M}_u) q_R 
+ {\bar q}_L U P_- {\cal F} ({\cal M}_d) q_R  + {\bar l}_L U P_- {\cal F} ({\cal M}_e) l_R 
,
\lbl{obs}
\\
{\bar\psi}_L L_\mu U {\cal F} ({\cal M}) \psi_R &=&
2 {\bar\psi}_L U T^a {\cal F} ({\cal M}) \psi_R \langle U T^a U^\dagger L_\mu \rangle
\nonumber\\
&=&
{\bar\psi}_L U P_{12} {\cal F} ({\cal M}) \psi_R \langle U P_{21} U^\dagger L_\mu \rangle
+ {\bar\psi}_L U P_{21} {\cal F} ({\cal M}) \psi_R \langle U P_{12} U^\dagger L_\mu \rangle
+ 2 {\bar\psi}_L U T^3 {\cal F} ({\cal M}) \psi_R \langle \tau_L L_\mu \rangle
\nonumber\\
&=&
{\bar\psi}_L U P_{12} {\cal F} ({\cal M}) \psi_R \langle U P_{21} U^\dagger L_\mu \rangle
+ {\bar\psi}_L U P_{21} {\cal F} ({\cal M}) \psi_R \langle U P_{12} U^\dagger L_\mu \rangle
\nonumber\\
&&
+ {\bar\psi}_L U P_+ {\cal F} ({\cal M}) \psi_R \langle \tau_L L_\mu \rangle
- {\bar\psi}_L U P_- {\cal F} ({\cal M}) \psi_R \langle \tau_L L_\mu \rangle
\nonumber\\
&=&
{\bar q}_L U P_{12} {\cal F} ({\cal M}_d) q_R \langle U P_{21} U^\dagger L_\mu \rangle
+ {\bar q}_L U P_{21} {\cal F} ({\cal M}_u) q_R \langle U P_{12} U^\dagger L_\mu \rangle
+ {\bar l}_L U P_{12} {\cal F} ({\cal M}_e) l_R \langle U P_{21} U^\dagger L_\mu \rangle
\nonumber\\
&&
+ {\bar q}_L U P_+ {\cal F} ({\cal M}_u) q_R \langle \tau_L L_\mu \rangle
- {\bar q}_L U P_- {\cal F} ({\cal M}_d) q_R \langle \tau_L L_\mu \rangle
- {\bar l}_L U P_- {\cal F} ({\cal M}_e) l_R \langle \tau_L L_\mu \rangle
.
\eea
Using them one can rewrite $\Delta^{(1/2)} {\cal L}_{\psi^2 Uh D^2}$ as
\bea
\Delta^{(1/2)} {\cal L}_{\psi^2 Uh D^2} &=& 
- \frac{1}{16\pi^2} \frac{1}{d-4} \bigg\{ 
 {\bar q}_L U P_+
\left[ \frac{F{\cal B}}{2v^2} {\cal M}''_u -\frac{\kappa^2-1}{Fv^2} \left(\frac{F'}{2}{\cal M}'_u-{\cal M}_u\right) \right] q_R
\langle L^\mu L_\mu\rangle
\nonumber\\
&&
+ 
 {\bar q}_L U P_-
\left[ \frac{F{\cal B}}{2v^2} {\cal M}''_d -\frac{\kappa^2-1}{Fv^2} \left(\frac{F'}{2}{\cal M}'_d-{\cal M}_d\right) \right] q_R
\langle L^\mu L_\mu\rangle
\nonumber\\
&&
+ 
 {\bar l}_L U P_-
\left[ \frac{F{\cal B}}{2v^2} {\cal M}''_e -\frac{\kappa^2-1}{Fv^2} \left(\frac{F'}{2}{\cal M}'_e-{\cal M}_e\right) \right] l_R
\langle L^\mu L_\mu\rangle
\nonumber\\
&&
+\frac{2i \kappa'}{v^2}\partial^\mu\eta
\left[
{\bar q}_L U P_{12} (F^{-1/2}{\cal M}_d)' q_R \langle U P_{21} U^\dagger L_\mu \rangle
+ {\bar q}_L U P_{21} (F^{-1/2}{\cal M}_u)' q_R \langle U P_{12} U^\dagger L_\mu \rangle
\right.
\nonumber\\
&&
\left.
+ {\bar l}_L U P_{12} (F^{-1/2}{\cal M}_e)' l_R \langle U P_{21} U^\dagger L_\mu \rangle
+ {\bar q}_L U P_+ (F^{-1/2}{\cal M}_u)' q_R \langle \tau_L L_\mu \rangle
\right.
\nonumber\\
&&
\left.
- {\bar q}_L U P_- (F^{-1/2}{\cal M}_d)' q_R \langle \tau_L L_\mu \rangle
- {\bar l}_L U P_- (F^{-1/2}{\cal M}_e)' l_R \langle \tau_L L_\mu \rangle
\right]
\nonumber\\
&&
+\frac{3{\cal B}}{Fv^2} \partial^\mu\eta\partial_\mu\eta
\left[
\bar q_L U P_+
\left(\frac{F'}{2}{\cal M}'_u-{\cal M}_u\right)q_R 
+ 
\bar q_L U P_-
\left(\frac{F'}{2}{\cal M}'_d - {\cal M}_d\right)q_R 
\right.
\nonumber\\
&&
\left.
+ \bar l_L U P_-
\left(\frac{F'}{2}{\cal M}'_e - {\cal M}_e\right)l_R\right]
+ {\rm h.c.}
\!\bigg\}
.
\eea
These divergences are removed through the renormalization of the operators
${\cal O}_{\psi Si}$ (and ${\cal O}_{\psi Si}^\dagger$) of ${\cal L}_{\psi^2 UhD^2}$, 
with (in order of appearance in the previous formula) $i=1,2,7,12,13,17,10,11,16,14,15,18$. 
Explicitly, one has (notice that the functions $F_{\psi Si}$ are actually 
matrices in generation space)
\be 
\stackrel{{\mbox{\tiny{o}}}}{F}_{\psi Si} (\eta) = F_{\psi Si} (\eta ; \mu)  
+ \frac{\Gamma_{\psi Si} (\eta)}{16\pi^2} \mu^{d-4} \left[ \frac{1}{d-4} 
- \frac{1}{2} (\ln 4\pi - \gamma_E) \right]
,
\ee
with
\bea
\Gamma_{\psi S1} &=& \frac{F{\cal B}}{2v^2} {\cal M}''_u -\frac{\kappa^2-1}{Fv^2} \left(\frac{F'}{2}{\cal M}'_u-{\cal M}_u\right),
\nonumber\\
\Gamma_{\psi S2} &=& \frac{F{\cal B}}{2v^2} {\cal M}''_d -\frac{\kappa^2-1}{Fv^2} \left(\frac{F'}{2}{\cal M}'_d-{\cal M}_d\right),
\nonumber\\
\Gamma_{\psi S7} &=& \frac{F{\cal B}}{2v^2} {\cal M}''_e -\frac{\kappa^2-1}{Fv^2} \left(\frac{F'}{2}{\cal M}'_e-{\cal M}_e\right),
\nonumber\\
\Gamma_{\psi S10} &=& \frac{2i \kappa'}{v^2} (F^{-1/2}{\cal M}_u)',
\nonumber\\
\Gamma_{\psi S11} &=& - \frac{2i \kappa'}{v^2} (F^{-1/2}{\cal M}_d)',
\nonumber\\
\Gamma_{\psi S12} &=& \frac{2i \kappa'}{v^2} (F^{-1/2}{\cal M}_d)',
\nonumber\\
\Gamma_{\psi S13} &=& \frac{2i \kappa'}{v^2} (F^{-1/2}{\cal M}_u)',
\nonumber\\
\Gamma_{\psi S14} &=& \frac{3{\cal B}}{Fv^2} \left(\frac{F'}{2}{\cal M}'_u-{\cal M}_u\right),
\nonumber\\
\Gamma_{\psi S15} &=& \frac{3{\cal B}}{Fv^2} \left(\frac{F'}{2}{\cal M}'_d-{\cal M}_d\right),
\nonumber\\
\Gamma_{\psi S16} &=& - \frac{2i \kappa'}{v^2} (F^{-1/2}{\cal M}_e)',
\nonumber\\
\Gamma_{\psi S17} &=& \frac{2i \kappa'}{v^2} (F^{-1/2}{\cal M}_e)',
\nonumber\\
\Gamma_{\psi S18} &=& \frac{3{\cal B}}{Fv^2} \left(\frac{F'}{2}{\cal M}'_e-{\cal M}_e\right).
\label{gampsis}
\eea

\subsection{Renormalization of the counterterms in the class $\psi^4 Uh$}

The expression to start with reads
\bea
\Delta {\cal L}_{\psi^4 Uh} 
&=&
- \frac{1}{16\pi^2} \frac{1}{d-4} \bigg\{\frac{3F^{-2}}{2v^4}\left(\bar\psi_L U
\left(\frac{F'}{2}{\cal M}'-{\cal M}\right)\psi_R +{\rm h.c.}\right)^2
+\frac{1}{2v^4}\left(\bar\psi_L U{\cal M}''\psi_R +{\rm h.c.}\right)^2
\nonumber\\
&&+\frac{4}{v^4}\left(i\bar\psi_L U T^a
\left(F^{-1/2} {\cal M}\right)'\psi_R +{\rm h.c.}\right)^2 
\bigg\}
\nonumber\\
&=&
- \frac{1}{16\pi^2} \frac{1}{d-4} \bigg\{
\frac{3F^{-2}}{2v^4} \left(\bar\psi_L U \left(\frac{F'}{2}{\cal M}'-{\cal M}\right)\psi_R \right)^2
+
\frac{3F^{-2}}{2v^4} \left(\bar\psi_R \left(\frac{F'}{2}{\cal M}'-{\cal M}\right)^\dagger U^\dagger \psi_L \right)^2
\nonumber\\
&&
+ \,
\frac{3F^{-2}}{v^4} \left(\bar\psi_L U \left(\frac{F'}{2}{\cal M}'-{\cal M}\right)\psi_R \right)
\left(\bar\psi_R \left(\frac{F'}{2}{\cal M}'-{\cal M}\right)^\dagger U^\dagger \psi_L \right)
\nonumber\\
&&
+ \,
\frac{1}{2v^4} \left(\bar\psi_L U{\cal M}''\psi_R \right)^2
+
\frac{1}{2v^4} \left(\bar\psi_R {\cal M}^{\prime\prime\dagger} U^\dagger 
\psi_L\right)^2
+
\frac{1}{v^4} \left(\bar\psi_L U{\cal M}''\psi_R \right) 
\left(\bar\psi_R {\cal M}^{\prime\prime\dagger} U^\dagger \psi_L\right)
\nonumber\\
&&
- \,
\frac{4}{v^4} \left(\bar\psi_L U T^a \left(F^{-1/2} {\cal M}\right)'\psi_R \right) \left(\bar\psi_L U T^a \left(F^{-1/2} {\cal M}\right)'\psi_R \right)
\nonumber\\
&&
- \,
\frac{4}{v^4} \left(\bar\psi_R \left(F^{-1/2} {\cal M}^\dagger\right)' T^a U^\dagger \psi_L \right) 
\left(\bar\psi_R \left(F^{-1/2} {\cal M}^\dagger\right)' T^a U^\dagger \psi_L \right)
\nonumber\\
&&
+ \,
\frac{8}{v^4} \left(\bar\psi_L U T^a \left(F^{-1/2} {\cal M}\right)'\psi_R \right)
\left(\bar\psi_R \left(F^{-1/2} {\cal M}^\dagger\right)' T^a U^\dagger \psi_L \right)
.
\eea

We have to consider the following structures:
\be
(\bar\psi_L U {\cal F}({\cal M}) \psi_R)^2 + {\rm h.c.},\quad 
(\bar\psi_L U {\cal F}({\cal M}) \psi_R)(\bar\psi_R {\cal F}({\cal M})^\dagger 
U^\dagger \psi_L)
\ee
and
\be
(\bar\psi_L U T^a {\cal F}({\cal M}) \psi_R) (\bar\psi_L U T^a {\cal F}({\cal M}) \psi_R) + {\rm h.c.},
\quad (\bar\psi_L U T^a {\cal F}({\cal M}) \psi_R)(\bar\psi_R {\cal F}({\cal M})^\dagger T^a U^\dagger \psi_L)
,
\ee
and decompose them onto the operator basis of Ref. \cite{Buchalla:2013rka}.
Notice that the coefficients of the four-fermion operators are actually
rank-four tensors in generation space. We use the notation
\bea
{\cal F}^{(1)} \otimes {\cal F}^{(2)}\, {\cal O} &\equiv&
{\cal F}^{(1)}_{ij} {\cal F}^{(2)}_{kl}\, 
\bar\psi_i \ldots \psi_j\, \bar\psi_k \ldots \psi_l \nonumber\\
{\cal F}^{(1)}\,\, \tilde\otimes \,\, {\cal F}^{(2)}\, {\cal O} &\equiv&
{\cal F}^{(1)}_{il} {\cal F}^{(2)}_{kj}\, 
\bar\psi_i \ldots \psi_j\, \bar\psi_k \ldots \psi_l 
\eea
for an operator 
${\cal O}\equiv \bar\psi_i \ldots \psi_j\, \bar\psi_k \ldots \psi_l$,
where $i,j,k,l$ are generation indices.
For the basis operators ${\cal O}_{AB}$ to be used below, we follow the notation of \cite{Buchalla:2012qq}.

Using Eq. \rf{obs}, one obtains
\bea
(\bar\psi_L U {\cal F}({\cal M}) \psi_R)^2 &=& ({\bar q}_L U P_+ {\cal F} ({\cal M}_u) q_R )^2 + ({\bar q}_L U P_- {\cal F} ({\cal M}_d) q_R )^2
+ ({\bar l}_L U P_- {\cal F} ({\cal M}_e) l_R)^2
\nonumber\\
&&\!\!\!
+ \, 2 ({\bar q}_L U P_+ {\cal F} ({\cal M}_u) q_R ) ({\bar q}_L U P_- {\cal F} ({\cal M}_d) q_R )
+ 2 ({\bar q}_L U P_+ {\cal F} ({\cal M}_u) q_R ) ({\bar l}_L U P_- {\cal F} ({\cal M}_e) l_R)
\nonumber\\
&&\!\!\!
+ \, 2 ({\bar q}_L U P_- {\cal F} ({\cal M}_d) q_R ) ({\bar l}_L U P_- {\cal F} ({\cal M}_e) l_R)
\nonumber\\
&=&
{\cal F} ({\cal M}_u) \otimes {\cal F} ({\cal M}_u) {\cal O}_{FY1} + {\cal F} ({\cal M}_d) \otimes {\cal F} ({\cal M}_d) {\cal O}_{FY3}
+ {\cal F} ({\cal M}_e) \otimes {\cal F} ({\cal M}_e) {\cal O}_{FY10}
\nonumber\\
&&\!\!\!
+ \, 2 {\cal F} ({\cal M}_u) \otimes {\cal F} ({\cal M}_d) {\cal O}_{ST5} + 2 {\cal F} ({\cal M}_u) \otimes {\cal F} ({\cal M}_e) {\cal O}_{ST9}
+ 2 {\cal F} ({\cal M}_d) \otimes {\cal F} ({\cal M}_e) {\cal O}_{FY7}
,
\eea
and
\bea
(\bar\psi_L U T^a {\cal F}({\cal M}) \psi_R) (\bar\psi_L U T^a {\cal F}({\cal M}) \psi_R) &=& 
({\bar q}_L U T^a P_+ {\cal F} ({\cal M}_u) q_R ) ({\bar q}_L U T^a P_+ {\cal F} ({\cal M}_u) q_R ) 
\nonumber\\
&&\!\!\!
+ \, ({\bar q}_L U T^a P_- {\cal F} ({\cal M}_d) q_R ) ({\bar q}_L U T^a P_- {\cal F} ({\cal M}_d) q_R )
\nonumber\\
&&\!\!\!
+ \, ({\bar l}_L U T^a P_- {\cal F} ({\cal M}_e) l_R) ({\bar l}_L U T^a P_- {\cal F} ({\cal M}_e) l_R)
\nonumber\\
&&\!\!\!
+ \, 2 ({\bar q}_L U T^a P_+ {\cal F} ({\cal M}_u) q_R ) ({\bar q}_L U T^a P_- {\cal F} ({\cal M}_d) q_R )
\nonumber\\
&&\!\!\!
+ \, 2 ({\bar q}_L U T^a P_+ {\cal F} ({\cal M}_u) q_R ) ({\bar l}_L U T^a P_- {\cal F} ({\cal M}_e) l_R)
\nonumber\\
&&\!\!\!
+ \, 2 ({\bar q}_L U T^a P_- {\cal F} ({\cal M}_d) q_R ) ({\bar l}_L U T^a P_- {\cal F} ({\cal M}_e) l_R)
.
\eea
Next, since
\be
T^a \otimes T^a = \frac{1}{2} P_{12} \otimes P_{21} + \frac{1}{2} P_{21} \otimes P_{12}
+ \frac{1}{4} P_{+} \otimes P_{+} + \frac{1}{4} P_{-} \otimes P_{-} - \frac{1}{4} P_{+} \otimes P_{-} - \frac{1}{4} P_{-} \otimes P_{+} ,
\ee
one has
\be
T^a P_\pm \otimes T^a P_\pm = \frac{1}{4} P_\pm \otimes P_\pm , ~~ 
T^a P_+ \otimes T^a P_- = \frac{1}{2} P_{21} \otimes P_{12} - \frac{1}{4} P_{+} \otimes P_{-} , ~~ 
T^a P_- \otimes T^a P_+ = \frac{1}{2} P_{12} \otimes P_{21} - \frac{1}{4} P_{-} \otimes P_{+} ,
\ee
so that
\bea
(\bar\psi_L U T^a {\cal F}({\cal M}) \psi_R) (\bar\psi_L U T^a {\cal F}({\cal M}) \psi_R) &=& 
\frac{1}{4} ({\bar q}_L U  P_+ {\cal F} ({\cal M}_u) q_R ) ({\bar q}_L U  P_+ {\cal F} ({\cal M}_u) q_R ) 
\nonumber\\
&&\!\!\!\!\!
+ \, \frac{1}{4} ({\bar q}_L U  P_- {\cal F} ({\cal M}_d) q_R ) ({\bar q}_L U  P_- {\cal F} ({\cal M}_d) q_R )
\nonumber\\
&&\!\!\!\!\!
+ \, \frac{1}{4} ({\bar l}_L U  P_- {\cal F} ({\cal M}_e) l_R) ({\bar l}_L U  P_- {\cal F} ({\cal M}_e) l_R)
\nonumber\\
&&\!\!\!\!\!
+ \, ({\bar q}_L U P_{21} {\cal F} ({\cal M}_u) q_R ) ({\bar q}_L U P_{12} {\cal F} ({\cal M}_d) q_R )
\nonumber\\
&&\!\!\!\!\!
- \, \frac{1}{2} ({\bar q}_L U P_+ {\cal F} ({\cal M}_u) q_R ) ({\bar q}_L U  P_- {\cal F} ({\cal M}_d) q_R )
\nonumber\\
&&\!\!\!\!\!
+ \, ({\bar q}_L U P_{21} {\cal F} ({\cal M}_u) q_R ) ({\bar l}_L U P_{12} {\cal F} ({\cal M}_e) l_R )
\nonumber\\
&&\!\!\!\!\!
- \, \frac{1}{2} ({\bar q}_L U P_+ {\cal F} ({\cal M}_u) q_R ) ({\bar l}_L U  P_- {\cal F} ({\cal M}_e) l_R )
\nonumber\\
&&\!\!\!\!\!
+ \, \frac{1}{2} ({\bar q}_L U  P_- {\cal F} ({\cal M}_d) q_R ) ({\bar l}_L U  P_- {\cal F} ({\cal M}_e) l_R)
\nonumber\\
&=&
\frac{1}{4} {\cal F} ({\cal M}_u) \otimes {\cal F} ({\cal M}_u) {\cal O}_{FY1}
+ \frac{1}{4} {\cal F} ({\cal M}_d) \otimes {\cal F} ({\cal M}_d) {\cal O}_{FY3}
\nonumber\\
&&\!\!\!\!\!
+ \, \frac{1}{4} {\cal F} ({\cal M}_e) \otimes {\cal F} ({\cal M}_e) {\cal O}_{FY10}
+ {\cal F} ({\cal M}_u) \otimes {\cal F} ({\cal M}_d) {\cal O}_{ST6}
\nonumber\\
&&\!\!\!\!\!
- \, \frac{1}{2} {\cal F} ({\cal M}_u) \otimes {\cal F} ({\cal M}_d) {\cal O}_{ST5}
+ {\cal F} ({\cal M}_u) \otimes {\cal F} ({\cal M}_e) {\cal O}_{ST10}
\nonumber\\
&&\!\!\!\!\!
- \, \frac{1}{2} {\cal F} ({\cal M}_u) \otimes {\cal F} ({\cal M}_e) {\cal O}_{ST9}
+ \frac{1}{2} {\cal F} ({\cal M}_d) \otimes {\cal F} ({\cal M}_e) {\cal O}_{FY7}
.
\eea

In turn, the structure
$(\bar\psi_L U {\cal F}({\cal M}) \psi_R)(\bar\psi_R {\cal F}({\cal M})^\dagger U^\dagger \psi_L)$
can be simplified using Eq. \rf{obs}:
\bea
(\bar\psi_L U {\cal F}({\cal M}) \psi_R)(\bar\psi_R {\cal F}({\cal M})^\dagger U^\dagger \psi_L)
&=&
[({\bar q}_L U P_+ {\cal F}({\cal M}_u) q_R) + ({\bar q}_L U P_- {\cal F}({\cal M}_d) q_R)
+ ({\bar l}_L U P_- {\cal F}({\cal M}_e) l_R)]
\nonumber\\
&&
\times
[({\bar q}_R {\cal F}({\cal M}_u)^\dagger P_+ U^\dagger q_L) + ({\bar q}_R {\cal F}({\cal M}_d)^\dagger P_- U^\dagger q_L)
+ ({\bar l}_R {\cal F}({\cal M}_e)^\dagger P_- U^\dagger l_L)]
\nonumber\\
&=&
({\bar q}_L U P_+ {\cal F}({\cal M}_u) q_R) ({\bar q}_R {\cal F}({\cal M}_u)^\dagger P_+ U^\dagger q_L)
\nonumber\\
&&\!\!\!\!\!
+ \,
({\bar q}_L U P_- {\cal F}({\cal M}_d) q_R)({\bar q}_R {\cal F}({\cal M}_d)^\dagger P_- U^\dagger q_L)
\nonumber\\
&&\!\!\!\!\!
+ \,
({\bar l}_L U P_- {\cal F}({\cal M}_e) l_R)({\bar l}_R {\cal F}({\cal M}_e)^\dagger P_- U^\dagger l_L)
\nonumber\\
&&\!\!\!\!\!
+ \,
({\bar q}_L U P_+ {\cal F}({\cal M}_u) q_R) ({\bar q}_R {\cal F}({\cal M}_d)^\dagger P_- U^\dagger q_L)
\nonumber\\
&&\!\!\!\!\!
+ \,
({\bar q}_L U P_- {\cal F}({\cal M}_d) q_R) ({\bar q}_R {\cal F}({\cal M}_u)^\dagger P_+ U^\dagger q_L)
\nonumber\\
&&\!\!\!\!\!
+ \,
({\bar q}_L U P_+ {\cal F}({\cal M}_u) q_R) ({\bar l}_R {\cal F}({\cal M}_e)^\dagger P_- U^\dagger l_L)
\nonumber\\
&&\!\!\!\!\!
+ \,
({\bar q}_R {\cal F}({\cal M}_u)^\dagger P_+ U^\dagger q_L) ({\bar l}_L U P_- {\cal F}({\cal M}_e) l_R) 
\nonumber\\
&&\!\!\!\!\!
+ \,
({\bar q}_L U P_- {\cal F}({\cal M}_d) q_R) ({\bar l}_R {\cal F}({\cal M}_e)^\dagger P_- U^\dagger l_L)
\nonumber\\
&&\!\!\!\!\!
+ \,
({\bar q}_R {\cal F}({\cal M}_d)^\dagger P_- U^\dagger q_L) ({\bar l}_L U P_- {\cal F}({\cal M}_e) l_R) 
.
\eea
The terms obtained this way can be decomposed on the operator basis 
through the use of the Fierz identities
\bea
({\bar q}_L U P_+ {\cal F}({\cal M}_u) q_R) ({\bar q}_R {\cal F}({\cal M}_u)^\dagger P_+ U^\dagger q_L)
&=&
- {\cal F}({\cal M}_u) \tilde\otimes {\cal F}({\cal M}_u)^\dagger
\left[
\frac{1}{12} {\cal O}_{LR1} + \frac{1}{2} {\cal O}_{LR2} + \frac{1}{6} {\cal O}_{LR10} +  {\cal O}_{LR11}
\right]
,
\nonumber\\
({\bar q}_L U P_- {\cal F}({\cal M}_d) q_R)({\bar q}_R {\cal F}({\cal M}_d)^\dagger P_- U^\dagger q_L)
&=&
- {\cal F}({\cal M}_d) \tilde\otimes {\cal F}({\cal M}_d)^\dagger
\left[
\frac{1}{12} {\cal O}_{LR3} + \frac{1}{2} {\cal O}_{LR4} - \frac{1}{6} {\cal O}_{LR12} -  {\cal O}_{LR13}
\right]
,
\nonumber\\
({\bar l}_L U P_- {\cal F}({\cal M}_e) l_R)({\bar l}_R {\cal F}({\cal M}_e)^\dagger P_- U^\dagger l_L)
&=&
- {\cal F}({\cal M}_e) \tilde\otimes {\cal F}({\cal M}_e)^\dagger
\left[ \frac{1}{4} {\cal O}_{LR8} - \frac{1}{2} {\cal O}_{LR17} \right]
,
\nonumber\\
({\bar q}_L U P_- {\cal F}({\cal M}_d) q_R) ({\bar l}_R {\cal F}({\cal M}_e)^\dagger P_- U^\dagger l_L)
&=&
- {\cal F}({\cal M}_d) \tilde\otimes {\cal F}({\cal M}_e)^\dagger
\left[ \frac{1}{4} {\cal O}_{LR9} - \frac{1}{2} {\cal O}_{LR18} \right]
,
\eea
so that
\bea
(\bar\psi_L U {\cal F}({\cal M}) \psi_R)(\bar\psi_R {\cal F}({\cal M})^\dagger U^\dagger \psi_L)
&=&
- {\cal F}({\cal M}_u) \tilde\otimes {\cal F}({\cal M}_u)^\dagger
\left[
\frac{1}{12} {\cal O}_{LR1} + \frac{1}{2} {\cal O}_{LR2} + \frac{1}{6} {\cal O}_{LR10} +  {\cal O}_{LR11}
\right]
\nonumber\\
&&\!\!\!
- \,
{\cal F}({\cal M}_d) \tilde\otimes {\cal F}({\cal M}_d)^\dagger
\left[
\frac{1}{12} {\cal O}_{LR3} + \frac{1}{2} {\cal O}_{LR4} - \frac{1}{6} {\cal O}_{LR12} -  {\cal O}_{LR13}
\right]
\nonumber\\
&&\!\!\!
- \,{\cal F}({\cal M}_e) \tilde\otimes {\cal F}({\cal M}_e)^\dagger
\left[ \frac{1}{4} {\cal O}_{LR8} - \frac{1}{2} {\cal O}_{LR17} \right]
\nonumber\\
&&\!\!\!
+ \,
{\cal F}({\cal M}_d) \otimes {\cal F}({\cal M}_u)^\dagger {\cal O}_{FY5}
+
{\cal F}({\cal M}_u) \otimes {\cal F}({\cal M}_d)^\dagger {\cal O}_{FY5}^\dagger
\nonumber\\
&&\!\!\!
+ \,
{\cal F}({\cal M}_e) \otimes {\cal F}({\cal M}_u)^\dagger {\cal O}_{FY9}
+
{\cal F}({\cal M}_u) \otimes {\cal F}({\cal M}_e)^\dagger {\cal O}_{FY9}^\dagger
\nonumber\\
&&\!\!\!
- \, 
{\cal F}({\cal M}_d) \tilde\otimes {\cal F}({\cal M}_e)^\dagger
\left[ \frac{1}{4} {\cal O}_{LR9} - \frac{1}{2} {\cal O}_{LR18} \right]
\nonumber\\
&&\!\!\!
- \, 
{\cal F}({\cal M}_e) \tilde\otimes {\cal F}({\cal M}_d)^\dagger  
\left[ \frac{1}{4} {\cal O}_{LR9}^\dagger - \frac{1}{2} {\cal O}_{LR18}^\dagger \right] .
\eea
Finally, one has
\bea
(\bar\psi_L U T^a {\cal F}({\cal M}) \psi_R)(\bar\psi_R {\cal F}({\cal M})^\dagger T^a U^\dagger \psi_L)
&=&
[({\bar q}_L U T^a P_+ {\cal F}({\cal M}_u) q_R) +
({\bar q}_L U T^a P_- {\cal F}({\cal M}_d) q_R) +
\nonumber\\
&&
+\,
({\bar l}_L U T^a P_- {\cal F}({\cal M}_e) l_R) ]
\times
[
({\bar q}_R {\cal F}({\cal M}_u)^\dagger P_+ T^a U^\dagger q_L)
\nonumber\\
&&
+ \,
({\bar q}_R {\cal F}({\cal M}_d)^\dagger P_- T^a U^\dagger q_L)
+
({\bar l}_R {\cal F}({\cal M}_e)^\dagger P_- T^a U^\dagger l_L)
]
\nonumber\\
&=&
({\bar q}_L U T^a P_+ {\cal F}({\cal M}_u) q_R) ({\bar q}_R {\cal F}({\cal M}_u)^\dagger P_+ T^a U^\dagger q_L)
\nonumber\\
&&\!\!\!\!\!
+ \,
({\bar q}_L U T^a P_- {\cal F}({\cal M}_d) q_R) ({\bar q}_R {\cal F}({\cal M}_d)^\dagger P_- T^a U^\dagger q_L)
\nonumber\\
&&\!\!\!\!\!
+ \,
({\bar l}_L U T^a P_- {\cal F}({\cal M}_e) l_R) ({\bar l}_R {\cal F}({\cal M}_e)^\dagger P_- T^a U^\dagger l_L)
\nonumber\\
&&\!\!\!\!\!
+ \,
({\bar q}_L U T^a P_+ {\cal F}({\cal M}_u) q_R) ({\bar q}_R {\cal F}({\cal M}_d)^\dagger P_- T^a U^\dagger q_L)
\nonumber\\
&&\!\!\!\!\!
+ \,
({\bar q}_L U T^a P_- {\cal F}({\cal M}_d) q_R) ({\bar q}_R {\cal F}({\cal M}_u)^\dagger P_+ T^a U^\dagger q_L)
\nonumber\\
&&\!\!\!\!\!
+ \,
({\bar q}_L U T^a P_+ {\cal F}({\cal M}_u) q_R) ({\bar l}_R {\cal F}({\cal M}_e)^\dagger P_- T^a U^\dagger l_L)
\nonumber\\
&&\!\!\!\!\!
+ \,
({\bar l}_L U T^a P_- {\cal F}({\cal M}_e) l_R) ({\bar q}_R {\cal F}({\cal M}_u)^\dagger P_+ T^a U^\dagger q_L)
\nonumber\\
&&\!\!\!\!\!
+ \,
({\bar q}_L U T^a P_- {\cal F}({\cal M}_d) q_R) ({\bar l}_R {\cal F}({\cal M}_e)^\dagger P_- T^a U^\dagger l_L)
\nonumber\\
&&\!\!\!\!\!
+ \,
({\bar l}_L U T^a P_- {\cal F}({\cal M}_e) l_R) ({\bar q}_R {\cal F}({\cal M}_d)^\dagger P_- T^a U^\dagger q_L) .
\eea
Using now
\be
T^a P_+ \otimes P_+ T^a  = \frac{1}{2} P_{21} \otimes P_{12} + \frac{1}{4} P_+ \otimes P_+ , ~~ 
T^a P_- \otimes P_- T^a  = \frac{1}{2} P_{12} \otimes P_{21} + \frac{1}{4} P_- \otimes P_- , ~~ 
T^a P_\pm \otimes P_\mp T^a = - \frac{1}{4} P_{\pm} \otimes P_{\mp} , 
\ee
together with the previous Fierz identities and the following ones:
\bea
({\bar q}_L U P_{12} {\cal F}({\cal M}_d) q_R) ({\bar q}_R {\cal F}({\cal M}_d)^\dagger P_{21} U^\dagger q_L)
&=&
- {\cal F}({\cal M}_d) \tilde\otimes {\cal F}({\cal M}_d)^\dagger
\left[
\frac{1}{12} {\cal O}_{LR3} + \frac{1}{2} {\cal O}_{LR4} + \frac{1}{6} {\cal O}_{LR12} +  {\cal O}_{LR13}
\right]
,
\nonumber\\
({\bar q}_L U P_{21} {\cal F}({\cal M}_u) q_R)({\bar q}_R {\cal F}({\cal M}_u)^\dagger P_{12} U^\dagger q_L)
&=&
- {\cal F}({\cal M}_u) \tilde\otimes {\cal F}({\cal M}_u)^\dagger
\left[
\frac{1}{12} {\cal O}_{LR1} + \frac{1}{2} {\cal O}_{LR2} - \frac{1}{6} {\cal O}_{LR10} -  {\cal O}_{LR11}
\right]
,
\nonumber\\
({\bar l}_L U P_{12} {\cal F}({\cal M}_e) l_R)({\bar l}_R {\cal F}({\cal M}_e)^\dagger P_{21} U^\dagger l_L)
&=&
- {\cal F}({\cal M}_e) \tilde\otimes {\cal F}({\cal M}_e)^\dagger
\left[ \frac{1}{4} {\cal O}_{LR8} + \frac{1}{2} {\cal O}_{LR17} \right]
,
\nonumber\\
({\bar q}_L U P_{12} {\cal F}({\cal M}_d) q_R) ({\bar l}_R {\cal F}({\cal M}_e)^\dagger P_{21} U^\dagger l_L)
&=&
- {\cal F}({\cal M}_d) \tilde\otimes {\cal F}({\cal M}_e)^\dagger
\left[ \frac{1}{4} {\cal O}_{LR9} + \frac{1}{2} {\cal O}_{LR18} \right]
,
\eea
one obtains
\bea
&& (\bar\psi_L U T^a {\cal F}({\cal M}) \psi_R)(\bar\psi_R {\cal F}({\cal M})^\dagger T^a U^\dagger \psi_L)
=
\nonumber\\
&&\ \ \ \ \ \ \
\frac{1}{2} ({\bar q}_L U P_{21} {\cal F}({\cal M}_u) q_R) ({\bar q}_R {\cal F}({\cal M}_u)^\dagger P_{12} U^\dagger q_L)
+ \,
\frac{1}{4} ({\bar q}_L U P_+ {\cal F}({\cal M}_u) q_R) ({\bar q}_R {\cal F}({\cal M}_u)^\dagger P_+ U^\dagger q_L)
\nonumber\\
&&\ \ \ \
+ \,
\frac{1}{2} ({\bar q}_L U P_{12} {\cal F}({\cal M}_d) q_R) ({\bar q}_R {\cal F}({\cal M}_d)^\dagger P_{21} U^\dagger q_L)
+ \,
\frac{1}{4} ({\bar q}_L U P_- {\cal F}({\cal M}_d) q_R) ({\bar q}_R {\cal F}({\cal M}_d)^\dagger P_- U^\dagger q_L)
\nonumber\\
&&\ \ \ \ 
+ \,
\frac{1}{2} ({\bar l}_L U P_{12} {\cal F}({\cal M}_e) l_R) ({\bar l}_R {\cal F}({\cal M}_e)^\dagger P_{21} U^\dagger l_L)
+ \,
\frac{1}{4} ({\bar l}_L U P_- {\cal F}({\cal M}_e) l_R) ({\bar l}_R {\cal F}({\cal M}_e)^\dagger P_- U^\dagger l_L)
\nonumber\\
&&\ \ \ \
- \,
\frac{1}{4} ({\bar q}_L U P_+ {\cal F}({\cal M}_u) q_R) ({\bar q}_R {\cal F}({\cal M}_d)^\dagger P_- U^\dagger q_L)
- \,
\frac{1}{4} ({\bar q}_L U P_- {\cal F}({\cal M}_d) q_R) ({\bar q}_R {\cal F}({\cal M}_u)^\dagger P_+ U^\dagger q_L)
\nonumber\\
&&\ \ \ \
- \,
\frac{1}{4} ({\bar q}_L U P_+ {\cal F}({\cal M}_u) q_R) ({\bar l}_R {\cal F}({\cal M}_e)^\dagger P_- U^\dagger l_L)
- \,
\frac{1}{4} ({\bar l}_L U P_- {\cal F}({\cal M}_e) l_R) ({\bar q}_R {\cal F}({\cal M}_u)^\dagger P_+ U^\dagger q_L)
\nonumber\\
&&\ \ \ \
+ \,
\frac{1}{2} ({\bar q}_L U P_{12} {\cal F}({\cal M}_d) q_R) ({\bar l}_R {\cal F}({\cal M}_e)^\dagger P_{21} U^\dagger l_L)
+ \,
\frac{1}{4} ({\bar q}_L U P_- {\cal F}({\cal M}_d) q_R) ({\bar l}_R {\cal F}({\cal M}_e)^\dagger P_- U^\dagger l_L)
\nonumber\\
&&\ \ \ \
+ \,
\frac{1}{2} ({\bar l}_L U P_{12} {\cal F}({\cal M}_e) l_R) ({\bar q}_R {\cal F}({\cal M}_d)^\dagger P_{21} U^\dagger q_L)
+ \,
\frac{1}{4} ({\bar l}_L U P_- {\cal F}({\cal M}_e) l_R) ({\bar q}_R {\cal F}({\cal M}_d)^\dagger P_- U^\dagger q_L)
\nonumber\\
&=&
- \, {\cal F}({\cal M}_u) \tilde\otimes {\cal F}({\cal M}_u)^\dagger
\left[
\frac{1}{24} {\cal O}_{LR1} + \frac{1}{4} {\cal O}_{LR2} - \frac{1}{12} {\cal O}_{LR10} - \frac{1}{2} {\cal O}_{LR11}
\right]
\nonumber\\
&&
- \, {\cal F}({\cal M}_u) \tilde\otimes {\cal F}({\cal M}_u)^\dagger
\left[
\frac{1}{48} {\cal O}_{LR1} + \frac{1}{8} {\cal O}_{LR2} + \frac{1}{24} {\cal O}_{LR10} + \frac{1}{4} {\cal O}_{LR11}
\right]
\nonumber\\
&&
- \, {\cal F}({\cal M}_d) \tilde\otimes {\cal F}({\cal M}_d)^\dagger
\left[
\frac{1}{24} {\cal O}_{LR3} + \frac{1}{4} {\cal O}_{LR4} + \frac{1}{12} {\cal O}_{LR12} + \frac{1}{2} {\cal O}_{LR13}
\right]
\nonumber\\
&&
- \, {\cal F}({\cal M}_d) \tilde\otimes {\cal F}({\cal M}_d)^\dagger
\left[
\frac{1}{48} {\cal O}_{LR3} + \frac{1}{8} {\cal O}_{LR4} - \frac{1}{24} {\cal O}_{LR12} - \frac{1}{4} {\cal O}_{LR13}
\right]
\nonumber\\
&&
- \, {\cal F}({\cal M}_e) \tilde\otimes {\cal F}({\cal M}_e)^\dagger
\left[ \frac{1}{8} {\cal O}_{LR8} + \frac{1}{4} {\cal O}_{LR17} \right]
- \, {\cal F}({\cal M}_e) \tilde\otimes {\cal F}({\cal M}_e)^\dagger
\left[ \frac{1}{16} {\cal O}_{LR8} - \frac{1}{8} {\cal O}_{LR17} \right]
\nonumber\\
&&
- \, {\cal F}({\cal M}_d) \tilde\otimes {\cal F}({\cal M}_e)^\dagger
\left[ \frac{1}{8} {\cal O}_{LR9} + \frac{1}{4} {\cal O}_{LR18} \right]
- \frac{1}{4} {\cal F}({\cal M}_d) \otimes {\cal F}({\cal M}_u)^\dagger {\cal O}_{FY5}
\nonumber\\
&&
- \, {\cal F}({\cal M}_d) \tilde\otimes {\cal F}({\cal M}_e)^\dagger
\left[ \frac{1}{16} {\cal O}_{LR9} - \frac{1}{8} {\cal O}_{LR18} \right]
- \frac{1}{4} {\cal F}({\cal M}_u) \otimes {\cal F}({\cal M}_d)^\dagger {\cal O}_{FY5}^\dagger
\nonumber\\
&&
- \, {\cal F}({\cal M}_e) \tilde\otimes {\cal F}({\cal M}_d)^\dagger  
\left[ \frac{1}{8} {\cal O}_{LR9}^\dagger + \frac{1}{4} {\cal O}_{LR18}^\dagger \right]
- \frac{1}{4} {\cal F}({\cal M}_e) \otimes {\cal F}({\cal M}_u)^\dagger   
{\cal O}_{FY9}
\nonumber\\
&&
- \, {\cal F}({\cal M}_e) \tilde\otimes {\cal F}({\cal M}_d)^\dagger   
\left[ \frac{1}{16} {\cal O}_{LR9}^\dagger - \frac{1}{8} {\cal O}_{LR18}^\dagger \right]
- \frac{1}{4} {\cal F}({\cal M}_u) \otimes {\cal F}({\cal M}_e)^\dagger {\cal O}_{FY9}^\dagger
.
\nonumber\\
\eea

These divergences are removed through the renormalization of the operators
${\cal O}_{FYi}$, ${\cal O}_{STi}$, and ${\cal O}_{LRi}$ of ${\cal L}_{\psi^4 Uh}$. 
Explicitly, one has (recall that the functions $F_{FYi}$, $F_{STi}$, and 
$F_{LRi}$ are actually rank-four tensors in generation space)
\bea
\stackrel{{\mbox{\tiny{o}}}}{F}_{FYi} (\eta) &=& F_{FYi} (\eta ; \mu)  
+ \frac{\Gamma_{FYi} (\eta)}{16\pi^2} \mu^{d-4} \left[ \frac{1}{d-4} - \frac{1}{2} (\ln 4\pi - \gamma_E) \right]
,
\nonumber\\
\stackrel{{\mbox{\tiny{o}}}}{F}_{STi} (\eta) &=& F_{STi} (\eta ; \mu)  
+ \frac{\Gamma_{STi} (\eta)}{16\pi^2} \mu^{d-4} \left[ \frac{1}{d-4} - \frac{1}{2} (\ln 4\pi - \gamma_E) \right]
,
\nonumber\\
\stackrel{{\mbox{\tiny{o}}}}{F}_{LRi} (\eta) &=& F_{LRi} (\eta ; \mu)  
+ \frac{\Gamma_{LRi} (\eta)}{16\pi^2} \mu^{d-4} \left[ \frac{1}{d-4} - \frac{1}{2} (\ln 4\pi - \gamma_E) \right]
,
\eea
with
\bea
\Gamma_{FY1} &=& \frac{3F^{-2}}{2v^4} \left(\frac{F'}{2}{\cal M}_u'-{\cal M}_u\right) \otimes \left(\frac{F'}{2}{\cal M}_u'-{\cal M}_u\right)
+ \frac{1}{2 v^4} {\cal M}_u'' \otimes {\cal M}_u'' - \frac{1}{v^4} \left( F^{-1/2} {\cal M}_u  \right)' \otimes \left( F^{-1/2} {\cal M}_u  \right)',
\nonumber\\
\Gamma_{FY3} &=& \frac{3F^{-2}}{2v^4} \left(\frac{F'}{2}{\cal M}_d'-{\cal M}_d\right) \otimes \left(\frac{F'}{2}{\cal M}_d'-{\cal M}_d\right)
+ \frac{1}{2 v^4} {\cal M}_d'' \otimes {\cal M}_d'' - \frac{1}{v^4} \left( F^{-1/2} {\cal M}_d  \right)' \otimes \left( F^{-1/2} {\cal M}_d  \right)',
\nonumber\\
\Gamma_{FY5} &=& \frac{3F^{-2}}{v^4} 
\left(\frac{F'}{2}{\cal M}_d'-{\cal M}_d\right) \otimes
\left(\frac{F'}{2}{\cal M}_u^{\dagger\prime}-{\cal M}_u^\dagger\right)  
+ \frac{1}{v^4} {\cal M}_d'' \otimes {\cal M}_u^{\dagger\prime\prime}  
- \frac{2}{v^4} \left( F^{-1/2} {\cal M}_d  \right)' \otimes
\left( F^{-1/2} {\cal M}_u^\dagger  \right)',  
\nonumber\\
\Gamma_{FY7} &=& \frac{3F^{-2}}{v^4} \left(\frac{F'}{2}{\cal M}_d'-{\cal M}_d\right) \otimes \left(\frac{F'}{2}{\cal M}_e'-{\cal M}_e\right)
+ \frac{1}{v^4} {\cal M}_d'' \otimes {\cal M}_e'' - \frac{2}{v^4} \left( F^{-1/2} {\cal M}_d  \right)' \otimes \left( F^{-1/2} {\cal M}_e  \right)',
\nonumber\\
\Gamma_{FY9} &=& \frac{3F^{-2}}{v^4} 
\left(\frac{F'}{2}{\cal M}_e'-{\cal M}_e\right) \otimes
\left(\frac{F'}{2}{\cal M}_u^{\dagger\prime}-{\cal M}_u^\dagger\right) 
+ \frac{1}{v^4} {\cal M}_e'' \otimes {\cal M}_u^{\dagger\prime\prime}   
- \frac{2}{v^4} \left( F^{-1/2} {\cal M}_e  \right)' \otimes
\left( F^{-1/2} {\cal M}_u^\dagger  \right)',  
\nonumber\\
\Gamma_{FY10} &=& \frac{3F^{-2}}{2v^4} \left(\frac{F'}{2}{\cal M}_e'-{\cal M}_e\right) \otimes \left(\frac{F'}{2}{\cal M}_e'-{\cal M}_e\right)
+ \frac{1}{2 v^4} {\cal M}_e'' \otimes {\cal M}_e'' - \frac{1}{v^4} \left( F^{-1/2} {\cal M}_e  \right)' \otimes \left( F^{-1/2} {\cal M}_e  \right)',
\nonumber\\
\label{gamfy}
\eea

\bea
\Gamma_{ST5} &=& \frac{3F^{-2}}{v^4} \left(\frac{F'}{2}{\cal M}_u'-{\cal M}_u\right) \otimes \left(\frac{F'}{2}{\cal M}_d'-{\cal M}_d\right)
+ \frac{1}{v^4} {\cal M}_u'' \otimes {\cal M}_d'' + \frac{2}{v^4} \left( F^{-1/2} {\cal M}_u  \right)' \otimes \left( F^{-1/2} {\cal M}_d  \right)',
\nonumber\\
\Gamma_{ST6} &=& - \frac{4}{v^4} \left( F^{-1/2} {\cal M}_u  \right)' \otimes \left( F^{-1/2} {\cal M}_d  \right)',
\nonumber\\
\Gamma_{ST9} &=& \frac{3F^{-2}}{v^4} \left(\frac{F'}{2}{\cal M}_u'-{\cal M}_u\right) \otimes \left(\frac{F'}{2}{\cal M}_e'-{\cal M}_e\right)
+ \frac{1}{v^4} {\cal M}_u'' \otimes {\cal M}_e'' + \frac{2}{v^4} \left( F^{-1/2} {\cal M}_u  \right)' \otimes \left( F^{-1/2} {\cal M}_e  \right)',
\nonumber\\
\Gamma_{ST10} &=& - \frac{4}{v^4} \left( F^{-1/2} {\cal M}_u  \right)' \otimes \left( F^{-1/2} {\cal M}_e  \right)',
\label{gamst}
\eea

\bea
\Gamma_{LR1} &=& - \frac{F^{-2}}{4 v^4} 
\left(\frac{F'}{2}{\cal M}_u'-{\cal M}_u\right) \tilde\otimes
\left(\frac{F'}{2}{\cal M}_u^{\dagger\prime}-{\cal M}_u^\dagger\right)  
- \frac{1}{12 v^4} {\cal M}_u'' \tilde\otimes {\cal M}_u^{\dagger\prime\prime}
- \frac{1}{2 v^4} \left( F^{-1/2} {\cal M}_u \right)' \tilde\otimes
\left( F^{-1/2} {\cal M}_u^\dagger  \right)',  
\nonumber\\
\Gamma_{LR2} &=& - \frac{3F^{-2}}{2 v^4} 
\left(\frac{F'}{2}{\cal M}_u'-{\cal M}_u\right) \tilde\otimes
\left(\frac{F'}{2}{\cal M}_u^{\dagger\prime}-{\cal M}_u^\dagger\right)  
- \frac{1}{2 v^4} {\cal M}_u'' \tilde\otimes {\cal M}_u^{\dagger\prime\prime}   
- \frac{3}{v^4} \left( F^{-1/2} {\cal M}_u \right)' \tilde\otimes 
\left( F^{-1/2} {\cal M}_u^\dagger  \right)', 
\nonumber\\
\Gamma_{LR3} &=& - \frac{F^{-2}}{4 v^4} 
\left(\frac{F'}{2}{\cal M}_d'-{\cal M}_d\right) \tilde\otimes
\left(\frac{F'}{2}{\cal M}_d^{\dagger\prime}-{\cal M}_d^\dagger\right)  
- \frac{1}{12 v^4} {\cal M}_d'' \tilde\otimes {\cal M}_d^{\dagger\prime\prime}
- \frac{1}{2 v^4} \left( F^{-1/2} {\cal M}_d \right)' \tilde\otimes
\left( F^{-1/2} {\cal M}_d^\dagger  \right)',  
\nonumber\\
\Gamma_{LR4} &=& - \frac{3 F^{-2}}{2 v^4} 
\left(\frac{F'}{2}{\cal M}_d'-{\cal M}_d\right) \tilde\otimes
\left(\frac{F'}{2}{\cal M}_d^{\dagger\prime}-{\cal M}_d^\dagger\right)  
- \frac{1}{2 v^4} {\cal M}_d'' \tilde\otimes {\cal M}_d^{\dagger\prime\prime}  
- \frac{3}{v^4} \left( F^{-1/2} {\cal M}_d \right)' \tilde\otimes
\left( F^{-1/2} {\cal M}_d^\dagger  \right)', 
\nonumber\\
\Gamma_{LR8} &=& - \frac{3 F^{-2}}{4 v^4} 
\left(\frac{F'}{2}{\cal M}_e'-{\cal M}_e\right) \tilde\otimes
\left(\frac{F'}{2}{\cal M}_e^{\dagger\prime}-{\cal M}_e^\dagger\right)  
- \frac{1}{4 v^4} {\cal M}_e'' \tilde\otimes {\cal M}_e^{\dagger\prime\prime}   
- \frac{3}{2 v^4} \left( F^{-1/2} {\cal M}_e \right)' \tilde\otimes
\left( F^{-1/2} {\cal M}_e^\dagger  \right)',  
\nonumber\\
\Gamma_{LR9} &=& - \frac{3 F^{-2}}{4 v^4} 
\left(\frac{F'}{2}{\cal M}_d'-{\cal M}_d\right) \tilde\otimes
\left(\frac{F'}{2}{\cal M}_e^{\dagger\prime}-{\cal M}_e^\dagger\right)  
- \frac{1}{4 v^4} {\cal M}_d'' \tilde\otimes {\cal M}_e^{\dagger\prime\prime}   
- \frac{3}{2 v^4} \left( F^{-1/2} {\cal M}_d \right)' \tilde\otimes
\left( F^{-1/2} {\cal M}_e^\dagger  \right)',  
\nonumber\\
\Gamma_{LR10} &=& - \frac{F^{-2}}{2 v^4} 
\left(\frac{F'}{2}{\cal M}_u'-{\cal M}_u\right) \tilde\otimes
\left(\frac{F'}{2}{\cal M}_u^{\dagger\prime}-{\cal M}_u^\dagger\right)  
- \frac{1}{6 v^4} {\cal M}_u'' \tilde\otimes {\cal M}_u^{\dagger\prime\prime}   
+ \frac{1}{3 v^4} \left( F^{-1/2} {\cal M}_u \right)' \tilde\otimes
\left( F^{-1/2} {\cal M}_u^\dagger  \right)',  
\nonumber\\
\Gamma_{LR11} &=& - \frac{3 F^{-2}}{v^4} 
\left(\frac{F'}{2}{\cal M}_u'-{\cal M}_u\right) \tilde\otimes
\left(\frac{F'}{2}{\cal M}_u^{\dagger\prime}-{\cal M}_u^\dagger\right)  
- \frac{1}{v^4} {\cal M}_u'' \tilde\otimes {\cal M}_u^{\dagger\prime\prime}   
+ \frac{2}{v^4} \left( F^{-1/2} {\cal M}_u \right)' \tilde\otimes
\left( F^{-1/2} {\cal M}_u^\dagger  \right)',  
\nonumber\\
\Gamma_{LR12} &=& \frac{F^{-2}}{2 v^4} 
\left(\frac{F'}{2}{\cal M}_d'-{\cal M}_d\right) \tilde\otimes
\left(\frac{F'}{2}{\cal M}_d^{\dagger\prime}-{\cal M}_d^\dagger\right)  
+ \frac{1}{6 v^4} {\cal M}_d'' \tilde\otimes {\cal M}_d^{\dagger\prime\prime}   
- \frac{1}{3 v^4} \left( F^{-1/2} {\cal M}_d \right)' \tilde\otimes
\left( F^{-1/2} {\cal M}_d^\dagger  \right)',  
\nonumber\\
\Gamma_{LR13} &=& \frac{3 F^{-2}}{v^4} 
\left(\frac{F'}{2}{\cal M}_d'-{\cal M}_d\right) \tilde\otimes
\left(\frac{F'}{2}{\cal M}_d^{\dagger\prime}-{\cal M}_d^\dagger\right)  
+ \frac{1}{v^4} {\cal M}_d'' \tilde\otimes {\cal M}_d^{\dagger\prime\prime}   
- \frac{2}{v^4} \left( F^{-1/2} {\cal M}_d \right)' \tilde\otimes
\left( F^{-1/2} {\cal M}_d^\dagger  \right)', 
\nonumber\\
\Gamma_{LR17} &=&  \frac{3 F^{-2}}{2 v^4} 
\left(\frac{F'}{2}{\cal M}_e'-{\cal M}_e\right) \tilde\otimes
\left(\frac{F'}{2}{\cal M}_e^{\dagger\prime}-{\cal M}_e^\dagger\right)  
+ \frac{1}{2 v^4} {\cal M}_e'' \tilde\otimes {\cal M}_e^{\dagger\prime\prime}  
- \frac{1}{v^4} \left( F^{-1/2} {\cal M}_e \right)' \tilde\otimes
\left( F^{-1/2} {\cal M}_e^\dagger  \right)',  
\nonumber\\
\Gamma_{LR18} &=&  \frac{3 F^{-2}}{2 v^4} 
\left(\frac{F'}{2}{\cal M}_d'-{\cal M}_d\right) \tilde\otimes
\left(\frac{F'}{2}{\cal M}_e^{\dagger\prime}-{\cal M}_e^\dagger\right)  
+ \frac{1}{2 v^4} {\cal M}_d'' \tilde\otimes {\cal M}_e^{\dagger\prime\prime}  
- \frac{1}{v^4} \left( F^{-1/2} {\cal M}_d \right)' \tilde\otimes
\left( F^{-1/2} {\cal M}_e^\dagger  \right)'  
.
\label{gamlr}
\eea

\indent

\subsection{Renormalization group equations for the coefficients of ${\cal L}_4$}
\label{subsec:rgel4}

In the preceding subsections we obtained the counterterms of the
next-to-leading order Lagrangian ${\cal L}_4$, employing a
systematic decomposition into basis operators. We are now in
a position to derive the renormalization group equations for
the operator coefficients.

Summarizing the previous results, we may write
\begin{equation}\label{l4of}
{\cal L}_4 = \sum_i 
\stackrel{{\mbox{\tiny{o}}}}{{\cal O}}_i \stackrel{{\mbox{\tiny{o}}}}{F}_i =
\sum_i 
{\cal O}_i\left(F_i +\frac{\Gamma_i}{16\pi^2}\frac{1}{d-4}\right)\, \mu^{d-4},
\end{equation}
where the sum extends over all the NLO terms in our basis, comprising the
classes $\beta_1$, $UhD^4$, $\psi^2 UhD$, $\psi^2 UhD^2$, and $\psi^4 Uh$.
The quantities $F_i$ and $\Gamma_i$ are functions of $\eta=h/v$ and, for the
fermionic terms, tensors in generation space.

The second equality in (\ref{l4of}) expresses the unrenormalized coefficients
$\stackrel{{\mbox{\tiny{o}}}}{F}_i$ through their renormalized version plus
counterterms $\sim\Gamma_i$, written here in the MS scheme. 
The counterterms are equal and opposite in sign to the one-loop divergences,
displayed in (\ref{ldivnlo}).
Inspecting the latter, we note that all terms ${\cal O}_i \Gamma_i$ have
canonical dimension exactly $4$, in $d=4-2\varepsilon$ space-time dimensions, 
once in this context we take $g$, $g'$ and $v$ to mean $g\mu^\varepsilon$, 
$g'\mu^\varepsilon$ and $v\mu^{-\varepsilon}$, respectively.
Then their dimensions are $[g]=[g']=\varepsilon$, $[v]=1-\varepsilon$,
and $g$, $g'$ and $v$ become $\mu$-independent at tree level. It follows
that also $d\Gamma_i/d\ln\mu=0$ at tree level.
Since $[{\cal L}_4]=d$, $[{\cal O}_i \Gamma_i]=4$ implies the factor
$\mu^{d-4}$ shown on the right-hand side of (\ref{l4of}).
From the $\mu$-independence of $\stackrel{{\mbox{\tiny{o}}}}{F}_i$ we then
infer the renormalization group equations (in $4$ space-time dimensions)
\begin{equation}\label{rgefgamma}
16\pi^2 \frac{d}{d\ln\mu} F_i = -\Gamma_i .
\end{equation}
The various $\Gamma_i$ are given in the previous sections,
for $i=\beta_1$ in (\ref{gambet1}), $D1$, $D2$, $D7$, $D8$, $D11$ in (\ref{gamd1}), (\ref{gamd2}),
$\psi Vk$ in (\ref{gampsiv}), $\psi Sk$ in (\ref{gampsis}), 
$FYk$ in (\ref{gamfy}), $STk$ in (\ref{gamst}), and $LRk$ in (\ref{gamlr}).
The one-loop beta functions vanish for all couplings
not present in the preceding list.

\section{Comparison with SMEFT}
\label{sec:compsmeft}
\setcounter{equation}{0}

SMEFT \cite{Buchmuller:1985jz,Grzadkowski:2010es} is the effective field
theory formulation of the electroweak and strong interactions, where the 
operators are organized as an expansion according to the their canonical dimension. 
To lowest order (dimension 4), it coincides with the SM. 
Excluding lepton-number violating effects, the leading corrections are given by
operators of dimension 6. 

Even though SMEFT is organized differently than the EWChL, there is an
overlap \cite{Buchalla:2014eca}, which can be used as a cross-check of our calculation.
The one-loop renormalization of the SM at dimension 4 has already been
shown in \cite{Buchalla:2017jlu} to follow as a special case from the
renormalization of the EWChL.
Beyond that, the one-loop divergences of the EWChL 
discussed here also contain the renormalization of those dimension-6 terms in SMEFT
that have chiral dimension 2 and are thus contained in (\ref{l2}).
In the SMEFT basis of \cite{Grzadkowski:2010es}, these terms can be
expressed as
\begin{equation}\label{dl2dim6}
  \Delta{\cal L}_{2,d=6} = \frac{1}{\Lambda^2} \left( C_{\phi\Box} Q_{\phi\Box}
  + C_\phi Q_\phi + C^{rs}_{\psi\phi}  Q^{rs}_{\psi\phi}\right),
\end{equation}  
where
\begin{eqnarray}
  Q_{\phi\Box} &=& \phi^\dagger\phi\, \Box\, \phi^\dagger\phi \label{qbox}\\
  Q_\phi &=& (\phi^\dagger\phi)^3 \label{qphi}\\
  Q^{rs}_{\psi\phi} &=& \phi^\dagger\phi\,
                        (\bar\psi_L(\tilde\phi,\phi))^r\, \psi^s_R\label{qpsiphi}
                        .
\end{eqnarray}  
Here $r$, $s$ are fermion flavor indices and $\phi$ is the Higgs doublet.

By extracting the corresponding terms from the one-loop divergences of the 
EWChL, one should obtain the known one-loop renormalization of SMEFT
\cite{Jenkins:2013zja,Jenkins:2013wua,Alonso:2013hga,Alonso:2014zka}, coming from the
single insertion of the dimension-6 operators (\ref{qbox}) -- (\ref{qpsiphi}).
In order to do this, we need the relations between the non-linear and the
linear representation of the Higgs sector, in particular,
\begin{equation}\label{phieta}
  (\tilde\phi,\phi)=\frac{v}{\sqrt{2}}(1+\eta)U\, ,\qquad
  \phi^\dagger\phi = \frac{v^2}{2} (1+\eta)^2\, ,\qquad
  D^\mu\phi^\dagger D_\mu\phi = \frac{1}{2}\partial^\mu h\partial_\mu h
  +\frac{v^2}{4}\,  \langle L^\mu L_\mu \rangle\, (1+\eta)^2,
\end{equation}  
where $\phi$ is the complex Higgs doublet in the conventional linear
representation, and $\tilde\phi_i =\epsilon_{ij}\phi^*_j$.

The operators in (\ref{qbox}) -- (\ref{qpsiphi}) modify the $\eta$-dependent
functions $F$, $V$ and ${\cal M}$ from their SM form
\begin{equation}\label{fvmsm}
  F_{SM}=(1+\eta)^2\, ,\qquad
  V_{SM}=-\frac{m^2 v^2}{2} (1+\eta)^2 +\frac{\lambda v^4}{8}(1+\eta)^4\, ,
  \qquad {\cal M}_{SM} =\frac{v}{\sqrt{2}}{\cal Y} (1+\eta)\, ,
\end{equation}  
where
${\cal Y}= {\rm diag}\left({\cal Y}_u,{\cal Y}_d,{\cal Y}_\nu,{\cal Y}_e\right)$
collects the Yukawa matrices of the SM.

The simplest case is the renormalization of $C_\phi Q_\phi$. Here
$F$ and ${\cal M}$ take their SM values, while the potential becomes
$V=V_{SM}+\Delta V$, with
\begin{equation}\label{deltavphi}
\Delta V=- \frac{C_\phi}{\Lambda^2}\, \frac{v^6}{8} (1+\eta)^6\, .
\end{equation}  
This term only affects the contribution ${\cal L}^{(0)}_{\rm div}$
in (\ref{eq:Ldiv0}), from which we extract the term of first order in
$C_\phi$,
\begin{equation}\label{ldivphi}
  32\pi^2\varepsilon\, \Delta{\cal L}^\phi_{\rm div} =
  -\frac{C_\phi}{\Lambda^2}(\phi^\dagger \phi)^3\,
  \left[\frac{3}{2}(3 g^2 + g'^2)+54\lambda\right]
  +24\frac{m^2}{\Lambda^2}\, C_\phi\, (\phi^\dagger \phi)^2\, . 
\end{equation}
As discussed e.g. in \cite{Buchalla:2019wsc},
we still need to subtract 
$K_{(\phi)} = 6(3 g^2 + g'^2- \dl {\cal Y}^\dagger {\cal Y}\dr)$
from the term in square brackets, due to the renormalization of $\phi$ inside $Q_\phi$.
We then find 
\begin{equation}\label{betaphilam}
  \beta_\phi \supseteq \left[-\frac{9}{2}(3 g^2 + g'^2)+54\lambda
    + 6 \dl {\cal Y}^\dagger {\cal Y}\dr\right]\, C_\phi\, ,\qquad
  \beta_\lambda \supseteq 48\frac{m^2}{\Lambda^2} C_\phi
\end{equation}  
for the contribution of $C_\phi$ to the beta functions of SMEFT at
dimension 4 and 6, in agreement with 
\cite{Jenkins:2013zja,Jenkins:2013wua,Alonso:2013hga,Alonso:2014zka}
and the compilation in \cite{Celis:2017hod}.
We recall that the beta functions of coefficient $C_i$
are defined as $\beta_i =16\pi^2 dC_i/d\ln\mu$ and the operator
multiplying $\lambda$ is $Q^{(4)}_\lambda=-(\phi^\dagger\phi)^2/2$
\cite{Buchalla:2019wsc}.

We next consider the renormalization of the modified Yukawa term
$C^{rs}_{\psi\phi} Q^{rs}_{\psi\phi}$. In this case, $F$ and $V$ are
the same as in the SM, but ${\cal M}={\cal M}_{SM} + \Delta {\cal M}$,
with
\begin{equation}\label{deltamphi}
\Delta {\cal M} = -\frac{v^3}{2\sqrt{2}\Lambda^2}\, (1+\eta)^3\, C_{\psi\phi}.
\end{equation}  
Here
$C_{\psi\phi}={\rm diag}\left(C_{u\phi},C_{d\phi},C_{\nu\phi},C_{e\phi}\right)$,
where the entries are matrices in generation space.
Working out the terms to first order in $\Delta {\cal M}$ from
${\cal L}_{\rm div}$ in (\ref{eq:Ldiv_decomp}), and expressing the result
through the Higgs-doublet $\phi$, we obtain
\begin{align}\label{ldivpsiphi}
  32\pi^2\varepsilon\, \Lambda^2\, \Delta{\cal L}^{\psi\phi}_{\rm div} &=
4\eta_1 m^2 (\phi^\dagger\phi)^2 + 4(- \eta_1 \lambda + 
\dl C^\dagger_{\psi\phi}{\cal Y}{\cal Y}^\dagger{\cal Y} +{\rm h.c.}\dr)
(\phi^\dagger\phi)^3\nonumber\\
 &+ \Bigl\{ 6 m^2 \bar\psi_L (\tilde\phi,\phi)C_{\psi\phi}\psi_R
 -2\eta_1 \phi^\dagger\phi\, \bar\psi_L (\tilde\phi,\phi){\cal Y}\psi_R
  +4 i \eta_5 \phi^\dagger\phi\, \bar\psi_L (\tilde\phi,\phi){\cal Y}T_3\psi_R
  \nonumber\\
  &-\phi^\dagger\phi\, \bar\psi_L (\tilde\phi,\phi)\left[-6 C_F g^2_s C_{q\phi}
  +\left(\frac{3}{4}(3g^2 + g'^2)-6 Y_L Y_R g'^2 + 12\lambda\right)
  C_{\psi\phi}\right]\psi_R
 \nonumber\\
&-\phi^\dagger\phi\, \bar\psi_L (\tilde\phi,\phi)\left[
7 {\cal Y}{\cal Y}^\dagger C_{\psi\phi} +6 C_{\psi\phi}  {\cal Y}^\dagger{\cal Y}
+ 2 {\cal Y} C^\dagger_{\psi\phi} {\cal Y}                                        
- \langle C_{\psi\phi}  {\cal Y}^\dagger\rangle {\cal Y}             
- 2 \langle {\cal Y} C^\dagger_{\psi\phi}\rangle {\cal Y}                
-\frac{3}{2} \langle {\cal Y}{\cal Y}^\dagger\rangle C_{\psi\phi}\right]\psi_R
\nonumber\\
&+{\rm h.c.}\Bigr\},
\end{align}
where we defined   
$C_{q\phi}={\rm diag}\left(C_{u\phi},C_{d\phi},0,0\right)$
and \cite{Celis:2017hod}
\begin{equation}\label{eta15def}
  \eta_1\equiv
\frac{1}{2} \dl C^\dagger_{\psi\phi}{\cal Y}+{\cal Y}^\dagger C_{\psi\phi}\dr\, ,
\qquad
  i\eta_5\equiv
\dl (C^\dagger_{\psi\phi}{\cal Y}-{\cal Y}^\dagger C_{\psi\phi})T_3\dr .
\end{equation}  
Since the EWChL is formulated explicitly in the broken phase,
in contrast to SMEFT, terms vanishing as $v\to 0$ after expressing
all scalar fields through the doublet $\phi$ have to be omitted in deriving
(\ref{ldivpsiphi}).
The beta-function contributions proportional to $C_{\psi\phi}$
can be read off from (\ref{ldivpsiphi}), once the field renormalization
factor $K_{(\psi\phi)}=3(3 g^2 + g'^2- \dl {\cal Y}^\dagger {\cal Y}\dr)$ has
been subtracted from the coefficient of $-C^{rs}_{\psi\phi} Q^{rs}_{\psi\phi}$.
We find the entries
\begin{equation}\label{betalamyphi}
  \beta_\lambda \supseteq 8\eta_1\frac{m^2}{\Lambda^2}\, ,\qquad
  \beta_{\cal Y} \supseteq 6\frac{m^2}{\Lambda^2} C_{\psi\phi}\, ,\qquad
  \beta_\phi \supseteq  4\eta_1 \lambda
  - 4 \dl C^\dagger_{\psi\phi}{\cal Y}{\cal Y}^\dagger{\cal Y} +{\rm h.c.}\dr ,
\end{equation}  
\begin{align}\label{betapsiphi}
\beta_{\psi\phi} &\supseteq  2 \eta_1 {\cal Y} - 4i \eta_5 {\cal Y}T_3
-6 C_F g^2_s C_{q\phi}
+\left(-\frac{9}{4}(3g^2 + g'^2)-6 Y_L Y_R g'^2 + 12\lambda
  +3\dl {\cal Y}^\dagger {\cal Y}\dr \right) C_{\psi\phi}
\nonumber\\  
&+ 7 {\cal Y}{\cal Y}^\dagger C_{\psi\phi}
  +6 C_{\psi\phi}  {\cal Y}^\dagger{\cal Y}
 + 2 {\cal Y} C^\dagger_{\psi\phi} {\cal Y}                                
- \langle C_{\psi\phi}  {\cal Y}^\dagger\rangle {\cal Y}             
- 2 \langle {\cal Y} C^\dagger_{\psi\phi}\rangle {\cal Y}                
-\frac{3}{2} \langle {\cal Y}{\cal Y}^\dagger\rangle C_{\psi\phi} .
\end{align}  

Finally, we extract the one-loop divergences induced by a single insertion
of the operator $Q_{\phi\Box}$. This case is more complicated than the previous
ones, since $Q_{\phi\Box}$ does not match the canonical form of the
chiral Lagrangian ${\cal L}_2$ in (\ref{l2}). Employing a suitable field
redefinition the desired information can nevertheless be obtained.

The operator $Q_{\phi\Box}$ modifies the SM Lagrangian such that the
kinetic term of $h= v\eta$ becomes
\begin{equation}\label{dellphibox}
{\cal L}_h\equiv\frac{1}{2}\partial_\mu h \partial^\mu h +
\frac{C_{\phi\Box}}{\Lambda^2} Q_{\phi\Box}
=\left(1-2 \frac{C_{\phi\Box}}{\Lambda^2}\, v^2(1+\eta)^2\right)\,
\frac{1}{2} \partial_\mu h \partial^\mu h =
\frac{1}{2} \partial_\mu \tilde h \partial^\mu \tilde h\, ,
\end{equation}
with all other terms unchanged. In the second step, (\ref{qbox}),
(\ref{phieta}) and an integration by parts have been used.
In the last step, the canonical form of the kinetic term is
recovered by means of the field redefinition
\begin{equation}\label{etatilde}
  \eta \doteq \tilde\eta +\frac{C_{\phi\Box}}{\Lambda^2}
  \frac{v^2}{3} (1+\eta)^3\equiv \tilde\eta +\Delta\eta\, ,
\end{equation}  
to first order in $C_{\phi\Box}$.
The functions in (\ref{fvmsm}) can next be expressed as
\begin{equation}\label{fsmtilde}
  F_{SM}(\eta)=F_{SM}(\tilde\eta + \Delta\eta)\doteq F_{SM}(\tilde\eta) +
  \frac{C_{\phi\Box}}{\Lambda^2}\frac{v^2}{3} (1+\eta)^3\, F'_{SM}(\eta)
  \equiv F_{SM}(\tilde\eta) + \Delta F
\end{equation}  
and similarly for $V_{SM}$ and ${\cal M}_{SM}$.
The SM Lagrangian including the $Q_{\phi\Box}$ term can thus be brought
to the form of (\ref{l2}), now written with $\tilde\eta$ replacing $\eta$,
and with the functions
\begin{equation}\label{fvmphibox}
  F(\tilde\eta) = F_{SM}(\tilde\eta) +\Delta F\, ,\qquad
  V(\tilde\eta) = V_{SM}(\tilde\eta) +\Delta V\, ,\qquad
  {\cal M}(\tilde\eta) = {\cal M}_{SM}(\tilde\eta) +\Delta {\cal M}\, .
\end{equation}  
In order to find the one-loop divergences proportional to $C_{\phi\Box}$,
we extract from (\ref{eq:Ldiv_decomp})
\begin{equation}
  {\cal L}_{\rm div}(F_{SM}(\tilde\eta)+\Delta F,V_{SM}(\tilde\eta) +\Delta V,
  {\cal M}_{SM}(\tilde\eta) +\Delta {\cal M})
\end{equation}
the terms to first order in $\Delta F$, $\Delta V$ and $\Delta {\cal M}$.
In addition, we need to re-express the SM limit of the divergences
through the original field $\eta$, which introduces a further
contribution
\begin{equation}
  {\cal L}^{SM}_{\rm div}(\tilde\eta)={\cal L}^{SM}_{\rm div}(\eta-\Delta\eta)
  \doteq {\cal L}^{SM}_{\rm div}(\eta) +{\cal L}^{\Delta\eta}_{\rm div} .
\end{equation}
Adding all four contributions and expressing the scalar fields in terms of
$\phi$, we obtain
\begin{align}\label{ldivphibox}
32\pi^2\varepsilon\, \frac{\Lambda^2}{C_{\phi\Box}}\,
\Delta{\cal L}^{\psi\phi}_{\rm div} &= 8 m^4\phi^\dagger\phi
+ m^2\left(\frac{20}{3}g^2 -32\lambda\right)(\phi^\dagger\phi)^2
- 2 m^2 (\bar\psi_L(\tilde\phi,\phi){\cal Y}\psi_R + {\rm h.c.})
\nonumber\\
&+\left(40\lambda^2-\frac{20}{3}g^2\lambda\right)(\phi^\dagger\phi)^3
-\left(8 g^2 +\frac{8}{3}g'^2 + 12\lambda\right)Q_{\phi\Box}
-\frac{20}{3}g'^2\, Q_{\phi D}\nonumber\\
&+\phi^\dagger\phi\left(\bar\psi_L(\tilde\phi,\phi)\left[
    \left(-\frac{10}{3}g^2 + 2\lambda\right){\cal Y}
    +6{\cal Y} {\cal Y}^\dagger {\cal Y}\right]\psi_R + {\rm h.c.}\right)
\nonumber\\
&+\phi^\dagger i\overset{\text{\footnotesize{$\leftrightarrow$}}}{D}_\mu\phi
\, \bar\psi_R\left(-\frac{g'^2}{3}Y_R+{\cal Y}^\dagger {\cal Y}\tau_3\right)
\gamma^\mu\psi_R  - 2\left(\tilde\phi^\dagger i D_\mu\phi\,
  \bar u_R {\cal Y}^\dagger_u {\cal Y}_d \gamma^\mu d_R + {\rm h.c.}\right)
\nonumber\\
&+\phi^\dagger i\overset{\text{\footnotesize{$\leftrightarrow$}}}{D}_\mu\phi
\, \bar\psi_L\left(-\frac{g'^2}{3}Y_L -\frac{1}{2}
  \langle {\cal Y}{\cal Y}^\dagger\tau_3\rangle\right)\gamma^\mu\psi_L
+\phi^\dagger i\overset{\text{\footnotesize{$\leftrightarrow$}}}{{D^a_\mu}}\phi
\, \bar\psi_L\left(-\frac{g^2}{6} +\frac{1}{2}\langle
  {\cal Y}{\cal Y}^\dagger\rangle\right)\tau^a\gamma^\mu\psi_L .
\end{align}  
Here $\tau^a\equiv 2 T^a$ are the Pauli matrices, and
$Q_{\phi D}=D^\mu\phi^\dagger\phi\, \phi^\dagger D_\mu\phi$.
Subtracting the field renormalization term
$K_{(\phi\Box)}=12 g^2 +4 g'^2 - 4 \dl {\cal Y}^\dagger {\cal Y}\dr$
from the bracketed term in front of $Q_{\phi\Box}$, we read off the 
contribution of $C_{\phi\Box}$ to the SMEFT beta functions:
\begin{equation}\label{betalamybox}
  \beta_{m^2} \supseteq -8\frac{m^4}{\Lambda^2}C_{\phi\Box}\, ,\qquad
  \beta_\lambda \supseteq \left(\frac{40}{3}g^2 -64\lambda\right)
  \frac{m^2}{\Lambda^2}C_{\phi\Box}\, ,\qquad
  \beta_{\cal Y} \supseteq -2\frac{m^2}{\Lambda^2}{\cal Y} C_{\phi\Box}
\end{equation}  
\begin{equation}\label{betaboxbox}
\beta_\phi \supseteq
  \left(\frac{20}{3}g^2\lambda - 40\lambda^2\right) C_{\phi\Box}\, ,\qquad
  \beta_{\phi\Box} \supseteq \left(-4 g^2-\frac{4}{3}g'^2 + 12\lambda
    + 4 \dl {\cal Y}^\dagger {\cal Y}\dr\right) C_{\phi\Box}\, ,\qquad
  \beta_{\phi D} \supseteq \frac{20}{3} g'^2 C_{\phi\Box}
\end{equation}  
\begin{equation}\label{betapsiphibox}
  \beta_{\psi\phi} \supseteq
  \left[\left(\frac{10}{3}g^2 - 2\lambda\right){\cal Y}
     - 6{\cal Y} {\cal Y}^\dagger {\cal Y}\right] C_{\phi\Box}
\end{equation}
\begin{equation}\label{betaphipsi}  
\beta_{\phi\psi} \supseteq 
\left(\frac{g'^2}{3}Y_R -{\cal Y}^\dagger {\cal Y}\tau_3\right)C_{\phi\Box}
\, ,\qquad
\beta_{\phi ud} \supseteq 2 {\cal Y}^\dagger_u {\cal Y}_d C_{\phi\Box}
\end{equation}
\begin{equation}
\beta^{(1)}_{\phi\psi} \supseteq \left(\frac{g'^2}{3}Y_L + \frac{1}{2}
  \langle {\cal Y}{\cal Y}^\dagger\tau_3\rangle\right) C_{\phi\Box}\, ,\qquad
\beta^{(3)}_{\phi\psi} \supseteq \left(\frac{g^2}{6} -\frac{1}{2}\langle
  {\cal Y}{\cal Y}^\dagger\rangle\right) C_{\phi\Box} .
\end{equation}  
All terms are in agreement with the known SMEFT results, as summarized e.g. 
in \cite{Celis:2017hod}.

\section{Brief survey of related literature}
\label{sec:literature}
\setcounter{equation}{0}

Several groups computed subsets of the one-loop renormalization and RGEs of the EWChL in the past 
\cite{Delgado:2013hxa,Delgado:2014jda,Gavela:2014uta,Delgado:2015kxa,
  Guo:2015isa,Alonso:2015fsp}. In our previous paper \cite{Buchalla:2017jlu}, we compared and found
agreement with the one-loop divergences reported in \cite{Guo:2015isa}, which treated the subset
of divergences that arise from scalar loops and their corresponding beta
functions. Previously, \cite{Gavela:2014uta} 
had computed the scalar 1, 2, 3, and 4-point functions at one loop. They focused on the scalar sector,
not only regarding the particles running in the loop, but also regarding the set of operators. 
Since this set is not closed under the application of the equations of motion, there are ambiguities
that make the RGEs listed there hard to compare with our result. 
In addition, the operator $\partial h\partial h F_h(h)$ was considered, 
which is always redundant \cite{Buchalla:2013rka}. However, \cite{Gavela:2014uta} 
compared their results with \cite{Delgado:2013hxa} and claimed agreement in the corresponding limit.

Ref.~\cite{Delgado:2013hxa} and its follow-ups \cite{Delgado:2014jda,Delgado:2015kxa}
considered explicit $VV$ scattering processes and derived the RGEs for the couplings involved. 
These results were soon thereafter corroborated by \cite{Guo:2015isa}. 
In a different approach, \cite{Alonso:2015fsp} considered a geometric formulation of the scalar sector.
They also computed the divergence structure of scalar loops and found full agreement
with \cite{Guo:2015isa}.
This means that all results previous to \cite{Buchalla:2017jlu} have been cross-checked against 
each other, either directly or indirectly. Note that \cite{Guo:2015isa} also compared to the result
in the Higgsless limit
\cite{Longhitano:1980iz,Longhitano:1980tm,Appelquist:1980vg,Herrero:1993nc,Herrero:1994iu} 
and found agreement. 

Shortly after we published \cite{Buchalla:2017jlu}, work on the same subject was
described in \cite{Alonso:2017tdy}.
The authors claim in the journal version of \cite{Alonso:2017tdy}: 
"Gauge bosons have the SM renormalization plus an extra contribution". We find that, after basis
projection (see discussion at the beginning of our section \ref{subsec:decompl1}), 
this is not the case and the gauge couplings just have the one-loop beta functions of the SM. 
This result can be confirmed by looking at the SMEFT RGE (through dimension 6). 
Here the only BSM contribution to the gauge-beta functions comes from 
$C_{\phi X} Q_{\phi X}$, 
an operator structure that is absent in the leading-order EWChL. The absence of new contributions
to the one-loop gauge-beta functions can also be seen by comparing the SM limit with the 
Higgsless limit of our result. They both agree, indicating that only Goldstone-boson and
gauge loops contribute to the one-loop gauge-beta functions, while the Higgs couplings (modified or not)
do not. One can actually check that the contribution of Higgs one-loop diagrams does not lead
to divergences, only to finite pieces.

\section{Vacuum expectation value of $h$ and tadpole counterterm}
\label{sec:applications}
\setcounter{equation}{0}

The potential $V$ for the scalar field $h$ is chosen such that $V'(0)=0$,
and hence the vev of the scalar field $h$ vanishes, $<h>=0$ at lowest order.
This property is not maintained when loop corrections are considered, and
one needs to perform a finite renormalization of the potential in order to
enforce it. This finite renormalization is computed here to illustrate
the application of the EWChL at one loop. 

Employing the background field method, we separate the various fields into a classical or 
background part, and a quantum, fluctuating, part. 
The terms involving only the background fields provide 
the tree-level amplitudes. This is the case, in particular for all the terms
in ${\mathcal L}_4$. The part linear in the quantum fields does not matter,
it vanishes when the classical equations of motion are enforced on the
background fields, and terms with three or more quantum fields are not relevant
for the computation of one-loop amplitudes.
For general one-loop calculations, therefore, only terms exactly quadratic
in the quantum fields, but with arbitrary powers of background fields, are required.
As we are only interested in the Higgs potential here, all background fields,
except for the Higgs, may be dropped in the following.
Specialising to the tadpole contribution, eventually only the terms linear
in the background Higgs field need to be kept.

Hereafter, quantum fields are written with a caret, the fields without carets 
being background fields (except for the ghosts, which are always quantum fields).
The decomposition in terms of background and quantum fields is a linear one,
with the exception of the Goldstone fields, where the split is multiplicative.
Omitting background fields, the Goldstone field reduces to its quantum part, 
for which we use
\be
{\hat U} = e^{2 i {\hat\varphi} F^{-1/2}/v} , \quad {\hat\varphi} = {\hat\varphi}^a T^a
.
\ee

Finally, one also needs to fix the gauge and add the corresponding 
contributions from the ghost fields. For the former, a background-gauge
invariant choice \cite{Dittmaier:1995ee} is made, namely\footnote{At this stage,
the QCD part does not matter, and is therefore omitted for the time being.}
\be
{\mathcal L}_{\rm g.f.} = - \frac{1}{2} \left( \partial_\mu {\hat B}^\mu +
  \frac{g' v}{2} F^{1/2} {\hat\varphi}^3\right)^2
- \langle \left(D^\mu {\hat W}_\mu - \frac{gv}{2} F^{1/2} {\hat\varphi} \right)^2 \rangle
.
\ee
This means that the Faddeev-Popov ghost Lagrangian reads
\be
{\cal L}_{\rm ghosts}
=
 \partial^\mu {\bar c} \partial_\mu c 
+ \frac{g' v}{2} F {\bar c} \left( \frac{g v}{2} c^3 - \frac{g' v}{2} c \right)
+ \partial^\mu {\bar c}^a \partial_\mu c^a
 - \left(\frac{g v}{2}\right)^2  F ({\bar c}^1 c^1 + {\bar c}^2 c^2) 
- \frac{g v}{2} F {\bar c}^3 \left( \frac{g v}{2} c^3 - \frac{g' v}{2} c \right)
.
\ee

In order to compute the Higgs tadpole at one loop,
we first extract from
${\cal L}_2 +{\cal L}_{\rm g.f.} + {\cal L}_{\rm ghosts}$
the contributions quadratic in all quantum fields, linear in $h$, and without any other classical field.
This gives
\bea
{\cal L}_2 + && \hspace{-0.35cm}{\cal L}_{\rm g.f.} + {\cal L}_{\rm ghosts} \ =
\nonumber\\
&&=\ 
- \frac{1}{4} (\partial_\mu {\hat W}^a_\nu - \partial_\nu {\hat W}^a_\mu) 
(\partial^\mu {\hat W}^{a\nu} - \partial^\nu {\hat W}^{a\mu})
- \frac{1}{4} (\partial_\mu {\hat B}_\nu - \partial_\nu {\hat B}_\mu) 
(\partial^\mu {\hat B}^\nu - \partial^\nu {\hat B}^\mu)
\nonumber\\
&&
+ \frac{v^2}{4}  F  
\langle \left(\frac{2}{v} \partial^\mu ({\hat\varphi}F^{-1/2}) + g {\hat W}^\mu  -  g' {\hat B}^\mu T^3 \right) 
\left(\frac{2}{v} \partial_\mu ({\hat\varphi}F^{-1/2}) + g {\hat W}_\mu  - g' {\hat B}_\mu T^3 \right)\rangle
\nonumber\\
&&
+ \frac{1}{2} \partial^\mu {\hat h} \partial_\mu {\hat h} - v^2 V_2 {\hat h}^2 - 3 v V_3 h {\hat h}^2
+ {\bar{\hat\psi}} i\!\! \not\!\partial {\hat\psi} - {\bar{\hat\psi}} ( {\cal M} P_R + {\cal M}^\dagger P_L ) {\hat\psi}
\nonumber\\
&&
- \frac{1}{2} \left( \partial^\mu {\hat B}_\mu + \frac{g' v}{2} F^{1/2} {\hat\varphi}^3  \right)^2
- \frac{1}{2} \left( \partial^\mu {\hat W}^a_\mu - \frac{g v}{2} F^{1/2} {\hat\varphi}^a  \right)
\left( \partial^\nu {\hat W}^a_\nu - \frac{g v}{2} F^{1/2} {\hat\varphi}^a  \right)
\nonumber\\
&&
+ \partial^\mu {\bar c} \partial_\mu c 
+ \frac{g' v}{2} F {\bar c} \left( \frac{g v}{2} c^3 - \frac{g' v}{2} c \right)
+ \partial^\mu {\bar c}^a \partial_\mu c^a
 - \left(\frac{g v}{2}\right)^2  F ({\bar c}^1 c^1 + {\bar c}^2 c^2) 
- \frac{g v}{2} F {\bar c}^3 \left( \frac{g v}{2} c^3 - \frac{g' v}{2} c \right)
+ \cdots
\nonumber\\
&& =\ 
- \frac{1}{2} (\partial_\mu {\hat W}^a_\nu) (\partial^\mu {\hat W}^{a\nu})
+ \frac{F}{2} \left( \frac{gv}{2} \right)^2 {\hat W}^{a\mu}{\hat W}^{a}_\mu
- \frac{1}{2} (\partial_\mu {\hat B}_\nu) (\partial^\mu {\hat B}^{\nu})
+ \frac{F}{2} \left( \frac{g'v}{2} \right)^2 {\hat B}^{\mu}{\hat B}_\mu
- F \frac{g g' v^2}{4} {\hat W}^3_\mu {\hat B}^\mu
\nonumber\\
&&
+ \frac{1}{2} (\partial^\mu {\hat\varphi}^a) (\partial_\mu {\hat\varphi}^a) 
- \frac{F}{2} \left(  \frac{gv}{2} \right)^2 {\hat\varphi}^a {\hat\varphi}^a 
- \frac{F}{2} \left(  \frac{g'v}{2} {\hat\varphi}^3 \right)^2 
- F \frac{g'v}{2} \partial^\mu ({\hat B}_\mu {\hat\varphi}^3 F^{-1/2}) + F \frac{gv}{2} \partial^\mu ({\hat W}^a_\mu {\hat\varphi}^a F^{-1/2})
\nonumber\\
&&
+ \frac{1}{2} \partial^\mu {\hat h} \partial_\mu {\hat h} - v^2 V_2 {\hat h}^2  - 3 v V_3 h {\hat h}^2
- \frac{1}{4F} (\partial^\mu F) \partial_\mu ({\hat\varphi}^a {\hat\varphi}^a)
+ {\bar{\hat\psi}} i \!\!\not\!\partial {\hat\psi} - {\bar{\hat\psi}} ( {\cal M} P_R + {\cal M}^\dagger P_L ) {\hat\psi}
\nonumber\\
&&
+ \partial^\mu {\bar c} \partial_\mu c 
+ \frac{g' v}{2} F {\bar c} \left( \frac{g v}{2} c^3 - \frac{g' v}{2} c \right)
+ \partial^\mu {\bar c}^a \partial_\mu c^a
 - \left(\frac{g v}{2}\right)^2  F ({\bar c}^1 c^1 + {\bar c}^2 c^2) 
- \frac{g v}{2} F {\bar c}^3 \left( \frac{g v}{2} c^3 - \frac{g' v}{2} c \right)
+ \cdots
\nonumber\\
&& =\ 
- \frac{1}{2} (\partial_\mu {\hat W}^1_\nu) (\partial^\mu {\hat W}^{1\nu})
- \frac{1}{2} (\partial_\mu {\hat W}^2_\nu) (\partial^\mu {\hat W}^{2\nu})
+ \frac{F}{2} \left( \frac{gv}{2} \right)^2 ({\hat W}^{1\mu}{\hat W}^{1}_\mu + {\hat W}^{2\mu}{\hat W}^{2}_\mu)
\nonumber\\
&&
- \frac{1}{2} (\partial_\mu {\hat Z}_\nu) (\partial^\mu {\hat Z}^{\nu})
+ \frac{F}{2} \left(\frac{v}{2} \right)^2 (g^2 + g'^2) {\hat Z}^\mu {\hat Z}_\mu
- \frac{1}{2} (\partial_\mu {\hat A}_\nu) (\partial^\mu {\hat A}^{\nu})
+ \frac{1}{2} \partial^\mu {\hat h} \partial_\mu {\hat h} - v^2 V_2 {\hat h}^2 - 3 v V_3 h {\hat h}^2
\nonumber\\
&&
+ \frac{1}{2} (\partial^\mu {\hat\varphi}^1) (\partial_\mu {\hat\varphi}^1)
+ \frac{1}{2} (\partial^\mu {\hat\varphi}^2) (\partial_\mu {\hat\varphi}^2) 
- \frac{F}{2} \left(  \frac{gv}{2} \right)^2 ({\hat\varphi}^1 {\hat\varphi}^1 + {\hat\varphi}^2 {\hat\varphi}^2)
+ \frac{1}{2} (\partial^\mu {\hat\varphi}^3) (\partial_\mu {\hat\varphi}^3)  
- \frac{F}{2} \left(\frac{v}{2} \right)^2 (g^2 + g'^2) {\hat\varphi}^3 {\hat\varphi}^3
\nonumber\\
&&
- F \frac{g'v}{2} \partial^\mu ({\hat B}_\mu {\hat\varphi}^3 F^{-1/2}) + F \frac{gv}{2} \partial^\mu ({\hat W}^a_\mu {\hat\varphi}^a F^{-1/2})
+ \frac{F_1}{4v}  ({\hat\varphi}^a {\hat\varphi}^a) \Box h
+ {\bar{\hat\psi}} i \!\!\not\!\partial {\hat\psi} - {\bar{\hat\psi}} ( {\cal M} P_R + {\cal M}^\dagger P_L ) {\hat\psi}
\nonumber\\
&&
+ \partial^\mu {\bar c}_\gamma \partial_\mu c_\gamma 
+ \partial^\mu {\bar c}^1 \partial_\mu c^1 + \partial^\mu {\bar c}^2 \partial_\mu c^2
- \left(\frac{g v}{2}\right)^2 \!\! F ( {\bar c}^1 c^1 + {\bar c}^2 c^2)
+ \partial^\mu {\bar c_Z} \partial_\mu c_Z - \left(\frac{v}{2}\right)^2 \!\! (g^2 + g'^2) F {\bar c}_Z c_Z
+ \cdots
.~~
\nonumber\\
\label{l2quad}
\eea
In the first equality, the terms of the fourth line represent the contributions from
the gauge-fixing term and the fifth line the contributions from the 
corresponding Faddeev-Popov ghosts. Total derivatives have been dropped.
In the last step, we have introduced mass eigenstates in the gauge sector in the
usual way by defining the combinations
\bea
&&
A_\mu = \frac{g}{\sqrt{g^2 + g'^2}} B_\mu + \frac{g'}{\sqrt{g^2 + g'^2}} W^3_\mu
,\quad
Z_\mu = \frac{g}{\sqrt{g^2 + g'^2}} W^3_\mu - \frac{g'}{\sqrt{g^2 + g'^2}} B_\mu
,
\nonumber\\
&&
c_\gamma = \frac{g}{\sqrt{g^2 + g'^2}} c + \frac{g'}{\sqrt{g^2 + g'^2}} c^3
,\quad
c_Z = \frac{g}{\sqrt{g^2 + g'^2}} c^3 - \frac{g'}{\sqrt{g^2 + g'^2}} c
\, .
\eea
In addition, the last expression involves $\Box h$, where $h$ is the classical 
field, so that one may use its equation of motion:
\be
\Box h = - m^2_h h + \cdots ,\qquad  m^2_h = 2 v^2 V_2\, .
\ee

Taking the background field $h$ to zero in (\ref{l2quad}), we obtain
the kinetic terms for all the quantum fields. Those terms fix the
propagators entering the calculation.
In order to compute the vev of $h$ at one loop, we further need to extract from
(\ref{l2quad}) all the cubic terms that are both linear in the background field $h$
and quadratic in the quantum fields, i.e.
\bea
{\cal L}_2 +  {\cal L}_{\rm g.f.} + {\cal L}_{\rm ghosts} 
&&=\ \cdots + \frac{h}{v} F_1 \bigg[
- \frac{v^2}{2} V_2 ({\hat\varphi}^a {\hat\varphi}^a) 
- \frac{1}{2} \left(  \frac{gv}{2} \right)^2 ({\hat\varphi}^1 {\hat\varphi}^1 + {\hat\varphi}^2 {\hat\varphi}^2)
- \frac{1}{2} \left(\frac{v}{2} \right)^2 (g^2 + g'^2) {\hat\varphi}^3 {\hat\varphi}^3
- 3 v^2 \frac{V_3}{F_1} {\hat h}^2
\nonumber\\
&&\!\!\!
+ \frac{1}{2} \left(  \frac{gv}{2} \right)^2
({\hat W}^1_\mu {\hat W}^{1\mu} + {\hat W}^2_\mu {\hat W}^{2\mu})
+ \frac{1}{2} \left(\frac{v}{2} \right)^2 (g^2 + g'^2) {\hat Z}_\mu {\hat Z}^\mu
-  \frac{g'v}{2} \partial^\mu ({\hat B}_\mu {\hat\varphi}^3) +  \frac{gv}{2} \partial^\mu ({\hat W}^a_\mu {\hat\varphi}^a )
\nonumber\\
&&\!\!\!
- \frac{1}{F_1} {\bar{\hat\psi}} ( {\cal M}_1 P_R + {\cal M}_1^\dagger P_L ) {\hat\psi}
- \left(\frac{g v}{2}\right)^2 \!\!  ( {\bar c}^1 c^1 + {\bar c}^2 c^2) 
- \left(\frac{v}{2}\right)^2 \!\! (g^2 + g'^2) {\bar c}_Z c_Z
\bigg]
+ \cdots
.
\eea
Then one finds
\bea
< h >_{\rm loop} &=& \frac{1}{2v^2 V_2} \times \frac{i F_1}{2v} \int \frac{d^d q}{(2\pi)^d} \bigg[
2 \frac{- v^2 V_2 -  (gv/2)^2}{q^2 - (gv/2)^2} + \frac{- v^2 V_2 - v^2 (g^2 + g'^2)/4}{q^2 - v^2 (g^2 + g'^2)/4}
- 6 v^2 \frac{V_3}{F_1} \frac{1}{q^2 - 2 v^2 V_2}
\nonumber\\
&&
+ d \times 2 \left(\frac{gv}{2}\right)^2 \frac{-1}{q^2 - (gv/2)^2}
+ d \times \left(\frac{v}{2}\right)^2 (g^2 + g'^2) \frac{-1}{q^2 - v^2 (g^2 + g'^2)/4}
\nonumber\\
&&
- 4 (-1) \left(\frac{gv}{2}\right)^2 \frac{1}{q^2 - (gv/2)^2}
- 2 (-1) \left(\frac{v}{2}\right)^2 (g^2 + g'^2) \frac{1}{q^2 - v^2 (g^2 + g'^2)/4}
\nonumber\\
&&
- \frac{2}{F_1} (-1) {\rm tr} \Big[ \left( 
{\cal M}_1 P_R + {\cal M}^\dagger_1 P_L \right)
\left( \not\!q + {\cal M}_0 P_L + {\cal M}_0^\dagger P_R \right)
\Big(
\frac{1}{q^2 - {\cal M}_0 {\cal M}_0^\dagger} P_R
+ \frac{1}{q^2 - {\cal M}_0^\dagger {\cal M}_0 } P_L
\Big)
\Big]
\bigg]
\nonumber\\
&=&
\frac{i}{2v^2 V_2} \times \frac{F_1}{2v} \bigg[
2(1-d) \left(\frac{gv}{2}\right)^2 A \left( \frac{g^2v^2}{4} \right)
+ (1-d) \left(\frac{v}{2}\right)^2 (g^2 + g'^2) A \left( \frac{v^2}{4} (g^2 + g'^2) \right)
\nonumber\\
&&
- 2 v^2 V_2 A \left( \frac{g^2v^2}{4} \right) - v^2 V_2 A \left( \frac{v^2}{4} (g^2 + g'^2) \right)
- 6 v^2 \frac{V_3}{F_1} A \left( 2 v^2 V_2  \right)
\nonumber\\
&&
+ \frac{4}{F_1} \dl {\cal M}_1 {\cal M}_0^\dagger A({\cal M}_0 {\cal M}_0^\dagger) 
+ {\cal M}_1^\dagger {\cal M}_0 A({\cal M}_0^\dagger {\cal M}_0) \dr
\bigg]
.
\eea
The first equality shows, successively, the contributions from the scalar fields
(first line), from the gauge fields (second line), from the ghosts (third line), 
and from the fermion fields (fourth line). The second equality involves the 
dimensionally regularized one-point loop function

\be
A (m^2) = \int \frac{d^4 q}{(2\pi)^4} \frac{1}{q^2 - m^2} \to
\mu^{4-d} \int \frac{d^d q}{(2\pi)^d} \frac{1}{q^2 - m^2} = -i m^2 \frac{1}{16 \pi^2}
\left[ \frac{2}{d-4} - \ln 4\pi + \gamma_E + \ln \frac{m^2}{\mu^2} -1 \right]
.
\ee
On the other hand, in the ${\overline{\rm MS}}$ scheme,
the counterterm $\Delta V=V'\delta$ in (\ref{l2delta}) contributes to $<h>$ as
\bea
<h>_{\Delta V} &\equiv & \frac{1}{2v^2 V_2}\frac{\delta (-\Delta V)}{\delta h}\bigg\vert_{h = 0}
\nonumber\\
&=&
\frac{1}{2v^2 V_2}  \frac{1}{16\pi^2} \left[ \frac{1}{d-4} - \frac{1}{2} (\ln 4\pi - \gamma_E) \right]
\left[\frac{3}{2}(3 g^4 + 2 g^2 g'^2 + g'^4)\frac{v^3}{8} F_1  
+ (3 g^2 + g'^2)\frac{v^3}{4}F_1V_2 
\right.
\nonumber\\
&&
\left. 
+ 12 v^3 V_2 V_3
- \frac{4}{v} \dl {\cal M}_0^\dagger {\cal M}_0 ({\cal M}_1^\dagger {\cal M}_0 + {\cal M}_0^\dagger {\cal M}_1)\dr
\right].
\eea
One thus obtains the finite, but scale-dependent, result 
\bea
< h >_{\rm loop} &+& <h>_{\Delta V} \ =\ 
\frac{1}{2v^2 V_2} \times \frac{1}{16\pi^2}
\bigg[ 
- \frac{3v^3 F_1}{16} g^4 \left(  \ln\frac{g^2v^2}{4 \mu^2} - \frac{1}{3} \right)
- \frac{3v^3 F_1}{32}(g^2 + g'^2)^2 \left( \ln\frac{(g^2 + g'^2)v^2}{4\mu^2} - \frac{1}{3} \right)
\nonumber\\
&&
- \frac{v^3}{8}F_1V_2 \left( 2 g^2 \left( \ln\frac{g^2v^2}{4 \mu^2}-1\right) +
 (g^2+ g'^2) \left(\ln\frac{(g^2 + g'^2)v^2}{4\mu^2} -1\right) \right)
- 6 v^3 V_2 V_3 \left( \ln\frac{2v^2V_2}{\mu^2} - 1 \right)
\nonumber\\
&&
+ \frac{2}{v} \dl {\cal M}_1 {\cal M}_0^\dagger {\cal M}_0 {\cal M}_0^\dagger
\left(\ln\frac{{\cal M}_0 {\cal M}_0^\dagger}{\mu^2} - 1\right) 
+ {\cal M}_1^\dagger {\cal M}_0 {\cal M}_0^\dagger {\cal M}_0
\left(\ln\frac{{\cal M}_0^\dagger {\cal M}_0}{\mu^2} - 1\right) \dr \bigg]
,
\eea
where $(d-1) A(m^2) = 3 A(m^2) - im^2/(8\pi^2)$ has been used.

In order to ensure that $V'(0)$ vanishes at the one-loop level, one thus must
add to $\Delta V$ a tadpole part, $\Delta V \to \Delta V + h T(\mu)$, with 
$T(\mu)$ chosen such that $< h >_{\rm loop} + <h>_{\Delta V} + < h >_{T} =0$.
This requires
\bea
T(\mu) &=& - \frac{1}{16\pi^2}
\bigg[
\frac{3v^3 F_1}{16} g^4 \left(  \ln\frac{g^2v^2}{4 \mu^2} - \frac{1}{3} \right) +
\frac{3v^3 F_1}{32}(g^2 + g'^2)^2\left(\ln\frac{(g^2 + g'^2)v^2}{4\mu^2}-\frac{1}{3}\right)
\nonumber\\
&&
+ \frac{v^3}{8}F_1V_2 \left( 2 g^2 \left( \ln\frac{g^2v^2}{4 \mu^2}-1\right) +
 (g^2+ g'^2) \left(\ln\frac{(g^2 + g'^2)v^2}{4\mu^2} -1\right) \right)
+ 6 v^3 V_2 V_3 \left( \ln\frac{2v^2V_2}{\mu^2} - 1 \right)
\nonumber\\
&&
- \frac{2}{v} \dl {\cal M}_1 {\cal M}_0^\dagger {\cal M}_0 {\cal M}_0^\dagger
\left(\ln\frac{{\cal M}_0 {\cal M}_0^\dagger}{\mu^2} - 1\right)
+ {\cal M}_1^\dagger {\cal M}_0 {\cal M}_0^\dagger {\cal M}_0
\left(\ln\frac{{\cal M}_0^\dagger {\cal M}_0}{\mu^2} - 1\right) \dr \bigg]
.
\eea
We note that the tadpole term $T(\mu)$ corresponds to a finite shift of the
Higgs field, which can be expressed as a finite contribution $\delta_{\rm fin}$
to the divergent parameter $\delta$ introduced in (\ref{l2delta}),
\be
\label{dfintmu}
\delta\to \delta + \delta_{\rm fin}\, ,\qquad
\delta_{\rm fin}\equiv \frac{1}{2 v^3 V_2} T(\mu) \, .
\ee
According to (\ref{l2delta}) this leads to finite corrections in the
effective Lagrangian, which cancel the Higgs tadpole in
all amplitudes to one-loop order.

\section{Conclusions}
\label{sec:concl}
\setcounter{equation}{0}

In this paper we have worked out the one-loop renormalization group equations of the EWChL,
taking as starting point the one-loop divergences given in \cite{Buchalla:2017jlu}.
The transition between the divergent structures of the theory and its beta functions
is most conveniently done
if the results are projected onto a complete basis. In this paper we have used
the conventions adopted in \cite{Buchalla:2013rka} and worked out in detail the one-loop
renormalization of both the leading order and next-to-leading order EWChL. Besides the complete
list of the beta functions, we provide, for completeness, the explicit calculation of the
finite piece needed to enforce the no-tadpole condition at one loop.

One of the most interesting results of our analysis is that the one-loop beta functions of the
gauge couplings happen to be universal, i.e. they are not affected by potential deviations
of the Higgs couplings with respect to their SM values. In order to reach this conclusion in a
transparent way, it is crucial to reduce the divergent operator set to a minimal, non-redundant basis.
Only after this step is done it is possible to directly read off the gauge-beta functions
from the divergent piece in front of the kinetic terms without doing an additional calculation.

Our results have been cross-checked in a number of ways. All the renormalization group equations
correctly reduce to the SM ones in the appropriate limit. The one-loop renormalization of SMEFT has
also been used for comparison. The EWChL and SMEFT are different electroweak EFTs but their one-loop
divergences partly overlap. We have explicitly shown that our results are consistent in this
overlapping sector. Finally, the fact that the computation of the finite tadpole contribution yields
actually a finite result is yet another meaningful cross-check. 

The computation presented in this paper is of relevance for the analyses of Higgs interactions
at the LHC. The EWChL is the right tool to implement consistently the $\kappa$ formalism into
an EFT language. The one-loop renormalization presented in this paper is necessary if one wants
to extend the $\kappa$ formalism to study differential distributions in Higgs processes.        
Various processes of interest will be considered in the future.
The framework is now available to extend their treatment to one loop.

\section*{Acknowledgements}

The work of G.B. has been supported in part by the 
Deutsche Forschungsgemeinschaft (DFG, German Research Foundation)
under grant BU 1391/2-2 (project number 261324988) and by the
DFG under Germany's Excellence Strategy -- EXC-2094 -- 390783311.
The work of O.C. has been supported in part by the Bundesministerium
for Bildung und Forschung (BMBF FSP-105) and by
the Deutsche Forschungsgemeinschaft (DFG, German Research Foundation) under grants FOR 1873 and
396021762 - TRR 257 "Particle Physics Phenomenology after the Higgs Discovery".
The work of M.K. has received partial support from the OCEVU Labex 
(ANR-11-LABX-0060) and the A*MIDEX project (ANR-11-IDEX-0001-02) funded by 
the ``Investissements d'Avenir'' French government program managed by the ANR.
C.K. is supported by the Alexander von Humboldt Foundation and by the
Spanish Government and ERDF funds from the EU Commission (Grants No.
FPA2017-84445-P and SEV-2014-0398). This manuscript has been authored by
Fermi Research Alliance, LLC under Contract No. DE-AC02-07CH11359 with
the U.S. Department of Energy, Office of Science, Office of High Energy
Physics.
For cross-checks of our calculations, the programs FeynCalc \cite{Mertig:1990an,Shtabovenko:2016sxi}
and Mathematica \cite{wolframresearch} proved useful,
as well as the compilation of formulas in \cite{Borodulin:2017pwh}.

\indent 

\indent

\appendix

\setcounter{equation}{0}
\setcounter{section}{0}
\def\theequation{\Alph{section}.\arabic{equation}}

\section{Equations of motion}
\label{sec:eom}

The equations of motion (eom), derived from the leading-order Lagrangian
in (\ref{l2}), are needed in particular to reduce NLO terms
to a set of basis operators (see Appendix \ref{sec:basis}).
Here we collect the eom for the gauge fields $B_\mu$, $W_\mu$,
$G_\mu$, the scalars $\eta=h/v$, $\varphi$, and
the fermions $\psi_{L,R}$:

\begin{eqnarray}
\partial^\mu B_{\mu\nu} &=& g'\left[\bar\psi_L\gamma_\nu Y_L\psi_L +
 \bar\psi_R\gamma_\nu Y_R\psi_R +\frac{v^2}{2}F\langle\tau_L L_\nu\rangle\right] 
\label{eomb}\\
D^\mu W^a_{\mu\nu} &=& g\left[\bar\psi_L\gamma_\nu T^a\psi_L 
 -\frac{v^2}{2}F\langle T^a L_\nu\rangle\right] 
\label{eomw}\\
D^\mu G^A_{\mu\nu} &=& g_s \bar q\gamma_\nu T^A q 
\label{eomg}\\ 
v^2\Box\eta &=& -V'+\frac{v^2}{4}\langle L_\mu L^\mu\rangle F'
                      -\bar\psi m'\psi 
\label{eomh}\\
F\, D^\mu L_\mu &=& -F'\, \partial^\mu\eta L_\mu - \frac{4}{v^2} UT^aU^\dagger\, 
    \left(\bar\psi_L U i T^a{\cal M}\psi_R +{\rm h.c.}\right) 
\label{eoml}\\
i\!\not\!\! D\psi_L &=& U{\cal M}\psi_R \label{eompl}\\
i\!\not\!\! D\psi_R &=& {\cal M}^\dagger U^\dagger\psi_L  \label{eompr}
\end{eqnarray}

\section{Basis of NLO operators}
\label{sec:basis}

In this section we list a basis of NLO operators for
the EWChL, following 
\cite{Buchalla:2013rka,Buchalla:2013eza}.
The NLO terms are the independent operators of chiral
dimension 4, with the field
content and the symmetries of the 
Lagrangian in (\ref{l2}).

In the following we will assume that custodial symmetry breaking takes place
at the electroweak scale. This means that the spurions of custodial symmetry breaking
$\sim\tau_L$ carry chiral dimension. Accordingly, terms of chiral dimension 4 with extra
factors of $\tau_L$, which were kept in the general analysis of NLO operators 
in \cite{Buchalla:2013rka}, will be omitted.
We further assume that tensor currents, e.g. $\bar q\sigma_{\mu\nu}Ur$,
only arise with a chiral dimension larger than 2. This eliminates
operators with tensor currents in \cite{Buchalla:2013rka}
from the list of NLO terms to be considered here.

The NLO operators can then be divided into the classes $UhD^2$,
$UhD^4$, $X^2Uh$, $XUhD^2$, $\psi^2UhD$, $\psi^2UhD^2$ and $\psi^4Uh$.
A well-defined subset of these operators represents the counterterms
necessary to renormalize the one-loop divergences of the EWChL 
calculated in this paper.

All operators arising at chiral dimension 4 in the classes
$UhD^2$, $UhD^4$, $X^2Uh$, $XUhD^2$, $\psi^2UhD$ and $\psi^2UhD^2$
are listed below. All the operators in the classes
$UhD^2$, $UhD^4$, $\psi^2UhD$ and $\psi^2UhD^2$
are needed as counterterms. On the other hand, no operator in class $X^2Uh$ or $XUhD^2$
is required to absorb one-loop divergences.
Within the class $\psi^4Uh$, we only list the operators that actually
appear as (divergent) counterterms.

\subsection*{Class \boldmath $UhD^2$:}

\begin{equation}\label{uhd2}
{\cal O}_{\beta_1}= - v^2\,\langle\tau_L L_\mu\rangle \langle\tau_L L^\mu\rangle .
\end{equation}
This operator has only two derivatives,
but its coefficient comes with two powers of the weak coupling $g'$,
related to custodial symmetry breaking.
In total, the term has chiral dimension 4 and enters at
NLO in the EFT.

\subsection*{Class \boldmath $UhD^4$:}

\begin{equation}\label{uhd4}
\begin{array}{lll}
{\cal O}_{D1} = \l L_\mu L^\mu \r^2 ,
&{\cal O}_{D2} = \l L_\mu L_\nu\r \ \l L^\mu L^\nu\r   
& \\
\rule{0cm}{0.6cm}
{\cal O}_{D7} = \l L_\mu L^\mu \r 
  \ \partial_\nu \eta\, \partial^\nu \eta,\hspace*{0.4cm} 
&{\cal O}_{D8} = \l L_\mu L_\nu \r 
  \ \partial^\mu \eta\, \partial^\nu \eta,\hspace*{0.4cm}
&{\cal O}_{D11} = (\partial_\mu \eta\, \partial^\mu \eta)^2. 
\end{array}
\end{equation}

All these operators are CP even.

\subsection*{Class \boldmath $X^2Uh$:}

The CP-even operators are 
\begin{align}\label{xhev}
{\cal{O}}_{Xh1}&=g^{\prime 2} B_{\mu\nu} B^{\mu\nu} \, F_{Xh1}(h)\nonumber\\
{\cal{O}}_{Xh2}&=g^2 \langle W_{\mu\nu} W^{\mu\nu}\rangle \, F_{Xh2}(h)\nonumber\\
  {\cal{O}}_{Xh3}&=\frac{g^2_s}{2}\,
                   G^\alpha_{\mu\nu} G^{\alpha\,\mu\nu} \, F_{Xh3}(h).
\end{align}
 
The CP-odd operators read
\begin{align}\label{xhod}
{\cal{O}}_{Xh4}&=g^{\prime 2} \varepsilon_{\mu\nu\lambda\rho}
B^{\mu\nu} B^{\lambda\rho} \, F_{Xh4}(h)\nonumber\\
{\cal{O}}_{Xh5}&=g^2 \varepsilon_{\mu\nu\lambda\rho}
\langle W^{\mu\nu} W^{\lambda\rho}\rangle \, F_{Xh5}(h)\nonumber\\
{\cal{O}}_{Xh6}&=\frac{g^2_s}{2}\, \varepsilon^{\mu\nu\lambda\rho}
  G^\alpha_{\mu\nu} G^\alpha_{\lambda\rho}\, F_{Xh6}(h).
\end{align}

Here
\begin{equation}\label{fxuh}
F_{Xi}(h)=\sum^\infty_{n=1}f_{Xi,n}\left(\frac{h}{v}\right)^n .
\end{equation}

\subsection*{Class \boldmath $XUhD^2$:}

CP-even operators: 
\begin{align}\label{xuev}
{\cal{O}}_{XU1}&=g^{\prime}gB_{\mu\nu}\langle W^{\mu\nu}\tau_L\rangle
\, (1+F_{XU1}(h))\nonumber\\
{\cal{O}}_{XU7}&=ig^{\prime}B_{\mu\nu}\langle\tau_L[L^{\mu},L^{\nu}]
\rangle \, F_{XU7}(h)\nonumber\\
{\cal{O}}_{XU8}&=ig\langle W_{\mu\nu}[L^{\mu},L^{\nu}]
\rangle \, F_{XU8}(h).
\end{align}
CP-odd operators:
\begin{align}\label{xuod}
{\cal{O}}_{XU4}
&=g^{\prime}g\varepsilon_{\mu\nu\lambda\rho}\langle \tau_L W^{\mu\nu}\rangle
B^{\lambda\rho}\, (1+F_{XU4}(h))\nonumber\\
{\cal{O}}_{XU10}
&=ig^{\prime}\varepsilon_{\mu\nu\lambda\rho}B^{\mu\nu}\langle\tau_L[L^{\lambda},L^{\rho}]
\rangle \, F_{XU10}(h)\nonumber\\
{\cal{O}}_{XU11}
&=ig\varepsilon_{\mu\nu\lambda\rho}\langle W^{\mu\nu}[L^{\lambda},L^{\rho}]
\rangle \, F_{XU11}(h),
\end{align}
with $F_{Xi}(h)$ as in (\ref{fxuh}).

\subsection*{Class \boldmath $\psi^2UhD$:}

\begin{equation}\label{psi2uhd}
\begin{array}{ll}
{\cal O}_{\psi V1}=-\bar q_L\gamma^\mu q_L\ \l \tau_L L_\mu \r
\quad & {\cal O}_{\psi V2}=-\bar q_L\gamma^\mu \tau_L q_L\ \l\tau_L L_\mu \r\\
\rule{0cm}{0.6cm}
{\cal O}_{\psi V3}=-\bar q_L\gamma^\mu U P_{12} U^\dagger q_L\ 
\l L_\mu U P_{21}U^\dagger\r \quad & {\cal O}_{\psi V4}=-\bar u_R\gamma^\mu u_R\ \l \tau_L L_\mu \r\\
\rule{0cm}{0.6cm}
{\cal O}_{\psi V5}=-\bar d_R\gamma^\mu d_R\ \l\tau_L L_\mu \r  \quad & {\cal O}_{\psi V6}=-\bar u_R\gamma^\mu d_R\ 
\l L_\mu UP_{21} U^\dagger\r\\
\rule{0cm}{0.6cm}
{\cal O}_{\psi V7}=-\bar l_L\gamma^\mu l_L\ \l\tau_L L_\mu \r
\quad & {\cal O}_{\psi V8}=-\bar l_L\gamma^\mu \tau_L  l_L\  \l\tau_L L_\mu \r\\
\rule{0cm}{0.6cm}
{\cal O}_{\psi V9}=-\bar l_L\gamma^\mu U P_{12} U^\dagger l_L\ 
\l L_\mu U P_{21} U^\dagger\r \quad & {\cal O}_{\psi V10}=-\bar e_R\gamma^\mu e_R\ \l\tau_L L_\mu \r ,
\end{array}
\vspace*{0.2cm}
\end{equation}
together with the hermitian conjugates ${\cal O}^\dagger_{\psi V3}$, ${\cal O}^\dagger_{\psi V6}$
and ${\cal O}^\dagger_{\psi V9}$.

\subsection*{Class \boldmath $\psi^2UhD^2$:}

\begin{equation}
\begin{array}{lll}
\label{psi2uhd2s}
\mathcal{O}_{\psi S1} = \bar{q}_L U P_{+} q_R \langle L_{\mu}L^{\mu}\rangle 
&\mathcal{O}_{\psi S2} = \bar{q}_L U P_{-} q_R \langle L_{\mu}L^{\mu}\rangle
&\mathcal{O}_{\psi S7} =\bar{l}_L U P_{-} l_R \langle L_{\mu}L^{\mu}\rangle \\
\rule{0cm}{0.6cm}
\mathcal{O}_{\psi S10} = \bar{q}_L U P_{+} q_R \langle\tau_{L}L_{\mu}\rangle 
\partial^{\mu}\eta
&\mathcal{O}_{\psi S11} = \bar{q}_L U P_{-} q_R \langle\tau_{L}L_{\mu}\rangle 
\partial^\mu \eta 
&\mathcal{O}_{\psi S12} = \bar{q}_L U P_{12} q_R 
\langle U P_{21}U^{\dagger}L_{\mu}\rangle \partial^{\mu}\eta\\
\rule{0cm}{0.6cm}
\mathcal{O}_{\psi S13} = \bar{q}_L U P_{21} q_R 
\langle U P_{12}U^{\dagger}L_{\mu}\rangle\partial^{\mu}\eta
&\mathcal{O}_{\psi S14} = \bar{q}_L U P_{+} q_R 
\partial_{\mu}\eta\, \partial^{\mu}\eta
&\mathcal{O}_{\psi S15} = \bar{q}_L U P_{-} q_R 
\partial_{\mu}\eta\, \partial^{\mu}\eta\\ 
\rule{0cm}{0.6cm}
\mathcal{O}_{\psi S16} = \bar{l}_L U P_{-} l_R \langle\tau_{L}L_{\mu}\rangle 
\partial^{\mu}\eta
&\mathcal{O}_{\psi S17} = \bar{l}_L U P_{12} l_R 
\langle U P_{21}U^{\dagger}L_{\mu}\rangle \partial^{\mu}\eta
&\mathcal{O}_{\psi S18} = \bar{l}_L U P_{-} l_R 
\partial_{\mu}\eta\, \partial^{\mu}\eta .
\end{array}
\end{equation}
For this class, hermitean conjugate versions have not been
listed separately.

\subsection*{Class $\psi^4Uh$:}

Here we list the four-fermion operators that are generated as one-loop
counterterms of the EWChL. The complete
basis can be found in \cite{Buchalla:2012qq}.
Note that some of the $ST$-type operators originally listed there are
redundant, as pointed out in Appendix A.4 of \cite{Krause:2018cwe}.
This redundancy doesn't affect the terms $ST5$, $ST6$, $ST9$, $ST10$
appearing below.
Generation indices are suppressed. $T^A$ denotes the generators of color $SU(3)$.

\begin{equation}
\begin{array}{ll}                                              
{\cal O}_{LR1}=\bar q_L\gamma^\mu q_L\, \bar u_R\gamma_\mu u_R\, ,\qquad \quad          
&{\cal O}_{LR2}=\bar q_L\gamma^\mu T^A q_L\, \bar u_R\gamma_\mu T^A u_R \\            
\rule{0cm}{0.6cm}                                            
{\cal O}_{LR3}=\bar q_L\gamma^\mu q_L\, \bar d_R\gamma_\mu d_R\, ,\qquad \quad      
&{\cal O}_{LR4}=\bar q_L\gamma^\mu T^A q_L\, \bar d_R\gamma_\mu T^A d_R \\           
\rule{0cm}{0.6cm}                               
{\cal O}_{LR8}=\bar l_L\gamma^\mu l_L\, \bar e_R\gamma_\mu e_R\, ,\qquad \quad 
&{\cal O}_{LR9}=\bar q_L\gamma^\mu l_L\, \bar e_R\gamma_\mu d_R \\             
\rule{0cm}{0.6cm}                                             
{\cal O}_{LR10}=
\bar q_L\gamma^\mu UT_3U^\dagger q_L\, \bar u_R\gamma_\mu u_R\, ,\qquad \quad 
&{\cal O}_{LR11}=\bar q_L\gamma^\mu T^A UT_3U^\dagger q_L\, \bar u_R\gamma_\mu T^A u_R\\
\rule{0cm}{0.6cm}                                            
{\cal O}_{LR12}=
\bar q_L\gamma^\mu UT_3U^\dagger q_L\, \bar d_R\gamma_\mu d_R\, ,\qquad \quad          
&{\cal O}_{LR13}=\bar q_L\gamma^\mu T^A UT_3U^\dagger q_L\, \bar d_R\gamma_\mu T^A d_R\\
\rule{0cm}{0.6cm}                                              
{\cal O}_{LR17}=
\bar l_L\gamma^\mu UT_3U^\dagger l_L\, \bar e_R\gamma_\mu e_R\, ,\qquad \quad 
&{\cal O}_{LR18}=\bar q_L\gamma^\mu UT_3U^\dagger l_L\, \bar e_R\gamma_\mu d_R     
\end{array}
\vspace*{0.25cm}
\end{equation}

\begin{equation}
\begin{array}{ll}                                           
{\cal O}_{ST5}=\bar q_L UP_+ q_R\, \bar q_L UP_- q_R\, ,\qquad \qquad \qquad \qquad \qquad
&{\cal O}_{ST6}=\bar q_L UP_{21} q_R\, \bar q_L UP_{12} q_R\\
\rule{0cm}{0.6cm}                                       
{\cal O}_{ST9}=\bar q_L UP_+ q_R\, \bar l_L UP_- l_R\, ,\qquad \qquad \qquad \qquad \qquad
&{\cal O}_{ST10}=\bar q_L UP_{21} q_R\, \bar l_L UP_{12} l_R
\end{array}
\vspace*{0.25cm}
\end{equation}

\begin{equation}
\begin{array}{ll}                                             
{\cal O}_{FY1}=\bar q_L UP_+ q_R\, \bar q_L UP_+ q_R\, ,\qquad \qquad \qquad \qquad \qquad
&{\cal O}_{FY3}=\bar q_L UP_- q_R\, \bar q_L UP_- q_R\\
\rule{0cm}{0.6cm}                                          
{\cal O}_{FY5}=\bar q_L UP_- q_R\, \bar q_R P_+ U^\dagger q_L\, ,\qquad \qquad \qquad \qquad \qquad
&{\cal O}_{FY7}=\bar q_L UP_- q_R\, \bar l_L UP_- l_R\\
\rule{0cm}{0.6cm}                                          
{\cal O}_{FY9}=\bar l_L UP_- l_R\, \bar q_R P_+ U^\dagger q_L\, ,\qquad \qquad \qquad \qquad \qquad
&{\cal O}_{FY10}=\bar l_L UP_- l_R\, \bar l_L UP_- l_R .
\end{array}
\vspace*{0.2cm}
\end{equation}

\section{One-loop divergences}
\label{sec:oneloopdiv}

This appendix gathers the explicit expressions for the
complete one-loop divergences of the EWChL
obtained in Ref. \cite{Buchalla:2017jlu}. In terms of the decomposition 
introduced in Eq. \rf{Ldiv_decomp}, they read:
\bea
{\cal L}_{\rm div}^{(1)} &=& 
- \frac{1}{16\pi^2} \frac{1}{d-4} \bigg\{
\frac{g^2_s}{2}\frac{11 N_c - 2 N_f}{3} \, G^{\alpha\mu\nu}G_{\mu\nu}^\alpha  
+ \frac{g^2}{3} \left[ 22 -\frac{\kappa^2+1}{4} - (N_c+1) N_{\rm g} \right] \langle W^{\mu\nu}W_{\mu\nu}\rangle
\nonumber\\
&&
\qquad\qquad\,\quad
- \, \frac{g'^2}{4} \left[ \left(\frac{22}{27} N_c + 2\right) N_{\rm g} + \frac{\kappa^2+1}{6} \right] B^{\mu\nu} B_{\mu\nu}
\nonumber\\
&&
\qquad\qquad\,\quad
+ \, \frac{\kappa^2-1}{6} gg' \langle\tau_L W^{\mu\nu}\rangle B_{\mu\nu}
-\frac{\kappa^2-1}{12}\left(ig\langle W^{\mu\nu}[L_\mu,L_\nu]\rangle
+ig' B^{\mu\nu}\langle\tau_L [L_\mu,L_\nu]\rangle\right)
\nonumber\\
&&
\qquad\qquad\,\quad
- \, \frac{\kappa\kappa'}{3}\partial^\mu\eta\left(g\langle W_{\mu\nu}L^\nu\rangle
-g' B_{\mu\nu}\langle\tau_L L^\nu\rangle\right)
\bigg\}
,
\lbl{Ldiv1}
\eea
\bea
{\cal L}_{\rm div}^{(0)} &=& 
- \frac{1}{16\pi^2} \frac{1}{d-4} \bigg\{
- \frac{1}{2} \left[ g'^2 (\kappa^2 + 3) \frac{v^2F}{4} + 3 g^2 (\kappa^2 + 1) \frac{v^2F}{2} 
+ (\kappa^2 - 1) \frac{F'V'}{Fv^2} - \frac{V'' F}{v^2}{\cal B} 
- 2 \dl {\cal M}^\dagger {\cal M}\dr \right] 
\langle L^\mu L_\mu\rangle
\nonumber\\
&&
+ \frac{1}{4} \left[ (3g^2+g'^2) v^2 ( F {\cal B} - 4 \kappa^2 )
+ 6 \frac{F'V'}{F v^2} {\cal B} \right] \partial^\mu\eta \partial_\mu\eta
+ 2\dl\partial^\mu{\cal M}^\dagger \partial_\mu{\cal M}\dr 
\nonumber\\
&&
+\frac{3}{2}(3 g^4 + 2 g^2 g'^2 + g'^4)\frac{v^4}{16} F^2  
+\frac{3 g^2 + g'^2}{8}F'V'+
\frac{3}{8}\left(\frac{F'V'}{F v^2}\right)^2
+ \frac{1}{2}\left(\frac{V''}{v^2}\right)^2
- 2 \dl ({\cal M}^\dagger {\cal M})^2\dr \nonumber\\
&&
+\frac{(\kappa^2-1)^2}{6} \langle L_\mu L_\nu\rangle \langle L^\mu L^\nu\rangle
+\left(\frac{(\kappa^2-1)^2}{12}+\frac{F^2{\cal B}^2}{8}\right)
\langle L^\mu L_\mu\rangle^2 
+\frac{2}{3}\kappa'^2\langle L_\mu L_\nu\rangle \partial^\mu\eta\partial^\nu\eta
\nonumber\\
&&-\left((\kappa^2-1){\cal B}+\frac{\kappa'^2}{6}\right)
\langle L^\mu L_\mu\rangle \partial^\nu\eta\partial_\nu\eta
+\frac{3}{2}{\cal B}^2(\partial^\mu\eta\partial_\mu\eta)^2
+ \frac{3}{4} g'^2 v^2 (1-\kappa^2)F\,\langle\tau_LL^\mu\rangle\langle\tau_LL_\mu\rangle
\nonumber\\
&&
+ 4i \dl (\partial^\mu {\mathcal M}^\dagger {\mathcal M}
- {\mathcal M}^\dagger \partial^\mu {\mathcal M}) T^3 \dr
\langle \tau_L L_\mu \rangle
\bigg\}
.
\lbl{Ldiv0}
\eea
Here $\dl\ldots\dr$ denotes the trace over isospin, as well as generation
and color indices, in distinction to $\langle\ldots\rangle$, which
refers to the trace over isospin indices only.

\bea
{\cal L}_{\rm div}^{(1/2)} &=& 
- \frac{1}{16\pi^2} \frac{1}{d-4} \bigg\{
\bar\psi_L\left(\frac{3}{2}g^2+2 g'^2 Y^2_L\right)i\!\not\!\! D\psi_L
+\bar\psi_R\, 2g'^2 Y^2_R i\!\not\!\! D\psi_R\nonumber\\
&&
+ 2 g^2_s C_F \, \bar q\left(i\!\not\!\! D 
  -4(U{\cal M}_q P_R+{\cal M}^\dagger_q U^\dagger P_L)\right)q\nonumber\\
&&+\frac{V''}{v^4}\left(\bar\psi_LU {\cal M}'' \psi_R +{\rm h.c.}\right)
-8 g'^2\left(\bar\psi_L Y_L U{\cal M}Y_R\psi_R +{\rm h.c.}\right)
\nonumber\\
&&+\left((3g^2+g'^2)\frac{v^2}{4} F+\frac{3}{2}\frac{F'V'}{F v^2}\right)
\frac{F^{-1}}{v^2}
\left(
\bar\psi_LU\left(\frac{F'}{2}{\cal M}'-{\cal M}\right)\psi_R +{\rm h.c.}
\right)
\nonumber\\
&&-\frac{8}{v^2}F^{-1}\left(\bar\psi_L U T^a{\cal M}{\cal M}^\dagger T^a
   {\cal M}\psi_R +{\rm h.c.}\right)
+\frac{2}{v^2}\left(\bar\psi_L U{\cal M}'{\cal M}^\dagger{\cal M}'\psi_R
   +{\rm h.c.}\right)
\nonumber\\
&&+\langle L^\mu L_\mu\rangle
\left[\frac{F{\cal B}}{2v^2}\bar\psi_L U{\cal M}''\psi_R
-\frac{\kappa^2-1}{Fv^2}
\bar\psi_L U\left(\frac{F'}{2}{\cal M}'-{\cal M}\right)\psi_R +{\rm h.c.}
\right]\nonumber\\
&&+\frac{2\kappa'}{v^2}\partial^\mu\eta\left(
i\bar\psi_L L_\mu U(F^{-1/2}{\cal M})'\psi_R +{\rm h.c.}\right)\nonumber\\
&&+\frac{3{\cal B}}{Fv^2} \partial^\mu\eta\partial_\mu\eta\left(\bar\psi_L U
\left(\frac{F'}{2}{\cal M}'-{\cal M}\right)\psi_R +{\rm h.c.}\right)
\nonumber\\
&&+\frac{3F^{-2}}{2v^4}\left(\bar\psi_L U
\left(\frac{F'}{2}{\cal M}'-{\cal M}\right)\psi_R +{\rm h.c.}\right)^2
+\frac{1}{2v^4}\left(\bar\psi_L U{\cal M}''\psi_R +{\rm h.c.}\right)^2
\nonumber\\
&&+\frac{4}{v^4}\left(i\bar\psi_L U T^a
\left(F^{-1/2} {\cal M}\right)'\psi_R +{\rm h.c.}\right)^2 
\nonumber\\
&&
+ 
\frac{4}{v^2}\bar\psi_LUT^a{\cal M}F^{-1/2} i\!\not\! \partial
\left({\cal M}^\dagger F^{-1/2}\right)T^a U^\dagger\psi_L\nonumber\\
&&
+\frac{4}{v^2}\bar\psi_L UT^a{\cal M}{\cal M}^\dagger F^{-1} 
   T^a U^\dagger i\!\not\!\! D\psi_L 
+\frac{1}{v^2}\bar\psi_L \not\!\! L 
  U{\cal M}{\cal M}^\dagger U^\dagger F^{-1}\psi_L \nonumber\\
&&
+\frac{1}{v^2}\bar\psi_L U{\cal M}' i\!\not\! \partial{\cal M}'^\dagger 
    U^\dagger \psi_L 
+\frac{1}{v^2}\bar\psi_L U{\cal M}'{\cal M}'^\dagger U^\dagger 
  (i\!\not\!\! D +\not\!\! L) \psi_L
\nonumber\\
&&-\frac{\kappa}{v^2}F^{-1/2}\left(\bar\psi_L U {\cal M}'{\cal M}^\dagger 
   U^\dagger\not\!\! L\psi_L +{\rm h.c.}\right) \nonumber\\
&&+\frac{3}{v^2}\bar\psi_R{\cal M}^\dagger F^{-1/2} 
  i\!\not\!\! D\left({\cal M}F^{-1/2}\psi_R\right)
+\frac{1}{v^2}\bar\psi_R{\cal M}'^\dagger 
  i\!\not\!\! D\left({\cal M}'\psi_R\right)
  \nonumber\\
&&
-\frac{F^{-1}}{v^2}\bar\psi_R{\cal M}^\dagger 
 U^\dagger \not\!\! L U{\cal M}\psi_R 
-\frac{1}{v^2}\bar\psi_R{\cal M}'^\dagger 
 U^\dagger \not\!\! L U{\cal M}'\psi_R\nonumber\\ 
&&+\frac{\kappa}{v^2}F^{-1/2}\left(\bar\psi_R{\cal M}^\dagger U^\dagger
   \not\!\! L U{\cal M}'\psi_R +{\rm h.c.}\right)
  \bigg\}
  .
\lbl{Ldivpsi}
\eea
In these expressions, one needs to separate the contributions that renormalize the lowest-order
Lagrangian ${\cal L}_2$ from those which require genuine next-to-leading order counterterms. 
Moreover, in the latter, one may use the equations of motion from ${\cal L}_2$. In the cases of
${\cal L}_{\rm div}^{(0)}$ and ${\cal L}_{\rm div}^{(1)}$, this is straightforward. In the
case of ${\cal L}_{\rm div}^{(1/2)}$, some additional manipulations are required. First, one 
may use the identity
\be
U T^a  {\cal M}{\cal M}^\dagger T^a U^\dagger = \frac{1}{2} \langle  {\cal M}{\cal M}^\dagger \rangle 
- \frac{1}{4} U {\cal M}{\cal M}^\dagger U^\dagger
.
\ee
Next, one needs to put ${\cal L}_{\rm div}^{(1/2)}$ into a manifestly hermitian form.
This can be done by adding total-derivative terms. Specifically, using the equations
of motion at leading order
\be
i\!\not\!\! D\psi_L = U {\cal M} \psi_R , \quad i\!\not\!\! D\psi_R = {\cal M}^\dagger U^\dagger \psi_L
,
\ee
one establishes the identities
\bea
\frac{3}{v^2}\bar\psi_R{\cal M}^\dagger F^{-1/2} i\!\not\!\! D\left({\cal M}F^{-1/2}\psi_R\right)
&=&
\frac{3i}{2 v^2} \partial_\mu \left( F^{-1} \bar\psi_R \gamma^\mu {\cal M}^\dagger {\cal M} \psi_R \right)
+
\frac{3 F^{-1}}{2 v^2}\bar\psi_R 
\left( {\cal M}^\dagger i\!\not\! \partial {\cal M} - i\!\not\! \partial {\cal M}^\dagger {\cal M} \right) \psi_R
\nonumber\\
&&
+ 
\frac{3 F^{-1}}{2 v^2} \left( \bar\psi_R {\cal M}^\dagger {\cal M} {\cal M}^\dagger U^\dagger \psi_L +
\bar\psi_L U {\cal M} {\cal M}^\dagger {\cal M} \psi_R \right)
,
\eea
\bea
\frac{1}{v^2} \bar\psi_R {\cal M}'^\dagger i\!\not\!\! D\left({\cal M}'\psi_R\right) &=&
\frac{i}{2 v^2} \partial_\mu \left( \bar\psi_R \gamma^\mu {\cal M}'^\dagger {\cal M}' \psi_R\right)
+
\frac{1}{2 v^2} \bar\psi_R \left( {\cal M}'^\dagger i\!\not\! \partial {\cal M}'
- i\!\not\! \partial {\cal M}'^\dagger {\cal M}' \right) \psi_R
\nonumber\\
&&
+
\frac{1}{2 v^2} \left( \bar\psi_R {\cal M}'^\dagger {\cal M}' {\cal M}^\dagger U^\dagger \psi_L
+ \bar\psi_L U {\cal M} {\cal M}'^\dagger {\cal M}' \psi_R \right)
,
\eea
\bea
\frac{1}{v^2} \bar\psi_L U{\cal M}' (i\!\not\! \partial{\cal M}'^\dagger ) U^\dagger \psi_L 
+\frac{1}{v^2} \bar\psi_L U{\cal M}'{\cal M}'^\dagger U^\dagger (i\!\not\!\! D + \not\!\! L) \psi_L
&=&
\frac{i}{2 v^2} \partial_\mu \left( \bar\psi_L U{\cal M}' {\cal M}'^\dagger U^\dagger \gamma^\mu \psi_L \right)
\nonumber\\
&&
+ \frac{1}{2 v^2}  \bar\psi_L U \left( {\cal M}' i\!\not\! \partial {\cal M}'^\dagger 
- i\!\not\! \partial{\cal M}' {\cal M}'^\dagger \right) U^\dagger \psi_L
\nonumber\\
&&
+
\frac{1}{2 v^2} 
\bar\psi_L \left(
U {\cal M}'{\cal M}'^\dagger U^\dagger \not\!\! L + \not\!\! L U {\cal M}'{\cal M}'^\dagger U^\dagger
\right) \psi_L
\\
&&
+
\frac{1}{2 v^2} \left(
\bar\psi_L U {\cal M}' {\cal M}'^\dagger {\cal M} \psi_R +
\bar\psi_R {\cal M}^\dagger {\cal M}' {\cal M}'^\dagger U^\dagger \psi_L
\right)
,
\nonumber
\eea
\bea
\frac{2 F^{-1/2}}{v^2} \bar\psi_L \langle {\cal M} i \not\!\partial ({\cal M}^\dagger F^{-1/2} ) \rangle \psi_L
+ \frac{2 F^{-1}}{v^2} \bar\psi_L \langle {\cal M} {\cal M}^\dagger \rangle i \not\!\!D \psi_L
&=&
\frac{i}{v^2} \partial_\mu \left( F^{-1}  \bar\psi_L \gamma^\mu \langle {\cal M} {\cal M}^\dagger \rangle \psi_L \right)
\nonumber\\
&&\!\!
+ \frac{F^{-1}}{v^2} \bar\psi_L \langle {\cal M} i \! \not\!\partial {\cal M}^\dagger 
- i \! \not\!\partial {\cal M} {\cal M}^\dagger \rangle \psi_L
\\
&&\!\!
+
\frac{F^{-1}}{v^2} \bar\psi_L U \langle {\cal M} {\cal M}^\dagger \rangle {\cal M} \psi_R
+
\frac{F^{-1}}{v^2} \bar\psi_R {\cal M}^\dagger \langle {\cal M} {\cal M}^\dagger \rangle U^\dagger \psi_L
,
\nonumber
\eea
\bea
- \frac{F^{-1/2}}{v^2} \bar\psi_L U {\cal M} i \not\!\partial ({\cal M}^\dagger F^{-1/2}) U^\dagger \psi_L
\!&-&\! \frac{F^{-1}}{v^2} \bar\psi_L U {\cal M}  {\cal M}^\dagger U^\dagger i \not\!\!D \psi_L
+ \frac{F^{-1}}{v^2} \bar\psi_L \not\!\!L  U {\cal M}  {\cal M}^\dagger U^\dagger \psi_L
\ =
\nonumber\\
&=&
- \frac{i}{2 v^2} \partial_\mu \left( F^{-1}  \bar\psi_L \gamma^\mu U {\cal M} {\cal M}^\dagger U^\dagger \psi_L \right)
\nonumber\\
&&
- \frac{F^{-1}}{2 v^2} \bar\psi_L U \left( {\cal M} i \! \not\!\partial {\cal M}^\dagger
- i \! \not\!\partial {\cal M} {\cal M}^\dagger\right) U^\dagger \psi_L
\nonumber\\
&&
+ \frac{F^{-1}}{2 v^2} \bar\psi_L \left( \not\!\!L U {\cal M} {\cal M}^\dagger U^\dagger
+ U {\cal M} {\cal M}^\dagger U^\dagger \not\!\!L \right) \psi_L
\nonumber\\
&&
- \frac{F^{-1}}{2 v^2} \bar\psi_L U {\cal M} {\cal M}^\dagger {\cal M} \psi_R 
- \frac{F^{-1}}{2 v^2} \bar\psi_R {\cal M}^\dagger {\cal M} {\cal M}^\dagger U^\dagger \psi_L
,
\eea
so that, upon dropping the total derivatives, ${\cal L}_{\rm div}^{(1/2)}$ may be rewritten as
\bea
{\cal L}_{\rm div}^{(1/2)} &=& 
- \frac{1}{16\pi^2} \frac{1}{d-4} \bigg\{
\bar\psi_L\left(\frac{3}{2}g^2+2 g'^2 Y^2_L\right)i\!\not\!\! D\psi_L
+\bar\psi_R\, 2g'^2 Y^2_R i\!\not\!\! D\psi_R\nonumber\\
&&
+ 2 g^2_s C_F \, \bar q\left(i\!\not\!\! D 
-4(U{\cal M}_q P_R+{\cal M}^\dagger_q U^\dagger P_L)\right)q
\nonumber\\
&&+\frac{V''}{v^4}\left(\bar\psi_LU {\cal M}'' \psi_R +{\rm h.c.}\right)
-8 g'^2\left(\bar\psi_L Y_L U{\cal M}Y_R\psi_R +{\rm h.c.}\right)
\nonumber\\
&&
+\left((3g^2+g'^2)\frac{v^2}{4} F+\frac{3}{2}\frac{F'V'}{F v^2}\right)
\frac{F^{-1}}{v^2}
\left(
\bar\psi_L U\left(\frac{F'}{2}{\cal M}'-{\cal M}\right)\psi_R +{\rm h.c.}
\right)
\nonumber\\
&&
+ \frac{3}{v^2}F^{-1}\left(\bar\psi_L U {\cal M}{\cal M}^\dagger {\cal M}\psi_R + {\rm h.c.}\right)
-\frac{3}{v^2}F^{-1}\left(\bar\psi_L U \langle{\cal M}{\cal M}^\dagger \rangle {\cal M}\psi_R + {\rm h.c.}\right)
\nonumber\\
&&
+
\frac{2}{v^2}\left(\bar\psi_L U{\cal M}'{\cal M}^\dagger{\cal M}'\psi_R + {\rm h.c.}\right)
+
\frac{1}{2 v^2} \left( \bar\psi_L U {\cal M} {\cal M}'^\dagger {\cal M}' \psi_R + {\rm h.c.} \right)
+
\frac{1}{2 v^2} \left(
\bar\psi_L U {\cal M}' {\cal M}'^\dagger {\cal M} \psi_R + {\rm h.c.} \right)
\nonumber\\
&&
+
\frac{3 F^{-1}}{2 v^2}\bar\psi_R 
\left( {\cal M}^\dagger i\!\not\! \partial {\cal M} - i\!\not\! \partial {\cal M}^\dagger {\cal M} \right) \psi_R
+
\frac{1}{2 v^2} \bar\psi_R \left( {\cal M}'^\dagger i\!\not\! \partial {\cal M}'
- i\!\not\! \partial {\cal M}'^\dagger {\cal M}' \right) \psi_R
\nonumber\\
&&
- \frac{F^{-1}}{2 v^2} \bar\psi_L U \left( {\cal M} i \! \not\!\partial {\cal M}^\dagger
- i \! \not\!\partial {\cal M} {\cal M}^\dagger\right) U^\dagger \psi_L
+ \frac{1}{2 v^2}  \bar\psi_L U \left( {\cal M}' i\!\not\! \partial {\cal M}'^\dagger 
- i\!\not\! \partial{\cal M}' {\cal M}'^\dagger \right) U^\dagger \psi_L
\nonumber\\
&&
+ \frac{F^{-1}}{v^2} \bar\psi_L \langle {\cal M} i \! \not\!\partial {\cal M}^\dagger 
- i \! \not\!\partial {\cal M} {\cal M}^\dagger \rangle \psi_L
-\frac{F^{-1}}{v^2}\bar\psi_R{\cal M}^\dagger 
 U^\dagger \not\!\! L U{\cal M}\psi_R 
-\frac{1}{v^2}\bar\psi_R{\cal M}'^\dagger 
 U^\dagger \not\!\! L U{\cal M}'\psi_R
\nonumber\\
&&
+\frac{\kappa}{v^2}F^{-1/2}\left(\bar\psi_R{\cal M}^\dagger U^\dagger
   \not\!\! L U{\cal M}'\psi_R +{\rm h.c.}\right)
-\frac{\kappa}{v^2}F^{-1/2}\left(\bar\psi_L U {\cal M}'{\cal M}^\dagger 
   U^\dagger\not\!\! L\psi_L +{\rm h.c.}\right)
\nonumber\\
&&
+
\frac{1}{2 v^2} 
\bar\psi_L \left(
U {\cal M}'{\cal M}'^\dagger U^\dagger \not\!\! L + \not\!\! L U {\cal M}'{\cal M}'^\dagger U^\dagger
\right) \psi_L
+ \frac{F^{-1}}{2 v^2} \bar\psi_L \left( \not\!\!L U {\cal M} {\cal M}^\dagger U^\dagger
+ U {\cal M} {\cal M}^\dagger U^\dagger \not\!\!L \right) \psi_L
\nonumber\\
&&+\langle L^\mu L_\mu\rangle
\left[\frac{F{\cal B}}{2v^2}\bar\psi_L U{\cal M}''\psi_R
-\frac{\kappa^2-1}{Fv^2}
\bar\psi_L U\left(\frac{F'}{2}{\cal M}'-{\cal M}\right)\psi_R +{\rm h.c.}
\right]\nonumber\\
&&+\frac{2\kappa'}{v^2}\partial^\mu\eta\left(
i\bar\psi_L L_\mu U(F^{-1/2}{\cal M})'\psi_R +{\rm h.c.}\right)\nonumber\\
&&+\frac{3{\cal B}}{Fv^2} \partial^\mu\eta\partial_\mu\eta\left(\bar\psi_L U
\left(\frac{F'}{2}{\cal M}'-{\cal M}\right)\psi_R +{\rm h.c.}\right)
\nonumber\\
&&+\frac{3F^{-2}}{2v^4}\left(\bar\psi_L U
\left(\frac{F'}{2}{\cal M}'-{\cal M}\right)\psi_R +{\rm h.c.}\right)^2
+\frac{1}{2v^4}\left(\bar\psi_L U{\cal M}''\psi_R +{\rm h.c.}\right)^2
\nonumber\\
&&+\frac{4}{v^4}\left(i\bar\psi_L U T^a
\left(F^{-1/2} {\cal M}\right)'\psi_R +{\rm h.c.}\right)^2 
\bigg\}
.
\lbl{Ldiv1/2}
\eea

\indent


\end{document}